\documentclass[11pt,oneside]{book}%if you are printing on two sides use %"twoside"
\usepackage{times}
\usepackage{amsmath}
\usepackage{amssymb}
\usepackage{mathrsfs}
\usepackage{array}
\usepackage{latexsym}
\usepackage{setspace}
\usepackage{subfigure}
\usepackage{epsf}
\usepackage{psfrag}
\usepackage{rotating}
\usepackage{url}
\usepackage{fancybox}
\usepackage{appendix}
\usepackage[usenames]{color}
\usepackage{index}
\usepackage{rotate}
\usepackage{lscape}

\newtheorem{theorem}{Theorem}[chapter]

\usepackage{graphicx}

\begin{document}
%%%%%%%%%%%%%%%%%%%%%%%%%%%%%%%%%%%%%%%%%%%%%%%%%%%%%%%%%%%%%%%%%%%%%%%%%%%%%%
%%%%%%%%%%%%%%%%%%%%%%%%%%%%%%%%%%%%%%%%%%%%%%%%%%%%%%%%%%%%%%%%%%%%%%%%%%%%%%
%%%%%%%%%%%%%%%%%%%%%%%%%%%%%%%%%%%%%%%%%%%%%%%%%%%%%%%%%%%%%%%%%%%%%%%%%%%%%%
\pagestyle{empty}
\begin{center}
\begin{doublespace}
   {\Huge \bf \textcolor{Black}{Characterization of entanglement in multiqubit systems via spin squeezing }}\\
\end{doublespace}
   \vspace{5ex}
   {\large{\bf Thesis submitted to Bangalore University \\
   in partial fulfillment for the degree of }}\\
   \vspace{7ex}
   {\huge{\bf{\color{Black}\bf Doctor  of  Philosophy}}}\\
   \vspace{5ex}
   {\huge{\bf{\color{Black}\bf in Physics}}}\\
   \vspace{7ex}
   { {\color{Black} by}}\\
   \vspace{.5cm}
      {\Large{\bf{\color{Black} Uma. M. S.}}}\\
   \vspace{7ex}
   {\color{Black} Under the supervision of}\\
   \vspace{3ex}
   {\large\bf
   {\color{Black} Dr. A. R. USHA DEVI}}\\
   Department of Physics,\\
   Bangalore University,\\
   Bangalore, India.\\ 
   
     \vspace{3cm}
	September 2008
	\end{center}
	\pagebreak

 %%%%%%%%%%%%%%%%%%%%%%%%%%%%%%%%%%%%%%%%%%%%%%%%%%%%%%%%%%%%%%%%%%%%%%%%%%%%%%
%%%%%%%%%%%%%%%%%%%%%%%%%%%%%%%%%%%%%%%%%%%%%%%%%%%%%%%%%%%%%%%%%%%%%%%%%%%%%%
%%%%%%%%%%%%%%%%%%%%%%%%%%%%%%%%%%%%%%%%%%%%%%%%%%%%%%%%%%%%%%%%%%%%%%%%%%%%%%
\newpage
\textcolor{White}{Theoretical studies on the non-classicality \\ of quantum state}
\vskip 1.2in
\newpage
\textcolor{White}{Theoretical studies on the non-classicality \\ of quantum state}
\vskip 1.2in
\begin{center}
 \textsf{{\LARGE DEDICATED\,\,TO\,\, MY}}\\ 
 \vskip .2in
  {\Huge\textsf{Parents,}\\
  \vskip.4in
  \,\,\textsl{Shri.M. Srinivas Rao}}\\
  \vskip .3in
 \textsf{{\LARGE AND}}\\
 \vskip.3in
  {\Huge\textsl{Smt. Annapoorna}}
  \end{center}
\newpage
%\rule{0mm}{30mm}
%\begin{center}
%\pagestyle{empty}
\textcolor{White}{Theoretical studies on the non-classicality \\ of quantum state}
\vskip 1.2in
%\rule{0mm}{30mm}

 \pagebreak
%%%%%%%%%%%%%%%%%%%%%%%%%%%%%%%%%%%%%%%%%%%%%%%%%%%%%%%%%%%%%%%%%%%%%%%%%%%%%%
%%%%%%%%%%%%%%%%%%%%%%%%%%%%%%%%%%%%%%%%%%%%%%%%%%%%%%%%%%%%%%%%%%%%%%%%%%%%%%
%%%%%%%%%%%%%%%%%%%%%%%%%%%%%%%%%%%%%%%%%%%%%%%%%%%%%%%%%%%%%%%%%%%%%%%%%%%%%%

\begin{flushleft}
\LARGE{ {\bf Acknowledgments}}
\end{flushleft}
It is my pleasure to thank all those who have enabled me to accomplish this dissertation.

In the first place, I would like to express my gratitude to my supervisor
Dr. A. R. Usha Devi, for giving me an opportunity to work under her guidance. I was totally new to the field of %%@
Quantum information theory and it was her valuable teaching
and supervision that helped me gain confidence to pursue my research work. 
I would like to thank her for sharing with me a lot of her expertise and research insight. She has devoted a lot of %%@
her precious time editing my thesis and has made many  suggestions which indeed helped me improve this thesis. She has %%@
supported and encouraged me through extremely difficult times during the course of my research and for that I am %%@
deeply indebted to her. I sincerely thank her for everything she did to me. 

I would like to thank my co-worker R. Prabhu for inspiring conversations and a fruitful collaboration. I really %%@
cherish the enthusiastic discussions we had during our research work.

I am grateful to Professor Ramani, Chairperson, Department of Physics, Bangalore University, Bangalore, for %%@
introducing me to my supervisor Dr. A. R. Usha Devi. 
I would also like to extend my thanks to the previous Chairmen of the Department - Professor M. C. Radha Krishna and %%@
Professor Puttaraja 
for providing me the necessary facilities.

I thank all the faculty members and research scholars of the Department of Physics, Bangalore University, Bangalore, %%@
for the help received from them during the different stages of my research work.

I specially thank my husband for his  patience, and cooperation during the course of my work. I have received an %%@
enormous support from him and his constant encouragement has helped me in more than many ways over the last four %%@
years. This work would not have been possible without his unconditional support.

My special gratitude is due to my  parents, my brother and my sister and all my family members who have helped me %%@
immensely during the my research period. I am indebted to my mother for her advice and encouragement during difficult %%@
times. She has always been a constant source of inspiration to me. My warmest thanks to her. A final word of thanks to %%@
my brother-In-law who has extended his support during the process of my thesis writing.
 
This research was made possible by the help and support of many people. I am grateful to them all.
 
I gratefully acknowledge CSIR, New Delhi for the award of Senior Research Fellowship. 

%%%%%%%%%%%%%%%%%%%%%%%%%%%%%%%%%%%%%%%%%%%%%%%%%%%%%%%%%%%%%%%%%%%%%%%%%%%%%%
%%%%%%%%%%%%%%%%%%%%%%%%%%%%%%%%%%%%%%%%%%%%%%%%%%%%%%%%%%%%%%%%%%%%%%%%%%%%%%
%%%%%%%%%%%%%%%%%%%%%%%%%%%%%%%%%%%%%%%%%%%%%%%%%%%%%%%%%%%%%%%%%%%%%%%%%%%%%%

{\bf \textcolor{white}{Declaration}}\\
{\bf \textcolor{white}{Declaration}}\\
{\bf \textcolor{white}{Declaration}}\\
\vskip 0.55in

\pagebreak
%%%%%%%%%%%%%%%%%%%%%%%%%%%%%%%%%%%%%%%%%%%%%%%%%%%%%%%%%%%%%%%%%%%%%%%%%%%%%%
%%%%%%%%%%%%%%%%%%%%%%%%%%%%%%%%%%%%%%%%%%%%%%%%%%%%%%%%%%%%%%%%%%%%%%%%%%%%%%
%%%%%%%%%%%%%%%%%%%%%%%%%%%%%%%%%%%%%%%%%%%%%%%%%%%%%%%%%%%%%%%%%%%%%%%%%%%%%%
\mainmatter
   \pagenumbering{roman}
   \pagestyle{plain}
   \doublespacing
 		\tableofcontents
          \clearpage
   \pagestyle{myheadings}
   \pagenumbering{arabic}

%%%%%%%%%%%%%%%%%%%%%%%%%%%%%%%%%%%%%%%%%%%%%%%%%%%%%%%%%%%%%%%%%%%%%%%%%%%%%%
%%%%%%%%%%%%%%%%%%%%%%%%%%%%%%%%%%%%%%%%%%%%%%%%%%%%%%%%%%%%%%%%%%%%%%%%%%%%%%
%%%%%%%%%%%%%%%%%%%%%%%%%%%%%%%%%%%%%%%%%%%%%%%%%%%%%%%%%%%%%%%%%%%%%%%%%%%%%%
\chapter{Introduction}
\label{Introduction}
\markboth{}{Introduction}
 
Quantum world opens up several puzzling aspects that are not amenable to classical\break
intuition. Correlation exhibited by subsystems of a composite quantum state is one such
striking feature and has been a source of philosophical debates - following the famous
Einstein-Podolsky-Rosen discussion on the foundational aspects of quantum theory. It is
now well established that entangled states play a crucial role in the modern quantum
information science, including quantum cryptography~\cite{Ek91}, quantum communication and quantum %%@
computation~\cite{Niel00,Bou00,Div95}.

Considerable interest has been evinced
recently~\cite{Wineland92,Kuz97,Kuz98,Pol99,Hald99,Kuz00,Duan00,Koz00,Sor101,
Ficek02,Duan03}
in producing, controlling and manipulating entangled multiqubit systems due 
to the possibility of applications in atomic interferometry~\cite{Yur86,Kit91}, 
high precession atomic clocks~\cite{Wineland94}, quantum computation and 
quantum information processing~\cite{Niel00}.  Multiqubit systems, which are
symmetric under permutation of the particles, allow for an elegant
description in terms of collective variables of the system.
Specifically, if we have $N$  qubits, each qubit may be
represented as a spin-$\frac{1}{2}$ system  and theoretical analysis in
 terms of collective spin operator
$\vec{J}=\frac{1}{2}\displaystyle\sum_{\alpha=1}^N\ \vec{\sigma}_\alpha$ \ \
($\vec{\sigma}_\alpha$ denote the Pauli spin operator of the $\alpha^{\rm th}$
qubit), leads to  reduction of the dimension of the Hilbert space from $2^N$ to
$(N+1)$, when the  multiqubit system respects exchange symmetry. A large number of
experimentally relevant multiqubit states exhibit symmetry under interchange of qubits,
facilitating a significant simplification in understanding the properties of
the physical system.
While complete characterization of multiqubit entanglement still remains a major task, collective behavior such as  %%@
{\em spin squeezing} 
~\cite{Wineland92,Kuz97,Kuz98,
Pol99,Hald99,Kuz00,Duan00,Koz00,Sor101,Kit93,Agarwal90,
Lukin00,Xwang01,Sor201,Orzel01,Sor02,Andre02,Usha103,Usha203,Ulam01,Xwang03}, exhibited by
 multiqubit systems, has been proposed as a signature of quantum correlation
between the atoms. A connection between spin squeezing and the nature of
quantum entanglement has been explored~\cite{Ulam01,Xwang03} and it is shown
that the presence of spin squeezing essentially reflects pairwise entanglement. However,
it is important to realize that spin squeezing serves only as a sufficient condition - not a necessary one - for %%@
pairwise entanglement. There will still be pairwise correlated states, which do not exhibit spin squeezing.  In a %%@
class of \break symmetric multiqubit states it has been shown~\cite{Xwang03}  that spin-squeezing and pairwise \break %%@
entanglement imply each other. Questions like``{\em Are there any other collective  signatures of pairwise %%@
entanglement?}\," are still being investigated. Recently, inequalities generalizing the concept of spin squeezing have %%@
been derived~\cite{Kor}. These inequalities are shown to provide necessary and sufficient conditions for pairwise %%@
entanglement and three-party entanglement in symmetric $N$-qubit states.

In this thesis, we have addressed the problem of characterizing pairwise entanglement in symmetric multiqubit systems %%@
in terms of two qubit \underbar{local invariants}~\cite{ARU1, ARU2, ARU3}. This is important because
quantum entanglement reflects itself through non-local correlations among the subsystems of a quantum system. These %%@
{\it non-local properties} remain unaltered by local manipulations on the subsystems and provide a characterization of %%@
quantum \break entanglement. 

Two composite quantum states $\rho_1$ and $\rho_2$ are said to be {\em equally entangled} if they are related to each %%@
other through {\em local unitary operations}, which merely imply a choice of bases in the spaces of the subsystems. %%@
One may define a polynomial invariant, which is by definition any real valued function of density operators, taking %%@
the same value for equally entangled density operators $\rho$.
\newpage
Basic issues of importance would then be 
\begin{itemize}
\item {to find {\em complete} set of polynomial entanglement invariants which assume identical values for density %%@
operators related to each other through local unitary operators.} 
\item {decide whether the set of separable states can be described in terms of a polynomial
invariant $f$, such that $f(\rho)\geq 0$ is equivalent to separability~\cite{werner}}.
\end{itemize}

In this context, Y. Makhlin ~\cite{Mak02} has studied the entanglement invariants of an arbitrary mixed state of %%@
two-qubits and has identified a complete set of 18 local invariants characterizing the system. A set of 8 polynomial %%@
invariants has been identified in the case of pure three qubit states~\cite{Sud01}. Linden et. al.~\cite{Lin99} have %%@
outlined a  general prescription to identify the invariants associated with a multi particle system \footnote %%@
{However, separability properties of two qubit states in terms of the local invariants, is not investigated in\break %%@
Ref.~\cite{Mak02,Sud01,Lin99}}. Here, we focus on constructing a complete set of local invariants characterizing %%@
symmetric two qubit systems and analyzing the pairwise entanglement properties like collective spin squeezing - %%@
exhibited by multiqubits - in terms of two qubit entanglement invariants. 

A brief Chapter wise summary of the thesis is given below.

\noindent {\bf Chapter 2: {\bf Symmetric two qubit local invariants}}

For an arbitrary two qubit mixed state, Makhlin~\cite{Mak02} has proposed a complete set of 18 polynomial invariants. %%@
In this Chapter, we show that the number of invariants reduces from 18  to 6 in the case of symmetric two qubit states %%@
owing to the exchange symmetry. We quantify entanglement in  symmetric two qubit states in terms of these complete set %%@
of six  invariants. More specifically, we prove that the {\em negative} values of some of the invariants serve as  %%@
signatures of  quantum entanglement in symmetric two qubit states. This leads us to identify sufficient conditions for %%@
non-separability in terms of entanglement invariants~\cite{ARU1}. Further, these conditions on invariants are shown %%@
here to be both {\em necessary and sufficient} for entanglement in a class of symmetric two qubit states.\\
\\
\\

\noindent {\bf Chapter 3: {\bf Characterization of pairwise entanglement in symmetric multiqubit systems}} 

As discussed in Chapter 2, some of the symmetric two qubit invariants reflect nonseparability~\cite{ARU1}.
In this Chapter, we focus on the characterization and classification of pairwise entanglement in symmetric multi-qubit %%@
systems, via local invariants associated with a random pair of qubits drawn from the collective systems. In other %%@
words, we investigate {\em collective} signatures of pairwise entanglement in symmetric N-qubit states as implied by %%@
the associated non-positive values of the two qubit invariants. More specifically, we identify here that a symmetric %%@
multi-qubit system is spin squeezed {\em iff}\,  one of the entanglement invariant is {\em negative}. An explicit %%@
classification, based on the {\em structure} of local invariants for pairwise entanglement in symmetric $N$-qubit %%@
states is given~\cite{ARU2}. We show that our characterization gets related to the {\em generalized spin squeezing %%@
inequalities} of Korbicz et. al~\cite{Kor}. \\

\noindent {\bf Chapter 4: {\bf Analysis of few dynamical models}}

 In the light of our characterization of pairwise entanglement in symmetric multiqubit states discussed in Chapter 3, %%@
we analyze some of the  experimentally relevant $N$ qubit permutation symmetric states and explicitly demonstrate the %%@
non-separability of such states as exhibited through two qubit local invariants. In particular, we evaluate the two %%@
qubit local invariants and hence discuss the collective pairwise entanglement properties in the following multiqubit %%@
states: 
\begin{enumerate}
\item{Dicke states~\cite{Dic54,Xi02}}
 \item{Kitagawa-Ueda state generated by one axis twisting Hamiltonian~\cite{Kit93}} 
 \item{Atomic squeezed states~\cite{Agarwal90}}. \\ 
 \end{enumerate}

\noindent {\bf Chapter 5: {\bf Necessary and sufficient criterion in symmetric two qubit states}}

Continuous variable systems (CV)~\cite{Ade} i.e.,  systems associated with infinite
dimensional spaces are a focus of interest and attention due to their practical relevance in applications to quantum %%@
optics and quantum information science.
Moreover two mode Gaussian states, a special class of CV systems provide a clean framework for the investigation of %%@
nonlocal correlations. Consequently, most of the results on CV entanglement have been obtained for Gaussian states. %%@
Entanglement for two-mode Gaussian states is {\em completely} captured in its {\em covariance matrix} . It is %%@
desirable to look for an
analogous covariance matrix pattern in finite dimensional systems - in particular in multiqubits.

In this Chapter, we identify such a structural parallelism~\cite{ARU3} between  continuous variable %%@
states~\cite{Simon} and  symmetric two qubit systems by constructing covariance matrix of the latter. Pairwise %%@
entanglement between any two qubits of a symmetric $N$ qubit state is shown to be {\em completely} characterized by %%@
the  off-diagonal block of the two qubit covariance matrix. We establish the inseparability  constraints satisfied by %%@
the covariance 
matrix~\cite{ARU3} and identify that these are equivalent to the generalized spin squeezing inequalities~\cite{Kor} %%@
for pairwise entanglement. The interplay  between  two basic principles viz, the uncertainty principle and the %%@
nonseparability gets highlighted through the restriction on the covariance matrix of a quantum correlated two qubit %%@
symmetric state.
So, the collective pairwise entanglement properties of symmetric multiqubit states depends entirely on the off %%@
diagonal block of the covariance matrix.      
We further establish an equivalence between the Peres-Horodecki~\cite{Peres, horo96}
criterion and the negativity of the covariance matrix ${\cal C}$ showing that our condition is both necessary  and %%@
sufficient for  entanglement in symmetric two qubit states.
We continue to identify the constraints satisfied  by the collective correlation 
matrix $V^{(N)}$ of pairwise entangled symmetric N qubit states.
\newpage
In other words, the local invariant separability condition necessarily implies that

{\em The symmetric $N$ qubit system is pairwise entangled iff the least eigen value of the 
real symmetric matrix ~$~V^{(N)}~+~\frac{1}{N}~\,~ SS^T$~
is less than $N/4$.}\\
($V^{(N)}$ denotes the collective covariance matrix and $S$ corresponds to the collective
average spin of the symmetric N-qubit system.)\\
\noindent {\bf Chapter 6: {\bf Summary}}\\
In this Chapter, we  briefly summarize the important results obtained
in this thesis.
    
%%%%%%%%%%%%%%%%%%%%%%%%%%%%%%%%%%%%%%%%%%%%%%%%%%%%%%%%%%%%%%%%%%%%%%%%%%%%%%
%%%%%%%%%%%%%%%%%%%%%%%%%%%%%%%%%%%%%%%%%%%%%%%%%%%%%%%%%%%%%%%%%%%%%%%%%%%%%%
%%%%%%%%%%%%%%%%%%%%%%%%%%%%%%%%%%%%%%%%%%%%%%%%%%%%%%%%%%%%%%%%%%%%%%%%%%%%%%
\chapter{ Symmetric two qubit local invariants}
\label{c:c2twoqubit}
\markboth{}{Symmetric two qubit local invariants }

\section{Introduction}
 
Initiated by the celebrated Einstien-Podolsky-Rosen
criticism~\cite{EPR}, counterintuitive features of quantum correlations have retained the focus for more than seven %%@
decades now, and quantum entanglement has emerged as an essential ingredient in the rapidly developing area of quantum %%@
computation and quantum information processing~\cite{Niel00,Bou00,Div95}. Characterization and quantification of %%@
entanglement has been one of the central tasks of quantum information theory. In simple terms, a bipartite quantum %%@
system is entangled, if it is not separable
i.e., if the density matrix cannot be expressed as a convex mixture of product states,
\begin{equation}
\label{prod}
\rho=\sum_w p_w \,\rho_w^{(1)}  \otimes \rho_w^{(2)} \,\,\,\,{\rm where} \,\,0\leq p_w\leq 1\,\, {\rm and} \,\,\sum_w %%@
p_w=1. 
\end{equation}
Here, $\{\rho_w^{(1)}\}$ and $\{\rho_w^{(2)}\}$ denote a set of density operators associated with
quantum systems 1 and 2. It is a non-trivial task to check whether a given state is expressible as a mixture of %%@
product states (see Eq.~(\ref{prod})) or not.

Peres~\cite{Peres} has identified that the partial transpose of a separable bipartite state $\rho$ is positive %%@
definite (See Appendix~\ref{apdx02}) and therefore negative eigenvalues of a partially transposed density matrix imply %%@
non-separability of a quantum state. Further,
Horodecki et. al~\cite{horo96} proved that negativity under partial transpose
provides a necessary and sufficient condition for quantum entanglement in $2\otimes 2$ and $2\otimes 3$ systems only.

It is possible to quantify the amount of entanglement in a bipartite pure state 
$\vert \psi \rangle$ through the von Neumann entropy of either of the two subsystems \cite{entr}
\begin{equation}
\label{entropy}
E(\psi)=- {\rm Tr}(\rho_A {\rm log}_2 \rho_A) = -{\rm Tr} (\rho_B {\rm log}_2 \rho_B)   
\end{equation}
where,
\begin{equation*} 
\rho_A= {\rm Tr}_B(\vert \psi\rangle\langle\psi\vert)
\end{equation*}
 and
\begin{equation*}
\rho_B={\rm Tr}_A(\vert \psi\rangle\langle\psi\vert)
\end{equation*}
 denote subsystem density matrices.
$E(\psi)$ is referred to as {\em Entropy of entanglement}.\break
 $\lim_{n\rightarrow\infty} n E(\psi)$   gives the number of {\em maximally entangled states} that can be formed with %%@
$n$ copies of $\vert\psi\rangle,$ in the asymptotic limit.

The {\em entanglement of formation} of a mixed bipartite state
$\rho$ is defined as the minimum average entanglement 
\begin{equation}
\label{formation}
E(\rho)={\rm min}\sum_i p_i E(\psi_i),
\end{equation}
where ${\vert\psi_i\rangle}$ corresponds to all possible decompositions of the state through $$\rho=\sum_i p_i %%@
\vert\psi_i\rangle \langle\psi_i\vert.$$  {\em Entanglement of formation} $E(\rho)$ reduces to entropy of entanglement
$E(\psi)$ in the case of pure states and is zero iff the state is separable. 

An explicit 
analytical expression for the entanglement of formation has been derived for an arbitrary pair of %%@
qubits~\cite{hillwoot,woot98} and is given by:  
\begin{equation}
\label{Conc}
E(\rho)= h\left(\frac{1+\sqrt{1+C^2}}{2}\right),
\end{equation}
where 
$$h(x)= -x\,{\rm log}_2x - (1-x)\,{\rm log}_2(1-x).$$
In Eq.~(\ref{Conc}), $C,$ the {\em Concurrence}~\cite{woot98} is given by 
$$C={\rm max}(0, \lambda_1-\lambda_2-\lambda_3-\lambda_4),$$
with $\lambda_1^2,\lambda_2^2,\lambda_3^2,\lambda_4^2$
denoting the eigenvalues of $$\rho(\sigma_2\otimes\sigma_2)\rho^*
(\sigma_2\otimes\sigma_2)$$
in the decreasing order. Here, 
\begin{equation}
\label{sigma}
\sigma_2=\left(\begin{array}{ll}
          0 & -i \cr 
          i & 0 \cr 
         \end{array}\right)
\end{equation} 
is the standard Pauli matrix and $\rho$ is expressed in the standard two qubit basis set
$$\{\vert 0_1\,0_2\rangle, \vert 0_1\,1_2\rangle, \vert 1_1\,0_2\rangle,\vert 1_1\,1_2\rangle\}.$$
The Concurrence $C$ varies from zero to one and is monotonically
related to entanglement of formation $E(\rho),$ thus gaining the status of a measure of entanglement on its %%@
own~\cite{woot98}.

Entanglement properties of a quantum system remain unaltered when the subsystems are {\em locally manipulated} and two %%@
quantum states $\rho_1$ and $\rho_2$ are equally entangled if they are related to each other through local unitary %%@
transformations. Non-separability of a quantum state may thus be represented through a {\em complete set of local %%@
invariants} which contains functions of the quantum state that remain unchanged by local unitary operations on the %%@
subsystems. In this Chapter, we investigate a complete set of local invariants for arbitrary symmetric two qubit %%@
states. 

We identify that a set of six invariants is sufficient to characterize a symmetric two qubit system, provided the %%@
average spin $\vert\langle\vec{\sigma}\rangle\vert$ of the qubits is non-zero. If    
 $\vert\langle\vec{\sigma}\rangle\vert=0,$ only two entanglement invariants represent the nonseparability of 
 the system. 
 
 We further show that all the invariants associated with separable symmetric systems are positive. This allows us to %%@
identify criteria for non-separability in terms of invariants.  
\section{Arbitrary two qubit density matrix}
Density matrix (See Appendix~\ref{apdx01}) of an arbitrary two-qubit state in the Hilbert-Schmidt space ${\cal %%@
H}={\cal C}^2\otimes{\cal C}^2$ is given by 
\begin{equation}
\label{twoqubit}
\rho=\frac{1}{4}\left(I\otimes I+ \sum_{i=1}^3 s_i\, \sigma_{1i}+\sum_{i=1}^3\sigma_{2i}\, r_i
     +\sum_{i,j=1}^3 t_{ij}\,\sigma_{1i}\sigma_{2j}\right ) \, , 
 \end{equation}
where $I$ denotes the $2\times 2$ unit matrix. Here,
\begin{equation}
\label{sigma}
 \sigma_1=\left(\begin{array}{ll}
                    0 & 1 \cr 
                    1 & 0  \cr 
                    \end{array}\right), \,  
	\sigma_2=\left(\begin{array}{ll}
                    0 & -i \cr 
                    i & 0 \cr 
                    \end{array}\right), \, 
	\sigma_3=\left(\begin{array}{ll}
                    1 & 0 \cr 
                    0 & -1  \cr 
                    \end{array}\right) \,   				
\end{equation}			
{\rm are\, the\, standard\, Pauli\, spin\, matrices} and
\begin{eqnarray}
\sigma_{1i}= \sigma_i\otimes I\nonumber\\
\sigma_{2i}=  I\otimes \sigma_i.
\end{eqnarray}
The  average spins of the qubits $\vec{s}=(s_1,\,s_2,\,s_3)$ and $\vec{r}=(r_1,\,r_2,\,r_3)$
 are given by
\begin{eqnarray}
\label{avgspin}
s_i={\rm Tr}\,(\rho\, \sigma_{1i}) \nonumber\\
r_i={\rm Tr}\, (\rho\, \sigma_{2i})
\end{eqnarray}
and the two-qubit correlations are given by,
\begin{eqnarray} 
\label{corrt}
t_{ij}={\rm Tr}\,[\rho\, (\sigma_{1i}\sigma_{2j})].
\end{eqnarray}
It is convenient to express the two qubit correlations $t_{ij} (i,\,j=1,\,2,\,3),$ in the form \break 
of a  $3\times 3$ matrix as follows:
\begin{equation}
\label{corrmat}
T=\left(\begin{array}{ccc}
                     t_{11} & t_{12} & t_{13} \cr 
                     t_{21} & t_{22} & t_{23}\cr
					 t_{31} & t_{32} & t_{33} \cr
                    \end{array}\right).   
\end{equation}
The correlation matrix $T$ is real and in general, nonsymmetric.
  
\noindent{\bf Transformation of state parameters under local unitary operations:}\\
The 15 parameters $\{ s_i,\ r_i, \  t_{ij}\}; i,j=1,2,3,$ characterizing two qubit density matrix
of Eq.~(\ref{twoqubit})
exhibit the following transformation properties under local unitary operations: 
\begin{eqnarray} 
\label{trans1} 
s'_i&=&\sum_{j=1}^3 O^{(1)}_{ij}s_j, \nonumber\\
%\end{eqnarray}
%\begin{eqnarray}
%\label{trans2}
r'_i&=&\sum_{j=1}^3 O^{(2)}_{ij}r_j,
\end{eqnarray}
\begin{eqnarray}
\label{trans3}
t'_{ij}&=&\sum_{k,l=1}^3 O^{(1)}_{ik}\, O_{jl}^{(2)}t_{kl} \hskip .2in \ \ {\rm \ or \ } \ \ 
    T'=O^{(1)}\, T\,  O^{(2)\,^{\rm T}}\, ,  
\end{eqnarray}    
where $O^{(1)},\,O^{(2)} \in SO(3,R)$ are the $3\times 3$ rotation matrices, uniquely corresponding to the $2\times 2$ %%@
unitary matrices $U_i~\in~SU(2).$
The above transformation properties, facilitate the construction of polynomial functions 
of state parameters $\{ s_i,\ r_i, \  t_{ij}\}$ which remain invariant~\cite{Mak02}
under local operations on individual qubits. We devote the next section for discussion 
of  a {\em complete set of local invariants} associated with an arbitrary two qubit
density matrix-which was proposed by Makhlin~\cite{Mak02}. 
\section{Local invariants of an arbitrary two qubit system}
\label{local invariants}
As has been emphasized earlier, genuine nonlocal properties should be described in terms of physical quantities that %%@
are invariant under local unitary operations. Makhlin~\cite{Mak02} investigated such local invariant properties  
of mixed states of two-qubit system.
{\em Two density matrices $\rho_1$ and $\rho_2$ are called locally equivalent if one can be transformed into the other %%@
by local operations}
$$\rho_2=(U_1\otimes U_2)\rho_1(U_1\otimes U_2)^\dagger.$$
A useful tool for verification of local equivalence of two states is a complete set of
invariants that distinguishes all inequivalent states:

 {\em If each invariant from
the set has equal values on two states $\rho_1,\, \rho_2$,  their local equivalence is
guaranteed.}\\
A complete set of invariants for an arbitrary two qubit system as given by Makhlin~\cite{Mak02} is listed in %%@
Table~\ref{tab:1}.\\
\begin{table}
\centering
\begin{tabular}{||l||l||}
\hline
\hline
 & \\
$I_1={\rm det}$ T & $I_{10}=\epsilon_{ijk}\, s_i\, (T\, T^{\rm T}\, s)_j\,
([T\,T^{\rm T}]^2\,s)_k$ \\
 & \\
$I_2={\rm Tr}\, (T^{\rm T}\, T)$ & $I_{11}=\epsilon_{ijk}\, r_i\, 
(T^{\rm T}\, T\, r)_j\, ([T^{\rm T}\,T]^2\,r)_k$\\
& \\
$I_3={\rm Tr}\, (T^{\rm T}\, T)^2$ & $I_{12}=s^{\rm T}\,T\, r$\\
& \\
$I_4=s^{\rm T}\, s$ & $I_{13}=s^{\rm T}\,T\,T^{\rm T}\,T\, r$\\
& \\
$I_5=s^{\rm T}\, T\, T^{\rm T}\, s$ & $I_{14}=\epsilon_{ijk}\,
\epsilon_{lmn}\, s_i\,r_l\, t_{jm}\, t_{kn}$\\
& \\
$I_6=s^{\rm T}\, (T\, T^{\rm T})^2\, s$ &  $I_{15}=\epsilon_{ijk}\, s_i\,
(T\, T^{\rm T} s)_j\, (T\,r)_k$\\
& \\
$I_7=r^{\rm T}\, r$ & $I_{16}=\epsilon_{ijk}\, (T^{\rm T}\, s)_i\, r_j\, 
(T^{\rm T}\, T\, r)_k$\\
& \\
$I_8= r^{\rm T}\, T\, T^{\rm T}\, r$ & $I_{17}=\epsilon_{ijk}\, (T^{\rm T} \, s)_i\, 
(T^{\rm T}\, T\, T^{\rm T}\, s)_j\, r_k$\\
& \\
$I_9=r^{\rm T}\, (T\, T^{\rm T})^2\, r$ & $I_{18}=\epsilon_{ijk}\, s_i\, 
(T\, r)_j\, (T\, T^{\rm T}\, T\, r)_k$\\
& \\
\hline
\hline
\end{tabular}
\caption{Complete set of 18 polynomial invariants for an arbitrary two qubit state.}
\label{tab:1}
\end{table}
It is clear that all the invariants $I_k,\ k=1,2,\ldots \,18,$ listed in Table~\ref{tab:1}, are invariant under local %%@
unitary transformations as can be verified by explicitly substituting Eqs.~(\ref{trans1}),~(\ref{trans3})
for transformed state parameters.
For example consider the invariant $I_4$ which, under local unitary operation, transform as, 
\begin{eqnarray}
\label{invI4}
I_4&=&s'^{\,^{\rm T}} s'=s^{\rm T}\,O^{(1)\, ^{\rm T}}\,O^{(1)}\,s \nonumber\\
&=&s^{\rm T}\,s \hskip.5in {\rm since}\,\,\,\, \,\,\,\,O^{(1)\,^{\rm T}}\,O^{(1)}=1.
\end{eqnarray}
Similarly, it is easy to identify that
\begin{eqnarray}
\label{invI12}
I_{12}&=&s'^{\,^{\rm T}}\,T'\, r'
=s^{\rm T}\,O^{(1)\, ^{\rm T}}\,O^{(1)}\, T\,  O^{(2)\, ^{\rm T}}\,O^{(2)}\,r \nonumber\\
&=&s^{\rm T}\,T\, r.\, 
\end{eqnarray}\\
\newpage
Makhlin~\cite{Mak02} has given  an  explicit procedure 
to find local unitary operations that transform any equivalent density matrices to a specific canonical form,  %%@
uniquely determined by the set of 18 invariants given in Table \ref{tab:1}. Further, it has been shown that when %%@
$T^{\rm T}\,T$ is nondegenerate, the entire set of
invariants $I_{1-{18}}$ is required to completely specify the canonical form of locally equivalent
(See Appendix~\ref{apdx03}) density matrices. However, when $T^{\rm T}\,T$ is degenerate, only a subset of 18 %%@
invariants would suffice for the complete specification of density matrices which are locally related to each 
other. In particular, (i) when two of the eigenvalues of $T^{\rm T}\,T$ are equal then only a subset of nine %%@
invariants $\{I_{4-9},I_{{12}-{14}}\}$ are required and
(ii) when all the three eigenvalues of $T^{\rm T}\,T$ are equal, the subset $\{I_{4-9},I_{12}\}$
containing  six invariants determines the canonical form of the density matrices. 

In the next section, we identify that the number of invariants required to
characterize an arbitrary symmetric two-qubit system reduces from  
18 (as proposed by Makhlin~\cite{Mak02}) to 6.
Moreover, we consider a specific case of symmetric two-qubit system, and show that
a subset of three independent invariants is sufficient to determine the non-local properties completely.  
\section{Invariants for symmetric two-qubit states}
\label{symmtwoqubit}
Symmetric two qubit states $\rho_{\rm sym},$ which obey exchange symmetry,
are defined by, 
$$\Pi_{12}\, \rho_{\rm sym}=\rho_{\rm sym}\Pi_{12}=\rho_{\rm sym},$$ 
where $\Pi_{12}$ denotes the permutation operator.

Quantum states of symmetric two qubits get confined to a three dimensional
subspace\footnote{The collective angular momentum basis states $\left\{\vert  \frac{N}{2},\ M\rangle %%@
;M=-\frac{N}{2}\leq M \leq \frac{N}{2} \right\}$ span the Hilbert space ${\cal H_{\rm sym}}=(C^2\otimes C^2\ldots %%@
\otimes C^2~)_{\rm sym}$ of symmetric $N$ qubit system. i.e., the 
dimension of the Hilbert space gets reduced from $2^N$ to $(N+1)$ for symmetric $N$
qubit states.} of the Hilbert space.
Explicitly the angular momentum states\break 
$\left\{\vert\frac{N}{2}=1,\ M=1\rangle, \vert\frac{N}{2}=1,\ M=0\rangle, \vert\frac{N}{2}=1,\ M=-1\rangle\right\}$  %%@
are related to the standard two qubit states 
 $\left\{\vert 0_1,0_2 \rangle,\, \vert 0_1,1_2 \rangle,\vert 1_1,0_2 \rangle,\vert 1_1,1_2 \rangle\right\}$ as %%@
follows:
\begin{eqnarray}
\label{angbasis}
\left\vert\frac{N}{2}=1,\ M=1\right\rangle&=& \vert 0_1,\,0_2\rangle, \nonumber\\
\left\vert\frac{N}{2}=1,\ M=0\right\rangle&=& \frac{1}{\sqrt 2}\,(\vert 0_1,\,1_2\rangle+\vert 1_1,\,0_2\rangle),  %%@
\nonumber\\
\hskip .2in\left\vert\frac{N}{2}=1,\ M=-1\right\rangle&=& \vert 1_1,\,1_2\rangle.
\end{eqnarray}
An arbitrary symmetric two qubit density matrix $\rho_{\rm sym}$ has the form:
\begin{equation}
\label{symrho}
\rho_{\rm sym}=\frac{1}{4}\left(I\otimes I+ \sum_{i=1}^3\,s_i\,(\sigma_{1i}+\sigma_{2i})\,+\sum_{i,j=1}^3 %%@
t_{ij}\,\sigma_{1i}\sigma_{2j}\right), 
 \end{equation}
\hskip 1.3in where,
\begin{eqnarray}
\label{symm}
{\rm Tr}(\rho_{\rm sym}\sigma_{1i})&=&{\rm Tr}(\rho_{\rm sym}\sigma_{2i}),\nonumber\\
{\rm or}\,\,\,\,\,\,\,\,\,\,\, r_i&=&s_i, 
\end{eqnarray}
\hskip 1.3in and
\begin{eqnarray}
\label{Tsymm}
{\rm Tr}[\rho_{\rm sym} (\sigma_{1i} \sigma_{2j})]&=&{\rm Tr}[\rho_{\rm sym} (\sigma_{1j}\sigma_{2i})],\nonumber\\
{\rm or}\,\,\,\,\,\ t_{ij}&=&t_{ji}\,\,\, \Longrightarrow \,\,T^{\rm T}=T .
\end{eqnarray} 
(Please see Eqs.~(\ref{twoqubit})~-~(\ref{corrt})) for a comparison with the density matrix of an arbitrary two qubit %%@
system.) 

Obviously, the number of state parameters for a symmetric two qubit system gets
reduced from 15 (for an arbitrary two qubit state) to 9 owing to symmetry constraints
given in Eqs.~(\ref{symm}),~(\ref{Tsymm}). Moreover, there is one more constraint on correlation matrix $T$ reducing %%@
the number of parameters required to characterize a symmetric two qubit system to 8 . To see this, let us consider the %%@
collective angular momentum of two qubit system which is given by
\begin{equation}
\label{angJ}
\vec{J}=\frac{1}{2}(\vec{\sigma}_{1} +\vec{\sigma}_{2}).
\end{equation}
We have,
\begin{equation}
\label{trJ}
{\rm Tr}[\,\rho_{\rm sym}(\vec{J}\cdot \vec{J})\,]=2,
\end{equation}
for symmetric two qubit states.
This is because, the symmetric two qubit system is confined to the maximum 
value of angular momentum $j=1$ in the addition of two spin $\frac{1}{2}$
(qubit) systems. (Note that in the addition of angular momentum of two spin
$\frac{1}{2}$ particles the total angular momentum can take values 0
and 1. The two qubit states with 0 net angular momentum are antisymmetric
under interchange - whereas those with maximum angular momentum 1  are symmetric 
under interchange of particles.)
Using Eq.~(\ref{angJ}) we obtain,
\begin{eqnarray}
\label{angmom}
\vec{J}\cdot \vec{J}&=&\frac{1}{4}[\,(\vec{\sigma}_{1}+\vec{\sigma}_{2})\cdot
(\vec{\sigma}_{1}+\vec{\sigma}_{2})\,],\nonumber\\
&=&\frac{1}{4}[\,(\vec{\sigma}_{1}\cdot\vec{\sigma}_{1})+
(\vec{\sigma}_{2}\cdot\vec{\sigma}_{2})+2\,(\vec{\sigma}_{1}\cdot\vec{\sigma}_%%@
{2})\,].
\end{eqnarray}
Since $(\vec{\sigma}_{1}\cdot\vec{\sigma}_{1})=(\vec{\sigma}_{2}\cdot\vec{\sigma}_{2%%@
})=3I,$
we can rewrite the above equation as,
\begin{eqnarray}
\label{trT} 
\vec{J}\cdot \vec{J}=\frac{1}{4}[6I\,+2\,
(\vec{\sigma}_{1}\cdot\vec{\sigma}_{2})].
\end{eqnarray}
In other words, we have,
\begin{eqnarray}
{\rm Tr}\,[\,\rho_{\rm sym}(\vec{J}\cdot\vec{J})\,]&=&\frac{1}{2}(3\,{\rm Tr}(\rho_{\rm sym}\,I)+{\rm Tr}[\rho_{\rm %%@
sym}\,
(\vec{\sigma}_{1}\cdot\vec{\sigma}_{2})])\nonumber\\
&=&\frac{1}{2}[3+{\rm Tr}(T)].
\end{eqnarray}
From Eq.~(\ref{trJ})  we have, 
\begin{equation} 
\frac{1}{2}[3+{\rm Tr}(T)]=2, \nonumber \\
\end{equation}
for symmetric two qubit systems.
This in turn leads to the constraint
\begin{equation}
\label{traceT=1}
 {\rm Tr}\,(T)=1.
\end{equation}
Thus only {\em eight} real state parameters viz.,  {\em three} real parameters $s_i$ and
{\em five} parameters $t_{ij},$ (completely specifying the $3\times 3$ real symmetric two qubit correlation matrix $T$ %%@
with unit trace) determine a symmetric two qubit system. 

We give below an explicit $4 \times 4$ matrix form 
 (in the standard  two-qubit basis \break $|0_1\, 0_2\rangle\, , |0_1\, 1_2\rangle\, , |1_1\, 0_2\rangle\, , |1_1\, %%@
1_2\rangle$) of
an arbitrary symmetric two qubit density matrix:
\begin{equation}
\label{symmetric two}
\rho_{\rm sym}=\frac{1}{4}\left( \begin{array}{cccc} 1+2\, s_3 + t_{33} & A^*  & A^* 
  & (t_{11}-t_{22})-2i\, t_{12} \cr 
A  & (t_{11}+t_{22}) & (t_{11}+t_{22}) & B^*   \cr 
A  & (t_{11}+t_{22}) & (t_{11}+t_{22}) & B^*   \cr 
(t_{11}-t_{22})+2\, i t_{12}& B & B & 1 - 2\, s_3 + t_{33}   
\end{array}  \right),  
\end{equation}
where $A=(s_1+i\, s_2)+ (t_{13}+i\, t_{23})$ and $B=(s_1+i\, s_2)- (t_{13}+i\, t_{23}).$\\
In the following discussion, we show that the number of local invariants required to determine a symmetric two qubit %%@
density matrix also reduces, owing to the symmetry constraints on the state parameters %%@
Eqs.~(\ref{symm}),~(\ref{Tsymm}).

\noindent {\bf Symmetric two qubit local invariants:}\\
In the case of symmetric two qubit system, it is easy to see that the
18 polynomial invariants (for an arbitrary two qubit system) given in Table.~\ref{tab:1} in terms of the qubit average %%@
$\vec{s}$ and two qubit correlations $T$ reduces to twelve:
\begin{table}[h]
\centering
\begin{tabular}{|l|l|}
\hline
 & \\
 & \\
 $\,\,\,\, I_1={\rm det}$ T  \,\,\,\, & \,\,\,\,$I_4=I_7=s^{\rm T}\, s,\,\,\,\,$ \\
 & \\
 & \\
$\,\,\,\,I_2={\rm Tr}(T^2)\,\,\,\,$ & \,\,\,\,$I_5=I_8=s^{\rm T}\, T\, T^{\rm T}\, s$\\
& \\
& \\
$\,\,\,\,I_3={\rm Tr}(T^2)^2\,\,\,\,$ & \,\,\,\,$I_6=I_9=s^{\rm T}\, (T\, T^{\rm T})^2\, s$\\
& \\
& \\
$\,\,\,\,I_{12}=s^{\rm T}\,T\, s,\,\,\,\,$ &  \,\,\,\,$I_{10}=I_{11}=\epsilon_{ijk}\, s_i\, (T\, T^{\rm T}\, %%@
s)_j\,([T\,T^{\rm T}]^2\,s)_k\, $\\
& \\
& \\
$\,\,\,\,I_{13}=s^{\rm T}\,T\,T^{\rm T}\,T\, s\,\,\,\,$ & \,\,\,\,$I_{15}=I_{16}=\epsilon_{ijk}\, s_i\,(T\, T^{\rm T} %%@
s)_j\, (T\,s)_k$\\
& \\
& \\
$\,\,\,\,I_{14}=\epsilon_{ijk}\,
\epsilon_{lmn}\, s_i\,s_l\, t_{jm}\, t_{kn}\,\,\,\,$ & \,\,\,\,$I_{17}=I_{18}=\epsilon_{ijk}\, s_i\, 
(T\, s)_j\, (T\, T^{\rm T}\, T\, s)_k$\\
& \\
& \\
\hline
\end{tabular}
\label{tab:2}
\caption{Invariants for a symmetric two qubit state.}
\end{table}
\\
We now proceed to identify that a set containing {\em six} local invariants 
$\{{\cal I}_{1-6}\}$ is sufficient to determine the canonical form of locally equivalent symmetric two qubit density %%@
matrices. 
\newpage
\begin{theorem}
All equally entangled symmetric two-qubit states have identical values for the
local invariants
$\{ {\cal I}_1 \, -\, {\cal I}_6 \}$ given below:
\begin{eqnarray}
\label{inv}
&{\cal I}_1={\rm det}\, T\, ,\ \     {\cal I}_2={\rm Tr}\, (T^2)\, ,& \nonumber
\\
& {\cal I}_3=s^T\, s\, , \ \  \    {\cal I}_4=s^T\, T\, s\, , \nonumber \\
&{\cal I}_{5} = \epsilon_{ijk}\,\epsilon_{lmn}\, s_i\,s_l\, t_{jm}\, t_{kn} \,
, \nonumber \\
& {\cal I}_{6}=\epsilon_{ijk}\, s_i\, (T\, s)_j\, (T^2\,s)_k \, ,
\end{eqnarray}
where $\epsilon_{ijk}$ denotes Levi-Civita symbol; $s\ (s^T)$ is a column (row)
with
$s_1,\, s_2$ and $s_3$ as elements.
\end{theorem}
\noindent {\bf  Proof:}
Let us first note that the state parameters of a symmetric two-qubit density
matrix transform under {\em identical} local unitary operation\footnote{A symmetric state transforms into another %%@
symmetric state under {\em identical} local unitary transformation
on both the qubits.}  $U\otimes U$ as follows:
\begin{eqnarray}
\label{tran}
&s'_i=\displaystyle\sum_{j=1}^3 O_{ij}\, s_j   \hskip 0.1in   {\rm \ or \ \ }  s'=O\, s \,
, \hskip 0.9in    \nonumber \\
&t'_{ij}=\displaystyle\sum_{k,l=1}^3\, O_{ik}\, O_{jl}\, t_{kl} \hskip 0.2in   {\rm
or \ } \ \      T'=O\, T\,  O^T\, ,
\end{eqnarray}
where $O\in SO(3, R)$ denotes $3\times 3$ rotation matrix, corresponding
uniquely to the $2\times 2$ unitary matrix
$U\in SU(2)$.

To find the minimum number of local invariants required to characterize a symmetric two qubit system, we refer to a %%@
canonical form of two qubit symmetric density matrix which is
achieved by identical local unitary transformations $U\otimes U$ such that the correlation matrix $T$ is diagonal. %%@
This is possible because
the real, symmetric correlation matrix $T$ can be diagonalized through
identical local rotations\footnote{Note that ${\rm Tr}(T)$ is preserved by identical local operations $U\otimes U$ %%@
i.e., we have $t_1+t_2+t_3=1.$}:
\begin{equation}
T^d=OTO^{\rm T}\\
=\left(\begin{array}{ccc}
                     t_1 & 0 & 0 \cr 
                      0 & t_2 & 0\cr
					 0 & 0 & t_3 \cr
                    \end{array}\right).   
\end{equation}
It is clear that the invariants \footnote{For an arbitrary two qubit state, the diagonal elements  $(t_1,\,t_2,\,t_3)$  %%@
of $T^d$ are not the eigenvalues of the correlation matrix $T$ (see Appendix~\ref{apdx03}). Hence, to determine %%@
$(t_1^2,\,t_2^2,\,t_3^2),$ (the eigenvalues of $T\,T^{\rm T}\,(T\,T^{\rm T})$ local invariants $I_{1-3}$ are required. %%@
However, in the case of a symmetric two qubit
state,  we have $T^{\rm T}=T.$ Therefore only two polynomial invariants
${\rm det} T,$ ${\rm Tr}(T^2)$ suffice to determine the eigenvalues of $T.$} ${\cal I}_1$ and ${\cal I}_2$ 
\begin{eqnarray}
\label{twoinv}
{\cal I}_1&=&{\rm det}\, T=t_1\,t_2\,t_3,\nonumber\\
{\cal I}_2&=&{\rm Tr}\, (T^2)=t_1^2+t_2^2+t_3^2,
\end{eqnarray} 
along with the unit trace condition
\begin{eqnarray}
\label{tracet} 
{\rm Tr}\,(T)=t_1+t_2+t_3=1,
\end{eqnarray}
determine the eigenvalues $t_1,\ t_2$ and $t_3$ of the two-qubit correlation matrix $T$. 
Further, the absolute values of the state variables $s_1,\ s_2,\ s_3,$ can be
evaluated using ${\cal I}_3,\ {\cal I}_4$ and ${\cal I}_5$:
\begin{eqnarray}
\label{abss}
{\cal I}_3&=&s^{\rm T}\, s=s_1^2+s_2^2+s_3^2, \nonumber \\
{\cal I}_4&=&s^{\rm T}\,T\, s=s_1^2\, t_1+s_2^2\, t_2+s_3^2\, t_3, \nonumber \\
{\cal I}_5&=&\epsilon_{ijk}\,\epsilon_{lmn}\, s_i\,s_l\, t_{jm}\, t_{kn}=
 2\,(s_1^2\, t_2\,t_3+s_2^2\, t_1\, t_3+s_3^2\, t_1\, t_2) .
\end{eqnarray}
Having determined  $s_1^2,\,s_2^2,\,s_3^2$  and thus fixing the absolute values
of the components of the qubit orientation vector $\vec{s}$ - the overall 
sign of the product $s_1 s_2 s_3$ is then fixed by ${\cal I}_6$:
\begin{eqnarray}
\label{sign}
{\cal I}_6&=&\epsilon_{ijk}\, s_i\, (T\, T^{\rm T}\, s)_j\,([T\,T^{\rm T}]^2\,s)_k\nonumber\\
&=&s_1 s_2 s_3\, \left[t_1\, t_2\, (t_2-t_1)+t_2\, t_3\,
(t_3-t_2)+t_3\, t_1\, (t_1-t_3)\right].
\end{eqnarray}
It is important to realize that only the overall sign of $s_1\, s_2\, s_3$ -
not the individual signs - is a local invariant.
More explicitly, if $(+,+,+)$ denote the signs of $s_1,\, s_2\, {\rm \ and \ }
s_3,$ identical local rotation through an angle
$\pi$ about the axes $1,\, 2\, {\rm \ or}\ 3$ affect only the signs, not the
magnitudes of $s_1,\, s_2,\,  s_3,$ leading to the
possibilities  $(+,-,-)$, $(-,+,-)$, $(-,-,+).$ All these combinations
correspond to the `$+$' sign for the product $s_1\, s_2\, s_3$.
Similarly, the overall `$-$' sign for  the product  $s_1 s_2 s_3$ arises from the
combinations, $(-,-,-)$, $(-,+,+)$, $(+,-,+),$
$(+, +, -)$, which are all related to each other by $180^{\circ}$ local
rotations about the  $1,\ 2 {\rm\  or}\ 3$ axes.

Thus we have shown that every symmetric two qubit density matrix can be
transformed by identical local unitary transformation
$U\otimes U$ to a {\em canonical form}, specified completely by the set of
invariants $\{{\cal I}_{1-6}\}$. In other words,
symmetric two-qubit states are equally entangled iff $\{{\cal I}_{1-6}\}$ are same.
In the next section, we consider a special class of symmetric density matrices
and show that a subset of {\em three independent invariants} is sufficient to 
characterize the non-local properties completely. 
\section{Special class of two qubit states}
\label{special class}
Many physically interesting cases of symmetric two-qubit states like for e.g.,  even and odd spin states~\cite{Xi03}, %%@
Kitagawa - Ueda state generated by one-axis twisting Hamiltonian~\cite{Kit93},  
atomic spin squeezed states~\cite{Agarwal90},  exhibit a 
particularly simple structure 
\begin{eqnarray}
\label{cansym}
\varrho_{\rm sym}=\frac{1}{4}\left(\begin{array}{llll}
                     a & 0 & 0& b\cr 
                     0& c & c& 0\cr 
                     0& c & c& 0 \cr 
                     b & 0 & 0& d\cr 
              \end{array}\right) \, , 
\end{eqnarray}
of the density matrix (in the standard two-qubit basis $|0_1\, 0_2\rangle\, , |0_1\, 1_2\rangle\, , |1_1\, %%@
0_2\rangle\, , |1_1\, 1_2\rangle$) with
\begin{eqnarray}
\label{trrho1} 
{\rm Tr}(\varrho_{\rm sym})=a+2c+d=1.
\end{eqnarray}
The qubit orientation vector $\vec{s}$ for the density matrices of the form Eq.~(\ref{cansym}) has the following %%@
structure
 \begin{equation}
\label{symS}
\vec{s}=(0,0,(a-d)),
\end{equation}
and the real symmetric $3\times 3$ correlation matrix $T$ for this special class of density matrix has the form:
\begin{equation}
\label{symT}
T=\left(\begin{array}{ccc}
                     2(c+b) & 0 & 0 \cr 
                      0 & 2(c-b) & 0\cr
					 0 & 0 & (a+d-2c) \cr
                    \end{array}\right).   
\end{equation}
It will be interesting to analyze the non-local properties of such systems through local invariants. The  specific  %%@
structure  $\varrho_{\rm sym}$ given by equation Eq.~(\ref{cansym}) of the  two-qubit density matrix   further reduces %%@
the number of 
parameters essential for the problem. Entanglement invariants associated with the symmetric two-qubit system %%@
$\varrho_{\rm sym}$, 
% with a simple structure $\varrho_{\rm sym}$ given by  Eq.~(\ref{cansym}) for the %state,
 may  now be identified through a simple calculation to be 
(see Eqs.~(\ref{twoinv}),~(\ref{abss}),~(\ref{sign})) 
\begin{eqnarray} 
\label{invsym}
 {\cal I}1&=&t_1\,t_2\,t_3=(4\, c^2-4\,|b|^2)\, (1-4\, c), \nonumber \\ 
 {\cal I}_2&=&t_1^2+t_2^2+t_3^2=(2\, c+2\, |b|)^2+(2\, c-2\, |b|)^2+(1-4\, c)^2,  \nonumber \\ 
 {\cal I}_3&=&s_1^2+s_2^2+s_3^2=(a-d)^2,   \nonumber \\ 
 {\cal I}_4&=&s_1^2\, t_1+s_2^2\, t_2+s_3^2\, t_3=(a-d)^2\, (1-4\, c),  \nonumber \\ 
 {\cal I}_5&=&2\,(s_1^2\, t_2\,t_3+s_2^2\, t_1\, t_3+s_3^2\, t_1\, t_2)
   =8\, (a-d)^2\, (c^2-\, |b|^2), \nonumber \\  
 {\cal I}_6&=&s_1 s_2 s_3\, \left[t_1\, t_2\, (t_2-t_1)+t_2\, t_3\,
(t_3-t_2)+t_3\, t_1\, (t_1-t_3)\right]=0 .
\end{eqnarray}

In this special case, we can express the invariants ${\cal I}_1$ and ${\cal I}_2$  in terms of $({\cal I}_3,{\cal %%@
I}_4,{\cal I}_5)$  ({\rm provided}\   ${\cal I}_3\neq 0$) :  
\begin{eqnarray}
\label{specialinv}
{\cal I}_1=\frac{{\cal I}_5\, {\cal I}_4}{2\ {\cal I}_3^2},  & \ & {\cal I}_2=\frac{({\cal I}_3-{\cal I}_4)^2-{\cal %%@
I}_3\, {\cal I}_5+{\cal I}_4^2}{{\cal I}_3^2}.
\end{eqnarray} 
If ${\cal I}_3=0$, then the set containing six invariants  reduces to  the subset of two non-zero invariants   $({\cal %%@
I}_1,\ {\cal I}_2).$
Thus the non-local properties of symmetric two-qubit states - having  a specific structure $\varrho_{\rm sym}$ given %%@
by equation Eq.~(\ref{cansym})  for the density matrix 
are characterized by \\
(i) subset of three invariants $({\cal I}_3, {\cal I}_4, {\cal I}_5)$ when ${\cal I}_3\neq 0$
 or\\ (ii) subset of two invariants $({\cal I}_1, {\cal I}_2)$ when ${\cal I}_3=0.$

In the next section,  we  propose  criteria,  which
provide a characterization of non-separability (entanglement) in  symmetric two-qubit states in terms of the local %%@
invariants $\{{\cal I}_{1-6}\}.$  
\section{Characterization of entanglement in symmetric two qubit states}
\label{characterization} 
A separable symmetric two-qubit density matrix is an arbitrary convex combination of direct product of  identical %%@
single qubit states,
\begin{eqnarray}
\label{singlequ}
\rho_w=\frac{1}{2}\left(I+\displaystyle\sum_{i=1}^3\sigma_i\, s_{wi}\right)
\end{eqnarray}
 and is given by 
\begin{equation}
\label{sepsym}
\rho_{\rm (sym-sep)}=\sum_w\, p_{w}\, \rho_w\otimes\rho_w,  
\end{equation} 
where $\sum_w\, p_w=1.$ \\
Separable symmetric system is a classically correlated system, which can be prepared through classical communications %%@
between two parties. A symmetric two qubit state which cannot be represented in the form Eq.~(\ref{sepsym}) is called %%@
{\em entangled}.

For a separable symmetric two qubit state, the components of the average spin 
of the qubits are given by 
\begin{eqnarray}
\label{sepspin}
s_i&=&{\rm Tr}[\rho_{\rm (sym-sep)}\, \sigma_{1i}]\nonumber \\
&=&\displaystyle\sum_w\, p_w\, {\rm Tr}[(\rho_w\otimes\rho_w)\,( \sigma_{1i})]\nonumber \\
&=&\displaystyle\sum_w\, p_w\,{\rm Tr}(\rho_w\,\sigma_{1i})\nonumber\\
&=& \displaystyle\sum_w\, p_w\, s_{wi}, 
\end{eqnarray}
and the elements of the correlation matrix $T$ can be expressed as,
\begin{eqnarray}
\label{sepcorr}
t_{ij}&=&{\rm Tr}[\rho_{\rm (sym-sep)}\, \sigma_{1i}\,\otimes \sigma_{2j}]\nonumber \\
&=&\displaystyle\sum_w\, p_w\ {\rm Tr}[(\rho_w\otimes\rho_w)\,( \sigma_{1i}\,\otimes \sigma_{2j})]\nonumber \\
&=&\displaystyle\sum_w\, p_w\,{\rm Tr}(\rho_w\,\sigma_{1i}){\rm Tr}(\rho_w\,\sigma_{2j})\nonumber\\
&=& \displaystyle\sum_w\, p_w\ s_{wi}\, s_{wj}. 
\end{eqnarray}
One of the important goals of quantum information theory has been to identify and characterize inseparability. We look %%@
for such identifying criteria for 
separability, in terms of entanglement invariants, in the following theorem:
\begin{theorem}
\label{separability}
The invariants,  ${\cal I}_4,\ {\cal I}_5$ and a  combination  ${\cal I}_4-{\cal I}_3^2$  of the invariants, %%@
necessarily assume 
positive values  for a symmetric separable two-qubit state,  with ${\cal I}_{3}\neq 0.$ 
\end{theorem}
\noindent{\bf Proof:} 
\noindent(i)  The invariant ${\cal I}_4$ has the following structure for  a
separable state:
\begin{eqnarray}
\label{I4}
{\cal I}_4&=&s^{\rm T}\, T\, s=\displaystyle\sum_{i,j=1}^3 t_{ij}\, s_i\,
s_j\nonumber \\
&=& \displaystyle\sum_w p_w\, \left(\displaystyle\sum_{i=1}^3 s_i^{(w)}\,
s_i\right)
         \left(\displaystyle\sum_{j=1}^3 s_j^{(w)}\, s_j\right) \nonumber \\
&=& \displaystyle\sum_w p_w\, \left(\vec{s}\cdot\vec{s}^{(w)}\right)^2 \geq 0.
\end{eqnarray}

\noindent(ii) Now, consider  the invariant ${\cal I}_5$ for a separable
symmetric system:
\begin{eqnarray}
\label{I5}
{\cal I}_{5} &=& \epsilon_{ijk}\,\epsilon_{lmn}\, s_i\,s_l\, t_{jm}\,
t_{kn}\nonumber \\
             &=&\displaystyle\sum_{w,w'} p_w\, p_{w'}\, \left(\epsilon_{ijk}\, s_i\,
s_j^{(w)}\, s_k^{(w')}\right)
               \left(\epsilon_{lmn}\, s_l\, s_m^{(w)}\, s_n^{(w')}\right)
\nonumber \\
             &=& \displaystyle\sum_{w,w'} p_w\, p_{w'}\, \left[ \vec{s}\cdot \left(
\vec{s}^{(w)}\times \vec{s}^{(w')}\right) \right]^2 \geq 0.
\end{eqnarray}

\noindent(iii) For the combination ${\cal I}_4-{\cal I}_3^2$ we obtain,
\begin{equation}
\label{comb}
{\cal I}_4-{\cal I}_3^2= \displaystyle\sum_w p_w\,
\left(\vec{s}\cdot\vec{s}^{(w)}\right)^2
               -\left(\displaystyle\sum_w p_w\,
(\vec{s}\cdot\vec{s}^{(w)})\right)^2,
\end{equation}
which has the structure $\langle A^2\rangle -\langle A\rangle ^2$ and is
therefore, essentially non-negative.
{\em Negative value assumed by any of the invariants ${\cal I}_4,\
{\cal I}_5$ or the ${\cal I}_3-{\cal I}_4^2$, is a signature of pairwise entanglement.}

Further from  the structure of the invariants in  a symmetric separable state,
it is clear that ${\cal I}_3=\displaystyle\sum_w p_w\,
\left(\vec{s}\cdot\vec{s}^{(w)}\right)=0$ implies  $\vec{s}^{(w)}\equiv 0$
for all `$w$', leading in turn to ${\cal I}_4=\displaystyle\sum_w p_w\,
\left(\vec{s}\cdot\vec{s}^{(w)}\right)^2=0$ and ${\cal I}_5=0.$ 
Thus when ${\cal I}_3=0,$ the two qubit local invariant ${\cal I}_1$ characterizes pairwise entanglement as shown in %%@
the theorem given below:
\begin{theorem} \label{the3}
For a symmetric separable two-qubit state, the invariant ${\cal I}_1$ assumes positive value.
\end{theorem}
\noindent {\bf Proof :} Consider,
\begin{eqnarray}
\label{i1neg}
T&=&{\rm diag}(t_1,\,t_2,\, t_3)\nonumber\\
&=&{\rm diag}\left(\displaystyle\sum_w p_w\, \left(s_1^{(w)}\right)^2,\
\displaystyle\sum_w p_w\, \left(s_2^{(w)}\right)^2, \
\displaystyle\sum_w p_w\, \left(s_3^{(w)}\right)^2\right).
\end{eqnarray}
We therefore have,
\begin{equation}
\label{I1}
{\cal I}_1=\det T=t_1\, t_2\, t_3
=\displaystyle\prod_{i=1}^3\left(\displaystyle\sum_w p_w\,
\left(s_i^{(w)}\right)^2\right),
\end{equation}
which is obviously non-negative for all symmetric separable two qubit states.
Thus when ${\cal I}_3=0,$ ${\cal I}_1>0$ provides a sufficient criteria for separability.

A simple example illustrating our separability criterion in terms of two qubit local 
invariants is the two qubit bell state
\begin{eqnarray}
\label{qubitbell}
\vert \Phi \rangle= \frac{1}{\sqrt 2}(\vert 0\,1 \rangle +    \vert 1\,0 \rangle).
\end{eqnarray}
The density matrix for the two qubit bell state is written as, 
\begin{eqnarray}
\label{Bcansym}
\rho=\vert \Phi \rangle\langle \Phi \vert=\frac{1}{2}\left(\begin{array}{llll}
                     0& 0 & 0& 0\cr 
                     0& 1 & 1& 0\cr 
                     0& 1 & 1& 0 \cr 
                     0& 0 & 0& 0\cr 
              \end{array}\right).
\end{eqnarray}
It is easy to identify that the above density matrix has a structure similar 
to $\varrho_{\rm sym}$ (see Eq.~(\ref{cansym})) with the matrix elements given by
\begin{eqnarray}
\label{matele}
a=0, & b=0,\nonumber\\
c=\frac{1}{2}, &  d=0.
\end{eqnarray}
The invariant ${\cal I}_3$ associated with the density matrix of Eq.~(\ref{Bcansym}) is given by,
$${\cal I}_3=(a-d)^2=0,$$ 
which, in turn implies that   
\begin{eqnarray}
\label{bellinv} 
{\cal I}_4&=&{\cal I}_3(1-4c)=0,\nonumber\\
{\cal I}_5&=&{\cal I}_3(c^2-\vert b \vert^2)=0.
\end{eqnarray}
The invariant ${\cal I}_1$ has the structure, 
\begin{equation}
\label{bellinv1} 
{\cal I}_1=(4\, c^2-4\,|b|^2)\, (1-4\, c)=-1.
%{\cal I}_2&=&(2\, c+2\, |b|)^2+(2\, c-2\, |b|)^2+(1-4\, c)^2=3.
\end{equation}
Since ${\cal I}_1\leq 0$, we can conclude from Theorem ~\ref{the3}, that the 
given state Eq.~(\ref{qubitbell}) is entangled.

\section{Necessary and sufficient criterion for a class of symmetric two qubit states}
\label{necc and suff}
It would be interesting to explore how these constraints on the invariants, 
get related to the other well established criteria of entanglement.
For two qubits states, it is well known that Peres's PPT (positivity of partial transpose) criterion~\cite{Peres} is %%@
both necessary and sufficient for separability. We now proceed to show that in the case of symmetric states, given by %%@
Eq.~(\ref{cansym}), 
there exists a simple connection between the Peres's PPT criterion and
the non-separability constraints (see Eqs.~(\ref{I4})~-~~(\ref{comb})) on the invariants.\\
We may recall from Sec.~\ref{special class} that the density matrix
for the special class of symmetric two qubit states (see Eq.~(\ref{cansym}))  
has the following structure:
\begin{eqnarray}
\label{cansym2}
\varrho_{\rm sym}=\frac{1}{4}\left(\begin{array}{llll}
                     a & 0 & 0& b\cr 
                     0& c & c& 0\cr 
                     0& c & c& 0 \cr 
                     b & 0 & 0& d\cr 
              \end{array}\right).
\end{eqnarray}
The partial transpose of the matrix $\varrho_{\rm sym}$ has the form,
\begin{eqnarray}
\label{cansympt}
(\varrho_{\rm sym})^{PT}=\frac{1}{4}\left(\begin{array}{llll}
                     a & 0 & 0& c\cr 
                     0& c & b& 0\cr 
                     0& b & c& 0 \cr 
                     c & 0 & 0& d\cr 
              \end{array}\right).
\end{eqnarray}
The eigenvalues of the partially transposed density matrix $(\varrho_{\rm sym})^{PT}$ 
are given by
\begin{eqnarray}
\label{eigenneg} 
\lambda_{1}&=&\frac{1}{2}\left((a+d)-\sqrt{(a-d)^2+4c^2}
\right), \nonumber \\
\lambda_{2}&=&\frac{1}{2}\left((a+d)+\sqrt{(a-d)^2+4c^2}
\right), \nonumber \\
\lambda_{3}&=&c\,-\vert b\vert, \nonumber\\
\lambda_{4}&=&c\,+\vert b\vert,
\end{eqnarray}
of which $\lambda_1$ and $\lambda_3$ can assume negative values ($a,c,d$ are positive 
quantities since they are the diagonal elements of the density matrix).\\
(i) If $\lambda_1<0,$ then we have, 
\begin{equation}
\label{ineq1} 
(a+d)^2\,<\,(a-d)^2+4c^2.
\end{equation}
The above inequality can be expressed as
\begin{equation}
\label{ineq2} 
(a+d+2c)(a+d-2c)\,<\,(a-d)^2,
\end{equation}
which on using Eq.~(\ref{trrho1}), gets simplified to 
\begin{eqnarray}
\label{ineq2}
(1-4c)<(a-d)^2.
\end{eqnarray}
From Eq.~(\ref{invsym}), we have
\begin{equation}
{\cal I}_4-{\cal I}_3^2=(a-d)^2\left((1-4c)-(a-d)^2\right),
\end{equation}
associated with the two qubit symmetric system $\varrho_{\rm sym}.$
It is obvious that ${\cal I}_4-{\cal I}_3^2<0$ \break when $(1-4c)<(a-d)^2$  i.e.,
when $\lambda_1$ is negative. \\
(ii)  When $\lambda_3<0,$ we can easily see that the invariant (see Eq.~(\ref{invsym})),
\begin{equation}
\label{eigen3}
{\cal I}_{5}=8\,(a-d)^2\,(c+\vert b\vert)\,(c-\vert b\vert)
\end{equation}
associated with the special class of symmetric two qubit states (Eq.~(\ref{cansym}))
is also negative. Thus $\lambda_3<0 \Longrightarrow {\cal I}_{5}<0$.

We have thus established an equivalence between the Peres's partial transpose criterion and the constraints on the %%@
invariants for two qubit symmetric state of the form $\varrho_{\rm (sym)}.$
In other words, the nonseparability conditions ${\cal I}_4-{\cal I}_3^2,\,{\cal I}_{5}<0,$ are 
both necessary and sufficient for a class of two-qubit symmetric states given by Eq.~(\ref{cansym}). 
%In the next few Chapters, we show that the non separability constraints
%${\cal I}_4-{\cal I}_3^2,\,{\cal I}_{5}<0,$ on the invariants serve as both necessary
%and sufficient for an arbitrary two qubit system.
\newpage
\section{Conclusions}
\label{conclusion}
We have shown that the number of 18 invariants as proposed by Makhlin~\cite{Mak02} for an arbitrary two qubit system %%@
gets reduced to 12
due to symmetry constraints. Further, we realize that a canonical form of two qubit symmetric density matrix achieved %%@
by identical local unitary transformations $U\otimes U$ restricts the minimum number of local invariants to specify an %%@
arbitrary symmetric two qubit system to 6. In other words,  we show that a subset of six invariants $\left\{{\cal %%@
I}_{1-6}\right\}$ of a more general set of 18 invariants proposed by Makhlin~\cite{Mak02}, is sufficient to 
characterize the nonlocal properties of a symmetric two qubit states. For a special class of two
qubit  symmetric states, only 3 invariants are sufficient to characterize the system. The invariants ${\cal I}_{4},\ %%@
{\cal I}_{5}$ and ${\cal I}_{4}-{\cal I}_3^2$ of separable symmetric two-qubit states are shown to be {\em %%@
non-negative}. We have proposed sufficient conditions for identifying entanglement in symmetric two-qubit states, when %%@
the qubits have a non-zero 
value for the average spin. Moreover these conditions on the invariants are 
shown to be necessary and sufficient for a class of symmetric two qubit states.

 %%%%%%%%%%%%%%%%%%%%%%%%%%%%%%%%%%%%%%%%%%%%%%%%%%%%%%%%%%%%%%%%%%%%%%%%%%%%%%
%%%%%%%%%%%%%%%%%%%%%%%%%%%%%%%%%%%%%%%%%%%%%%%%%%%%%%%%%%%%%%%%%%%%%%%%%%%%%%
%%%%%%%%%%%%%%%%%%%%%%%%%%%%%%%%%%%%%%%%%%%%%%%%%%%%%%%%%%%%%%%%%%%%%%%%%%%%%%
\chapter{Collective signatures of entanglement in symmetric multiqubit systems}
\label{c:multiqubit states}
\markboth{}{Collective signatures of entanglement in symmetric multiqubit systems}

\section{Introduction}
Quantum correlated systems of macroscopic atomic ensembles~\cite{macro} have been
drawing considerable attention recently. This is especially in view of their possible 
applications in atomic interferometers~\cite{Yur86,Kit91}  and high precision atomic clocks~\cite{Wineland94} and also %%@
in quantum information and computation~\cite{Niel00}.
Spin squeezing~\cite{Wineland92,Kit93} has been established as a
standard {\em collective} method to detect entanglement in these multiatom (multiqubit) systems.

Spin squeezing, is defined as the reduction of quantum fluctuations
in one of the spin components orthogonal to the mean spin direction below the
fundamental noise limit $N/4.$ Spin squeezing of around $N\approx 10^7$ atoms 
is nowadays routinely achieved in laboratories. It has been shown in Ref.~\cite{Xwang03} 
that spin squeezing is directly related to pairwise entanglement in 
atomic ensembles - though it provides a sufficient condition for inseparability.  

In the original sense, spin squeezing
is defined for multiqubit states belonging to the maximum multiplicity
subspace of the collective angular momentum operator $\vec{J}.$ 
Multiqubit states with highest collective angular momentum value
$J=\frac{N}{2}$ exhibit symmetry under the interchange of particles. 
In other words, the concept of spin squeezing is defined for symmetric 
multiqubit systems and it is reflected through collective variables
associated with the system.

In this Chapter, we concentrate on symmetric multiqubit systems and 
connect the average values of the collective first and second order spin
observables in terms of the two qubit state parameters.\footnote{As a consequence
of exchange symmetry, density matrices characterizing any random pair of qubits 
drawn from a symmetric multiqubit system are all identical. So, the
average values of  qubit observables are identical for any random pair of 
qubits belonging to a symmetric $N$-qubit system.}
We examine how collective signatures of pairwise entanglement,
like spin squeezing, manifest themselves via {\em negative values of
two qubit local invariants}. This leads to a classification of pairwise
entanglement in symmetric multi-qubit states in terms of the associated two qubit local 
invariants.
\section{Collective spin observables in terms of two qubit variables}
The collective spin operator $\vec{J}$ for a $N$ qubit system
is defined by,
\begin{eqnarray}
\label{angmomJ}
\vec{J}=\frac{1}{2}\sum_{\alpha=1}^N\vec{\sigma}_{\alpha}.
\end{eqnarray}
Let us concentrate on symmetric multiqubit systems, which respect exchange symmetry:
$$\Pi_{\alpha\beta}\, \rho^{(N)}_{\rm sym}= \rho^{(N)}_{\rm sym}\,\Pi_{\alpha\beta}=\rho^{(N)}_{\rm sym}, $$
where $\Pi_{\alpha\beta}$ denotes the permutation  operator interchanging
$\alpha^{\rm th}$ and $\beta^{\rm th}$ qubits.

The expectation value of collective spin correlations $\langle J_i \rangle$ is given
by 
\begin{equation}
\label{vars}
\langle J_i\rangle=\frac{1}{2}\sum_{\alpha=1}^N \langle \sigma_{\alpha
i}\rangle ;\,\,\,\,\,\,i=1,2,3.
\end{equation}
Here, we denote the average value $\langle \ldots \rangle$ of any observable by
\begin{equation}
\label{expec}
\langle\ldots\rangle={\rm Tr}\,[\rho_{\rm sym}\, (\ldots)].
\end{equation}
It may be noted that the density matrix of the $\alpha^{\rm th}$ qubit extracted from
a system of $N$ qubits, which respect exchange symmetry, is given by
\begin{eqnarray}
\label{alpharho}
\rho^{(\alpha)}_{\rm sym}&=&{\rm Tr}_{1,2,\ldots\, \alpha-1,\,\alpha+1,\ldots N}\rho^{(N)}_{\rm sym},\nonumber\\
&=&\frac{1}{2}[1+\sigma_i^{(\alpha)}s_i],\\
{\rm where,}\hskip 1in \nonumber\\
s_i&=&\rm {Tr}[\rho^{(\alpha)}_{\rm sym}\,\sigma_i^{(\alpha)}]\nonumber\\
%\end{eqnarray}
{\rm and}\hskip 1in \nonumber\\
%\begin{equation}
%\label{alpsigma}
 \vec{\sigma}^{(\alpha)}&=&I\otimes I\otimes \ldots\otimes \vec{\sigma}\otimes I\otimes
 \ldots\otimes I
 \end{eqnarray}
 is the $\alpha^{\rm th}$  qubit spin operator, with $\vec{\sigma}$ appearing at $\alpha^{\rm th}$
 position. In Eq.~(\ref{alpharho}), the density matrix 
 $\rho^{(\alpha)}_{\rm sym}$ of 
 the $\alpha^{\rm th}$ qubit is obtained by tracing the multiqubit state
$(\rho^{(N)}_{\rm sym})$ over all the qubit indices, expect $\alpha.$\\
We emphasize that the qubit averages {\em are independent of the qubit index $\alpha:$}
\begin{equation}
\label{ecpecsi}
\langle \sigma_i^{(\alpha)}\rangle=s_i.
\end{equation}

Therefore the first moments of the collective spin $\langle J_i \rangle$ 
(see Eq.~(\ref{vars})) assume the form
\begin{equation}
\label{vars1}
\langle J_i\rangle=\frac{1}{2}\sum_{\alpha=1}^N \langle \sigma_{\alpha
i}\rangle =\frac{N}{2}s_i
\end{equation}
in terms of the single qubit state parameter $s_i.$

For our further discussion, we need to associate the second order moments
of the collective spin variables i.e., $\langle (J_iJ_j+J_jJ_i)\rangle$ with
the two qubit correlation parameters.\\
Using Eq.~(\ref{vars}), we obtain,
\begin{eqnarray}
\label{vart}
\frac{1}{2}\langle (J_iJ_j+J_jJ_i)\rangle&=& \frac{1}{8}\,
\displaystyle\sum_{\alpha,\beta=1}^N\left\langle (\sigma_{\alpha
i}\sigma_{\beta j}
+\sigma_{\beta i}\sigma_{\alpha j})\right\rangle\nonumber\\
%&=&\frac{1}{4}\displaystyle
%\sum_{\alpha,\beta=1}^N\left\langle (\sigma_{\alpha i}\sigma_{\beta
%j})\right\rangle,\nonumber \\
&=&\frac{N}{4}\, \delta_{i\, j}+\frac{1}{4}\sum_{\alpha\neq\beta=1}^N\left\langle (\sigma_{\alpha
i}\sigma_{\beta j})\right\rangle. 
%&=&\frac{N}{4}\, \delta_{i\, j}+\frac{N(N-1)}{4}\left\langle (\sigma_{1
%i}\sigma_{2 j})\right\rangle. 
\end{eqnarray}
Here, $\left\langle (\sigma_{\alpha
i}\sigma_{\beta j})\right\rangle (\alpha\neq\beta)$ are the spin correlations 
of a pair of qubits $\alpha,\,\beta$ drawn randomly from
a symmetric multiqubit system. 

The density matrix $(\rho^{(\alpha\beta)}_{\rm sym})$
of such a pair of qubits  obtained by taking a partial trace
over the remaining $(N-2)$ qubits is given by 
\begin{eqnarray}
\label{alphabetarho}
(\rho^{(\alpha\beta)}_{\rm sym})&=&{\rm Tr}_{1,2,\ldots 
{\rm except}\,(\alpha,\beta)}(\rho^{(N)}_{\rm sym}).\nonumber
\end{eqnarray}
The general form of such a two qubit density matrix in terms of 
8 (see Eq.~(\ref{symrho})) state variables is given by,
\begin{eqnarray}
\label{alphabetarho1}
\rho^{(\alpha\beta)}_{\rm sym}&=&\frac{1}{4}\left(I\otimes I+ %%@
\sum_{i=1}^N\,s_i\,(\sigma_{i}^{(\alpha)}+\sigma_{i}^{(\beta)})\,
+\sum_{\alpha,\beta=1}^N t_{ij}\,\sigma_{i}^{(\alpha)}\sigma_{j}^{(\beta)}\right), 
\end{eqnarray}
where,
\begin{eqnarray}
\label{tmatele}
t_{ij}&=&{\rm Tr}\,[\rho^{(\alpha\beta)}_{\rm sym}(\sigma_{i}^{(\alpha)}\sigma_{j}^{(\beta)})]\nonumber\\
&=&\langle(\sigma_{i}^{(\alpha)}\sigma_{j}^{(\beta)})\rangle,
\end{eqnarray}
irrespective of the qubit indices $\alpha, \beta.$

Substituting Eq.~(\ref{tmatele}) in Eq.~(\ref{vart}), we obtain
\begin{eqnarray}
\label{varcorrt}
\frac{1}{2}\langle (J_iJ_j+J_jJ_i)\rangle&=&\frac{N}{4}\, \delta_{i\, j}+
\frac{N(N-1)}{4}\left\langle (\sigma_{1i}\sigma_{2 j})\right\rangle\nonumber\\
&=&\frac{N}{4}\,\left[ \delta_{i\, j}+ (N-1)\, t_{ij}\, \right],  \hskip 0.5in
i,j=1,2,3.
\end{eqnarray}
From Eqs.~(\ref{vars1}),~(\ref{varcorrt}), it is evident that the collective spin 
observables (upto first and second order) can be expressed in terms of state parameters
of a pair of qubits  chosen arbitrarily from a symmetric $N$ qubit state.
Thus the collective pairwise entanglement behavior in symmetric multiqubits
results from the properties of the two qubit state parameters $\left\{s_i,\,t_{ij}\right\}.$ 

We now proceed to  identify collective  
criteria of pairwise entanglement in a symmetric multi-qubit state, in
terms of the two-qubit local invariants  $\{ {\cal I}_1\, -\, {\cal I}_6\}$.
%{\em spin squeezing} in $N$-qubit symmetric states. 
\section{Collective signatures of pairwise entanglement}
Collective phenomena,
reflecting pairwise entanglement of qubits, can be
expressed through two qubit local invariants as the first and second
moments $\langle J_i\rangle,\,\langle (J_iJ_j+J_jJ_i)\rangle$ are related
to two qubit state parameters $\left\{s_i,\,t_{ij}\right\}$ in symmetric multiqubit systems. Here, we
show that spin squeezing- which is one of the collective signatures of pairwise 
entanglement in symmetric multiqubit systems-gets
reflected through one of the separability criterion derived in Chapter~\ref{c:c2twoqubit}.

For the sake of continuity, the main result of Chapter~\ref{c:c2twoqubit} is summarized in the following sentence: 

 {\em The non-positive  values of the invariants
$\{I_{4},\  I_{5}$ or  $I_{4}-I_3^2\}$  serve  as  a signature of entanglement
and hence provide  sufficient conditions for non-separability of the quantum state}.

Let us now review spin squeezing criteria.  
Kitagawa and Ueda~\cite{Kit93} pointed out that a definition of spin squeezing,~\cite{wod}
based only
on the uncertainty relation,
\begin{eqnarray}
\label{uncertainty}
(\bigtriangleup J_1)^2 (\bigtriangleup J_2)^2 \geq \frac{|\langle J_3 \rangle|^2}{4}
\end{eqnarray}
exhibits co-ordinate frame dependence and does not arise from the 
quantum correlations among the elementary spins. They identified
a mean spin direction
\begin{eqnarray}
\label{spin squee}
\hat{n}_0=\frac{\langle \vec{J}\rangle}{\vert \langle \vec{J}\rangle \vert},
\end{eqnarray}
where, $\vert \langle \vec{J}\rangle \vert=\sqrt{\langle \vec{J}\rangle\cdot\langle \vec{J}\rangle}$
%\begin{eqnarray}
%\label{spin squee1}
%\end{eqnarray}
(The collective spin operator $\vec{J}$ for an $N$ qubit system is  given by Eq.~(\ref{vars})).

Associating a mutually orthonormal set $\left\{\hat{n}_{1\,\perp},\,\hat{n}_{2\,\perp},\,\hat{n}_0\right\},$
with the system, let us consider the following collective operators,
\begin{eqnarray}
\label{spin squee2}
J_{1\,\perp}=\vec{J}\cdot\hat{n}_{1\perp},\,\,\,J_{2\perp}=\vec{J}\cdot
\hat{n}_{2\,\perp}\,\,\,
{\rm and}\,\,\, J_{0}=\vec{J}\cdot\hat{n}_{0}
\end{eqnarray}
which satisfy the usual angular momentum commutation relations
\begin{eqnarray}
\label{commutation}
[J_{1\perp},J_{2\perp}]=i\,J_{0}.
\end{eqnarray}
Now, employing a collective spin component $J_\perp$ orthogonal to the mean spin 
direction $\hat{n}_0,$ given by, 
\begin{eqnarray}
\label{spin squee3}
J_\perp&=&\vec{J}\cdot\hat{n}_{\perp}\nonumber\\
&=&J_{1\perp} \cos\theta + J_{2\perp} \sin\theta,
\end{eqnarray}
minimization of the variance,
\begin{eqnarray}
\label{spinvariance}
(\bigtriangleup J_{\perp})^2=\langle J_\perp^2\rangle-\langle J_\perp\rangle^2
\end{eqnarray}
can be done over the angle $\theta.$ Kitagawa and Ueda~\cite{Kit93} proposed that a multiqubit
state can be regarded as spin squeezed if the minimum of $\Delta J_\perp$
is smaller than the standard quantum limit
$\frac{\sqrt N}{2}$ of the spin coherent state.

A spin squeezing parameter incorporating this feature is defined by~\cite{Kit93} 
\begin{equation}
\label{xi}
\xi= \frac{2\,(\Delta J_\perp)_{\rm min}}{\sqrt N}. 
\end{equation}
Symmetric multiqubit states with $\xi<1$ are spin squeezed.
We next proceed to show that the two qubit local invariant ${\cal I}_5$ and the spin squeezing parameter $\xi$ are %%@
related to each other. 
\subsection{Spin squeezing in terms of the local invariant {\bf ${\cal I}_5$}}
We now prove the following theorem. 
\begin{theorem}
 For all spin squeezed states, the local invariant ${\cal I}_5$ is negative.
\end{theorem}
\noindent {\bf  Proof.}
%To prove that ${\cal I}_5<0$ implies {\em spin squeezing}, 
%The invariant ${\cal I}_5$ is given by 
%$${\cal I}_{5} = \epsilon_{ijk}\,\epsilon_{lmn}\, s_i\,s_l t_{jm}\,
%t_{kn}$$
It is useful to evaluate the invariant ${\cal I}_5$ (see Eq.~(\ref{inv})),
after subjecting the quantum state to a identical local rotation $U\otimes U\otimes U\otimes \cdots$
on all the qubits,
which is designed to align the average spin vector
$\langle \vec{J}\rangle$ along the $3$-axis. {\em After this local rotation,}
orientation of the qubits would be along $3$ axis and the qubit orientation vector is given by 
$$\vec{s}\equiv (0,\ 0,\ s_0).$$ We may then express the local invariant ${\cal I}_5$ as,
\begin{eqnarray}
\label{ssI5}
{\cal I}_{5} &=&\epsilon_{ijk}\,\epsilon_{lmn}\, s_i\,s_l\, t_{jm}\, t_{kn}\nonumber\\ 
&=&\epsilon_{3jk}\,\epsilon_{3mn}\, s_0^2 \, t_{jm}\,t_{kn}\nonumber \\
             &=& 2\, s_0^2\, (t'_{11}t'_{22}-(t'_{12})^2)\nonumber \\
             &=& 2\, s_0^2\, \det T_{\perp},
\end{eqnarray}
where, 
\begin{equation}
\label{tperpen}
T_{\perp}
=\left(\begin{array}{cc}
                     t'_{11} & t'_{12}  \cr 
                      t'_{12} & t'_{22}\cr
					 \end{array}\right),   
\end{equation}
denotes the $2\times 2$ block of the correlation matrix in
the subspace  orthogonal to the qubit orientation direction i.e., 3-axis.

Now, we can still exploit the freedom of local rotations $O_{12}$ in the
$1-2$ plane, which leaves the average  spin $\vec{s}= (0,\ 0,\ s_0)$
unaffected. We use this to diagonalize
$T_{\perp}$: 
\begin{eqnarray}
O_{12} T_\perp O^T_{12}
=T^d_{\perp}&=&
\left(\begin{array}{ll}t_\perp^{(+)} & 0 \\ 0 & t_{\perp}^{(-)}\end{array}\right) 
\end{eqnarray} 
with the diagonal elements given by
$$t_{\perp}^{(\pm)}=\frac{1}{2}\left[(t'_{11}+t'_{22})\pm\sqrt{(t'_{11}
-t'_{22})^2+4\, (t'_{12})^2}\right].$$
We once again emphasize that local rotations on the qubits leave the invariants unaltered 
and here, we choose local operations to transform the two qubit state variables as,
\begin{equation}
\label{sz}
\vec{s}=(0, 0, s_0),
\end{equation}
{\rm and}
\begin{equation}
\label{tperp}
T=\left(\begin{array}{lll}t_\perp^{(+)} & 0 & t''_{13}\\
 0 & t_{\perp}^{(-)}& t''_{23}\\
 t''_{13} & t''_{23}& t'_{33}
\end{array}\right), 
\end{equation} 
so that the two qubit invariant ${\cal I}_5$ can be expressed as,
\begin{eqnarray}
\label{ss2I5}
{\cal I}_5&=& 2\, s_0^2\, \det T_{\perp},\nonumber\\
&=& 2\ s_{0}^2\ t_{\perp}^{(+)}\, t_{\perp}^{(-)}.
\end{eqnarray}

We now express the spin squeezing parameter $\xi$, given by Eq.~(\ref{xi}),
in terms of the two-qubit state parameters
\begin{eqnarray}
 \label{xi2}
\xi^2&=&\frac{4\,(\Delta J_\perp)^2_{\rm min}}{N}, \nonumber\\
&=&\frac{4\,\langle (\vec{J}\cdot\hat{n}_\perp)^2\rangle-\langle (\vec{J}\cdot\hat{n}_\perp)\rangle^2}{N}.
\end{eqnarray}
Writing the collective spin operator $\vec{J}$ for an $N$ qubit system
in terms of the two qubit operators (see Eq.~(\ref{angmomJ})), we obtain
\begin{eqnarray}
    \xi^2 &=& \frac{1}{N}\, \displaystyle\sum_{\alpha,\beta=1}^N \, \left\langle
(\vec{\sigma}_{\alpha}\cdot\hat{n}_{\perp})\,
         (\vec{ \sigma}_{\beta}\cdot\hat{n}_{\perp})\right\rangle_{\rm min}
\nonumber \\
     &=&1+\frac{1}{N} \displaystyle\sum_{\alpha=1}^N
\displaystyle\sum_{\beta\neq\alpha=1}^N
           \left\langle (\vec{\sigma}_{\alpha}\cdot\hat{n}_{\perp})\,
           (\vec{\sigma}_{\beta}\cdot\hat{n}_{\perp})\right\rangle_{\rm
min}\nonumber \\
      &=&  1+  \frac{2}{N}\, \displaystyle\sum_{\alpha=1}^N
\displaystyle\sum_{\beta>\alpha=1}^N
            \left(\displaystyle\sum_{i,j=1}^3 \, \left\langle (\sigma_{\alpha\,
i}\sigma_{\beta\, j})\right\rangle
           n_{\perp i}\, n_{\perp j}\right)_{\rm min}.
\end{eqnarray}
Since for a symmetric system, we have $\left\langle \sigma_{\alpha\, i}\sigma_{\beta\,
j}\right\rangle =t_{ij}$, we express the squeezing parameter as follows:
\begin{eqnarray}
\label{xi3}
\xi^2&=& 1+(N-1)\, \left(\displaystyle\sum_{i,j=1}^3 \, t_{ij}\, n_{\perp i}\,
n_{\perp j}\right)_{\rm min} \nonumber \\
     &=& 1+(N-1)\, (n_{\perp}^T\, T\, n_{\perp})_{\rm min}.
\end{eqnarray}
In Eq.~(\ref{xi3}), we have denoted  the row vector $n_{\perp}^T =\left(
n_{1\perp},\, n_{2\perp},\, 0\right)=\left(\cos\theta,\, \sin\theta,\, 0\right).$ The minimum value of the quadratic %%@
form
$(n_{\perp}^T\, T\, n_{\perp})_{\rm min}$ in Eq.~(\ref{xi3}) is fixed as
follows:
\begin{eqnarray}
\label{min}
(n_{\perp}^T\, T\, n_{\perp})_{\rm min}&=&\left(\begin{array}{c}{\rm min}\\
\theta\end{array}\right) \left(
t'_{11}\,\cos^2\theta+t'_{22}\,\sin^2\theta+t'_{12}
\sin\, 2\theta\right)\nonumber
\\
&=& \frac{1}{2}\left[(t'_{11}+t'_{22})-\sqrt{(t'_{11}
-t'_{22})^2+4\,
(t'_{12})^2}\right]\nonumber \\
&=& t_{\perp}^{(-)},
\end{eqnarray}
where $t_{\perp}^{(-)}$ is the least eigenvalue of $T_{\perp}$ (see
Eq.~(\ref{tperpen})). 

We finally obtain,
\begin{equation}
\label{xifinal}
\xi^2=\frac{4}{N}\, (\Delta\, J_{\perp})^2_{\rm min}=\left(1+(N-1)\,
t_{\perp}^{(-)}\right).
\end{equation}
Following similar lines we can also show that
\begin{equation}
\label{maxfluct}
\frac{4}{N}\, (\Delta\, J_{\perp})^2_{\rm max}=\left(1+(N-1)\,
t_{\perp}^{(+)}\right),
\end{equation}
which relates the eigenvalue $t_{\perp}^{(+)}$ of $T_{\perp}$ to the maximum
collective fluctuation  $(\Delta\, J_{\perp})^2_{\rm max}$
orthogonal to the mean spin direction.
Substituting Eqs.(\ref{xifinal}), (\ref{maxfluct}),  and expressing \break
$s_{0}=\frac{2}{N}\vert\langle J_3\rangle\vert$ in Eq.~(\ref{ss2I5}), we get,
\begin{equation}
\label{invsq}
{\cal I}_5= \frac{8\,
\vert\langle\vec{J}\rangle\vert^2}{\left(N(N-1)
\right)^2}\,
\left(\xi^2-1\right)\,
(\frac{4}{N}\, (\Delta\, J_{\perp})^2_{\rm max}-1).
\end{equation}
Having related the local invariant ${\cal I}_5$ to collective spin observables,
we now proceed to show  that
${\cal I}_5<0$  iff $\xi^2<1$ i.e.,  iff the state is spin
squeezed.

Note that the two qubit correlation parameters $t_{ij}$ are bound by
$$-1\leq t_{ij}\leq 1.$$ This bound, together
with the unit trace condition ${\rm Tr}\, (T)=1$ on the correlation matrix of a
symmetric two-qubit state, leads to the
identification that {\em only one} of the diagonal elements of $T$ can be
negative. This in turn implies that if the diagonal element $t^{(-)}_\perp$ is negative,
then the other diagonal element $t^{(+)}_\perp$ is necessarily positive. 
Thus, from Eq.~(\ref{maxfluct}), it is evident that
$$t_{\perp}^{(+)}=\left( \frac{4}{N}(\Delta\, J_{\perp})^2_{\rm
max}-1\right)\geq 0.$$
whenever $t^{(-)}_\perp<0.$ It is therefore clear (from Eq.~(\ref{invsq})) that
a symmetric multiqubit state is spin-squeezed  {\em iff} ${\cal I}_5<0$. 
In other words,
\begin{equation}
\label{I5spin} 
\xi^2<1 \Longleftrightarrow {\cal I}_5<0.
\end{equation}
Further, from the structure of the invariant ${\cal I}_5$ (Eq.~(\ref{inv})),
it is clear that ${\cal I}_5<0$ implies 
$$\,(s_1^2\, t_2\,t_3+s_2^2\, t_1\, t_3+s_3^2\, t_1\, t_2)<0,$$
i.e., one of the eigenvalues $t_1, \, t_2$ or $t_3$  of the correlation matrix $T$
must be negative. This in turn implies that the invariant
$${\cal I}_1=t_1t_2t_3<0.$$
In other words, when ${\cal I}_5<0,$ the invariant ${\cal I}_1$
is also negative. 

We now explore other collective signatures of pairwise entanglement, which are
manifestations of   {\em negative values}
of the invariants   ${\cal I}_4$ and  ${\cal I}_4-{\cal I}_{3}^2$.
\subsection{Collective signature in terms of {\bf ${\cal I}_4$}}
When the average spin is aligned along the $3$-axis through local rotations
such that \break $\vec{s}= (0,\ 0,\ s_0),$ the local invariant ${\cal I}_4$ (see Eq.~(\ref{inv}))
assumes the form,
\begin{eqnarray}
\label{I4colle}
{\cal I}_4=s^T\,T\,s=
 \left(\begin{array}{lll}0 & 0 & s_0\end{array}\right)
 \left(\begin{array}{lll}t_\perp^{(+)} & 0 & t''_{13}\\
 0 & t_{\perp}^{(-)}& t''_{23}\\
 t''_{13} & t''_{23}& t'_{33}\end{array}\right)\,
\left(\begin{array}{l}
0\\
 0 \\
 s_0\end{array}\right)\,
&=&s_0^2t'_{33}.  
\end{eqnarray}
It is evident from the above equation that 
$${\cal I}_4<0\,\,\,{\rm iff}\,\,\,t'_{33}<0.$$
Simplifying Eq.~(\ref{vars1}) and Eq.~(\ref{varcorrt}), we express $t'_{33}$ and $s_0$ in terms of the collective spin %%@
observables  i.e.,
\begin{eqnarray}
\label{colli4}
s_0&=&\frac{2}{N}\langle\vec{J}\cdot\hat{n}_0\rangle
=\frac{2}{N}|\langle \vec{J}\rangle|, \nonumber\\
t'_{33}&=&\frac{1}{(N-1)}\,
\left(\frac{4}{N}\langle(\vec{J}\cdot \hat{n}_0)^2\rangle -1\right),
\end{eqnarray}
leading further to the following structure for the invariant ${\cal I}_4$
\begin{equation}
\label{I42}
{\cal I}_{4}=\frac{4}{N^2\,(N-1)}\, \vert\langle\vec{J}\rangle \vert ^2\,
\left(\frac{4}{N}\langle (\vec{J}\cdot \hat{n}_0)^2\rangle -1\right),
\end{equation}
where $\hat{n}_0$ denotes a unit vector along the direction of mean spin. We
therefore read
from Eq.~(\ref{I42}), that {\em the average  of the squared spin component,
along the \underbar {mean spin direction}, reduced below the value  $N/4$,  signifies pairwise
entanglement in symmetric $N$-qubit system.}

Further, we may note that when  ${\cal I}_4\leq 0,$ the invariant ${\cal I}_5,$
which reflects spin squeezing is \underbar{{\em not negative}}.    
This is because $t'_{33}\leq 0\,\,\,\Longrightarrow\,\,\,t_\perp^{(\pm)}\geq 0$ as,

\noindent(i)\,\,\,$t_{\perp}^{(+)}+t_{\perp}^{(-)}+t'_{33}=1,$\,\,\,\,\,\,
 (unit trace condition),\,\, and\\
\noindent(ii)\,\,\,$-1\leq t_{\perp}^{(\pm)},\ t'_{33} \leq 1$.\\
Therefore, {\em spin squeezing and $\langle (\vec{J}\cdot
\hat{n}_0)^2\rangle\leq \frac{N}{4}$ are two mutually exclusive
criteria of pairwise entanglement}. 

However, from
the structure of the invariant ${\cal I}_{4}$, as given in
Eq.~(\ref{abss}), it is obvious that 
\begin{equation}
\label{i4<0}
 {\cal I}_4=s_1^2\, t_1+s_2^2\, t_2+s_3^2\, t_3\leq 0
 \end{equation}
 implying that one of the eigenvalues $t_1,\,t_2,\,t_3 $ of two qubit correlation matrix 
 must be negative. This leads to the identification that
 \begin{equation*}  
 {\cal I}_1=t_1\, t_2\,t_3\,\leq 0.
 \end{equation*}
We now continue 
to relate the combination of invariants ${\cal I}_4-{\cal I}_3^2$ to
the collective spin observables.
\subsection{{\bf ${\cal I}_4-{\cal I}_3^2$} in terms of the collective variables}
In terms of the two qubit state parameters, expressed
in a conveniently chosen local coordinate system 
(see Eqs.~(\ref{sz}),~(\ref{tperp})), the 
invariant combination ${\cal I}_4-{\cal I}_3^2$ can be written as,
\begin{equation}
\label{I4I31}
{\cal I}_4-{\cal I}_3^2=s^{\rm T}\,T\, s-s^{\rm T}\, s=s_0^2\, (t'_{33}-s_0^2). 
\end{equation}
Since the state parameters can be expressed in terms of the 
expectation values of the collective spin variables Eq.~(\ref{colli4}), the
invariant quantity  ${\cal I}_4-{\cal I}_3^2$ may be rewritten as 
\begin{eqnarray}
\label{I4I32}
{\cal I}_4-{\cal I}_3^2&=& \frac{4}{N^2}\, \vert \langle\vec{J}\rangle\vert^2\, 
\left[\frac{4}{N(N-1)}\,
\langle(\vec{J}\cdot \hat{n}_0)^2 \rangle -\frac{1}{(N-1)}-
\frac{4}{N^2}\,\vert\langle\vec{J}\rangle\vert^2
\right],\nonumber \\
\nonumber \\
&=& \frac{16}{N^3(N-1)}\, \vert \langle\vec{J}\rangle\vert^2\,
\left[\langle(\vec{J}\cdot\hat{n}_0)^2\rangle-\left(\frac{N}{4}\,+\frac{(N-1)}{N}\,
\vert\langle\vec{J}\rangle\vert^2\right)\right].
\end{eqnarray}  
 Negative value of the combination ${\cal I}_4-{\cal I}_3^2$ manifests
itself through 
$$\langle(\vec{J}\cdot \hat{n}_0)^2\rangle<  \frac{N}{4}+\frac{(N-1)}{N}\,
\vert\langle\vec{J}\rangle\vert^2.$$
From Eqs.~({\ref{I42}) and (\ref{I4I32}), we conclude that
{\em pairwise entanglement  resulting from
$${\cal I}_3\neq 0,\,\,\,\,\, {\cal I}_{4}> 0,\,\, {\rm  but}\,\,
{\cal I}_4-{\cal I}_3^2< 0,$$  
is realized, whenever}
$$\frac{N}{4}< \langle(\vec{J}\cdot \hat{n}_0)^2\rangle <
\frac{N}{4}+\frac{(N-1)}{N}\, \vert\langle\vec{J}\rangle\vert^2.$$
All the cases discussed above are valid when the average spin vector 
$\vert\langle\vec{J}\rangle\vert\neq 0$
 i.e., when $\vert\langle\vec{J}\rangle\vert$ is oriented along the 3-axis. In the special
case when $\vert\langle\vec{J}\rangle\vert=0,$ we show that pairwise entanglement
manifests itself through negative value of the local invariant ${\cal I}_1.$    
\subsection{Characterization of pairwise entanglement through 
{\bf ${\cal I}_1$}}
In the cases where the qubits have no preferred orientation, i.e., when 
$\vert\langle\vec{J}\rangle\vert=0$, it is evident from  Eq.~({\ref{inv}) that
the local  invariants  $\left\{{\cal I}_{3-6}\right\}$ are zero
\begin{eqnarray}
\label{invzero}
 {\cal I}_3&=&s^T\, s=0,  \nonumber \\
 {\cal I}_4&=&s^T\, T\, s=0, \nonumber \\
{\cal I}_{5}&=&\epsilon_{ijk}\,\epsilon_{lmn}\, s_i\,s_l\, t_{jm}\, t_{kn}=0, \nonumber \\
{\cal I}_{6}&=&\epsilon_{ijk}\, s_i\, (T\, s)_j\, (T^2\,s)_k=0.
\end{eqnarray}
The  remaining two nonzero invariants ${\cal I}_{1}$ and ${\cal I}_2$ are, 
\begin{eqnarray}
\label{invnonzero}
{\cal I}_1&=&{\rm det}\, T, \nonumber\\
{\cal I}_2&=&{\rm Tr}\, (T^2).
\end{eqnarray}
In such situations, i.e., when  $\vert\langle\vec{J}\rangle\vert=0,$
{\em pairwise entanglement manifests itself through ${\cal I}_1<0.$}

Writing the invariant ${\cal I}_1$ in terms of collective observables Eq.~(\ref{varcorrt}), we have,
\begin{equation}
{\cal I}_1={\rm det}\, T=t_1t_2 t_3=\left(\frac{4}{N(N-1)} \right)^3\, \displaystyle\prod_{i=1}^3
\left(\langle J_{i}^2\rangle-\frac{N}{4}\right).
\end{equation}
{\em Negative value of ${\cal I}_1$ shows up through 
$\langle J_{i}^2\rangle<\frac{N}{4}$}  along the  axes $i=1,\, 2\, $ or 3,
which are fixed by verifying   
$\langle (J_i J_j+J_j J_i)\rangle=0; \  i\neq j$, as $T$ is diagonal with such a choice of the axes.\\
Note that, 
$${\cal I}_1={\rm det}\, T<0\,\, \Longrightarrow \,\,{\cal I}_2={\rm Tr}\, (T^2)>1$$
since $t_1+t_2+t_3=1$ (unit trace condition) and $-1\leq t_1,\ t_2, t_{3}\leq 1.$ Therefore we have,
\begin{eqnarray}
\label{I1I2<0}
{\cal I}_1<0,\,\,\,\,{\cal I}_2>1,
\end{eqnarray}
both implying pairwise entanglement.\\
Thus, we have related the two qubit entanglement invariants Eq.~(\ref{inv})
to the collective spin observables and shown that the collective signatures of pairwise entanglement are manifested %%@
through the negative values of the invariants
${\cal I}_4,\,{\cal I}_5,\,{\cal I}_4-{\cal I}_3^2.$\\
In Sec.~\ref{equalities},
(a) we relate the invariant criteria with the recently proposed generalized spin squeezing inequalities for two %%@
qubits~\cite{Kor} and (b) propose a classification scheme for pairwise entanglement in symmetric multiqubit systems.
 
\section{Classification of pairwise entanglement}
\label{equalities}
Recently, Korbicz {\it et al.}~\cite{Kor} 
proposed generalized spin squeezing inequalities for pairwise entanglement, which provide necessary and sufficient %%@
conditions
for genuine 2-, or 3- qubit entanglement for symmetric states:
These generalized spin squeezing inequalities are given by~\cite{Kor} 
\begin{eqnarray}
\label{ineq}
\frac{4\langle\Delta J_k\rangle^2}{N}<1-\frac{4\langle J_k\rangle^2}{N^2}
\end{eqnarray}
where $J_k=\vec{J}\cdot\hat{k}$\,; with $\hat{k}$ denoting an arbitrary unit vector.\\
We now show that that the generalized spin squeezing inequality given in
Eq.~(\ref{ineq}) can be related to our invariant criteria.

We consider various situations as discussed below:\\
(i) Let  $\hat{k}=\hat{n}_\perp,$ a direction orthogonal to the mean spin vector $\langle\vec{J}\rangle$. The %%@
inequality given by
Eq.~(\ref{ineq}) reduces to
$$(\Delta J_{n_\perp})^2 < \frac{N}{4}.$$
Minimizing the variance $(\Delta J_{n_\perp})^2$ we obtain
$$\xi^2=\frac{4(\Delta J_{\perp})_{\rm min}^2}{N}<1.$$
\newpage
\noindent This is nothing but the conventional spin squeezing condition ${\cal I}_5<0$ in terms of the invariant.\\
(ii) If $\hat{k}$ is aligned along the mean spin direction i.e.,
$\hat{k}=\hat{n}_0$ with $\hat{n}_0=\frac{\langle\vec{J}\rangle}{\vert \langle\vec{J}\rangle \vert},$ the generalized %%@
spin squeezing inequalities (see Eq.~(\ref{ineq})) reduce to the form,
\begin{eqnarray}
\label{rineq}
\langle(\vec{J}\cdot\hat{n}_0)^2\rangle<\frac{N}{4}+\frac{(N-1)}{N}\vert \langle \vec{J}\rangle \vert^2.
\end{eqnarray}
From Eq.~(\ref{I42}) we have,
\begin{equation}
\label{I42a}
{\cal I}_{4}=\frac{4}{N^2\,(N-1)}\, \vert\langle\vec{J}\rangle \vert ^2\,
\left( \frac{4}{N}\langle (\vec{J}\cdot \hat{n}_0)^2\rangle -1\right).
\end{equation}
Now the condition ${\cal I}_4<0$ on the local invariant leads to the collective signature [see Table.1]
$$\langle(\vec{J}\cdot\hat{n}_0)^2\rangle<\frac{N}{4},$$
which is a stronger restriction than that given by Eq.~(\ref{rineq}).\\
Further, if ${\cal I}_4>0$ but ${\cal I}_4-{\cal I}_3^2<0$ we obtain the inequality [see Table.1]
$$ \frac{N}{4}<\langle(\vec{J}\cdot\hat{n}_0)^2\rangle<\frac{N}{4}+\frac
{(N-1)}{N}\vert \langle \vec{J}\rangle \vert^2,$$
which covers the remaining range of possibilities contained in the
generalized spin squeezing inequalities of Eq.~(\ref{rineq}) with $\hat{n}$ along the mean spin direction.
\\
(iii) If the average spin is zero for a given state i.e., we have $\langle J_k\rangle=0$  for all directions %%@
$\hat{k}$. The inequalities of Korbicz {\it et al.}~\cite{Kor} assume a simple form $$\langle %%@
J_k\rangle^2<\frac{N}{4}.$$ This case obviously corresponds to ${\cal I}_3=0$ and ${\cal I}_1<0.$ 
\newpage   
In the following table we summarize the results and prescribe a classification of pairwise entanglement in symmetric %%@
multiqubit states.

\begin{table}[ph] 
\centering  %Table~1
{\begin{tabular}{|c|c|c|} \hline
\multicolumn{2}{|c|}
 {} &  \\ 
\multicolumn{2}{|c|}
{Criterion of pairwise entanglement}  & Collective
behaviour to look for \\ 
\multicolumn{2}{|c|}
 {} &  \\ 
\hline
& &  \\
& & \\
& ${\cal I}_5\leq 0$ & $(\Delta J_\perp)^2_{\rm min} \leq \frac{N}{4}$\\
& & \\
& & \\
\cline{2-3}
& & \\
& & \\
 $\,\,\,{\cal I}_3\neq 0\,\,\,$ & ${\cal I}_4\leq 0$ &
$\langle(\vec{J}\cdot\hat{n}_0)^2\rangle \leq \frac{N}{4}$ \\
& & \\
& & \\
 \cline{2-3}
 & & \\
 & & \\
 & ${\cal I}_4 > 0,\  {\cal I}_4-{\cal I}_3^2 < 0$
  & $ \frac{N}{4}< \langle(\vec{J}\cdot\hat{n}_0)^2\rangle <
\frac{N}{4}+\frac{(N-1)}{N}\, \vert\langle\vec{J}\rangle\vert^2$ \\
& & \\
& & \\
\hline
& & \\
& & \\
 ${\cal I}_3 =0$ & ${\cal I}_1 < 0$  & $\langle J_i^2\rangle < \frac{N}{4}$\\
& & for any direction $i=1,\,2,\, 3,$ so that\\
& &  $\langle (J_i\, J_j+J_j\, J_i)\rangle=0;\ {\rm for\ } i\neq j$ \\
& & \\
\hline
\end{tabular}}
\\
\caption{Classification of pairwise entanglement in symmetric multi-qubit states
in terms of two-qubit local invariants.}
\label{tab1}
\end{table}
\pagebreak
\section{Conclusions}
In summary, we have shown that a set of six local invariants $\{ {\cal I}_1\ -\  {\cal I}_6\}$, associated
with the two-qubit partition of a symmetric multiqubit system, characterizes  the pairwise entanglement
properties of the collective state. 
We have proposed a detailed classification scheme,  for pairwise entanglement in symmetric
multiqubit system, based on  {\em negative} values of the  invariants
${\cal I}_1,\, {\cal I}_4,\, {\cal I}_5$ and ${\cal I}_4-{\cal I}_3^2$ .  Specifically, we have shown, collective spin %%@
squeezing in symmetric multi-qubit states is a manifestation of ${\cal I}_5<0$. Moreover,
we have related our criteria, which are essentially given in terms of invariants of the quantum state, to the recently %%@
proposed generalized spin squeezing inequalities~\cite{Kor} for two qubit entanglement.

 %%%%%%%%%%%%%%%%%%%%%%%%%%%%%%%%%%%%%%%%%%%%%%%%%%%%%%%%%%%%%%%%%%%%%%%%%%%%%%
%%%%%%%%%%%%%%%%%%%%%%%%%%%%%%%%%%%%%%%%%%%%%%%%%%%%%%%%%%%%%%%%%%%%%%%%%%%%%%
%%%%%%%%%%%%%%%%%%%%%%%%%%%%%%%%%%%%%%%%%%%%%%%%%%%%%%%%%%%%%%%%%%%%%%%%%%%%%%
\chapter{Dynamical models}
\label{c:c4dynamicalmodels}
\markboth{}{Dynamical models }

In the previous chapters, we have proposed separability criteria for
symmetric multiqubit states in terms of two qubit local invariants. 
In the light of our characterization for pairwise 
entanglement, we analyze few symmetric
multi-qubit dynamical models like, 
\begin{enumerate}
\item{Dicke states~\cite{Dic54,Xi02}}
 \item{Kitagawa-Ueda state generated by one axis twisting Hamiltonian~\cite{Kit93}} 
 \item{Atomic squeezed states~\cite{Agarwal90}}. 
  \end{enumerate}
\section{Dicke State}
Collective spontaneous emission from dense atomic systems has been of interest since the pioneering work of Dicke, who %%@
predicted that two-level atoms (or qubits) possess collective quantum states in which spontaneous emission is enhanced %%@
(superradiance) or suppressed (subradiance). Multiqubit Dicke states are of interest for quantum information %%@
processing because they are robust under qubit loss~\cite{stockton, rajgop} and stand as an example of %%@
decoherence-free subspaces. 
An N-qubit symmetric Dicke states  $\vert J=\frac{N}{2}, M \rangle\,  ;\ -J\leq M \leq J$,
with $(\frac{N}{2}-M)$ excitations (spin up) is defined as~\cite{stockton}
\begin{eqnarray}
\label{dicke qubits}
\left\vert \frac{N}{2}, M\right\rangle=
\displaystyle\left(\begin{array}{c}
N \cr
M \cr
\end{array}\right)^{-\frac{1}{2}}\,
\sum_k P_k\left(\vert 1_1,1_2,...1_m,0_{m+1},...0_N\rangle\right)
\end{eqnarray}
where $m=\frac{N}{2}-M$ and $\left\{P_k\right\}$ is the set of all distinct permutations of the spins. 
A well known example is the W-state given by,
\begin{eqnarray}
\label{wstate}
\left\vert \frac{N}{2},\, M=\frac{N}{2}-1\right\rangle= \frac{1}{\sqrt{N}}\left[\left\vert 1_10_20_3 \cdots %%@
0_N\right\rangle+\left\vert 0_11_20_3 \cdots 0_N\right\rangle+\cdots+\left\vert 0_10_20_3 \cdots %%@
1_N\right\rangle\right],
\end{eqnarray}
which is a collective spin state with one excitation.

Multiqubit Dicke states are symmetric under permutation of atoms and entanglement-robust against particle %%@
loss~\cite{stockton}.
They exhibit unique entanglement properties~\cite{Usha-Raj1} and are excellent candidates for experimental %%@
manipulation and characterization of genuine multipartite entanglement. It has been shown that a wide family of Dicke %%@
states  can be generated in an ion chain by single global laser pulses~\cite{Retzker}. Further, a selective technique %%@
that allow a collective manipulation of the ionic degrees of freedom inside the symmetric Dicke subspace has been %%@
proposed~\cite{Lopez}. An experimental scheme to reconstruct the spin-excitation number distribution of the collective %%@
spin states  i.e.,  tomographic reconstruction of the diagonal elements of the density matrix in the Dicke basis of %%@
macroscopic ensembles containing   atoms, with low mean spin excitations, has also been put forth~\cite{Koji-Usami}. %%@
More recently, Thiel et. al.~\cite{Thiel-GSA} proposed conditional detection of photons in a Lambda system, as a way %%@
to produce symmetric Dicke states. 

In order to analyze the pairwise entanglement properties of N-qubit Dicke states 
in terms of two qubit local invariants, we need to 
evaluate the first and second order moments  $\langle J_i\rangle,$ $\langle J_iJ_j+J_jJ_i\rangle$ of the collective %%@
spin observable. 
These moments in turn allow us to determine the two qubit state parameters
%(see Eqs.~(\ref{vars1}),~(\ref{varcorrt}))
associated with a random pair of qubits (atoms), drawn from a multiqubit Dicke state. 
\newpage
\noindent{\bf Average values of the collective spin observable $J$:}\\ 
It is easy to see that  
\begin{eqnarray}
\label{expecs1}
\langle J_1\rangle=\langle J,M \vert J_1 \vert J,M\rangle&=&\frac{1}{2}\left[\langle J,M \vert J_+ + J_- \vert %%@
J,M\rangle\right] \nonumber\\
&=&0\nonumber\\
%\nonumber\\
\langle J_2\rangle=\langle J,M \vert J_2 \vert J,M\rangle&=&\frac{1}{2i}\left[\langle J,M \vert J_+ - J_- \vert %%@
J,M\rangle\right] \nonumber\\
&=&0\nonumber\\
%\nonumber\\
\langle J_3\rangle=\langle J,M \vert J_3 \vert J,M\rangle&=&M [\langle J,M \vert J,M\rangle]\nonumber\\
&=&M.
\end{eqnarray}
The average values of second order collective spin correlations in N-qubit Dicke states are readily evaluated and are %%@
given by, 
\begin{eqnarray}
\label{diagonalele}
\langle J_1^2\rangle&=&\langle J,M \vert J_1^2 \vert J,M\rangle\nonumber\\
&=&\frac{1}{4}\langle J,M \vert (J_+ + J_-)^2 \vert J,M\rangle\nonumber\\
&=&\frac{1}{8}\left[(N^2+2N-4M^2)\right] \nonumber\\
&=&\frac{(N^2+2N-4M^2)}{8},\nonumber\\
\nonumber\\
\langle J_2^2\rangle&=&\langle J,M \vert J_2^2 \vert J,M\rangle\nonumber\\
&=&\frac{1}{4}\langle J,M \vert (J_+ - J_-)^2 \vert J,M\rangle\nonumber\\
&=&\frac{1}{8}\left[(N^2+2N-4M^2)\right]\nonumber\\
&=&\frac{(N^2+2N-4M^2)}{8},\nonumber\\
\nonumber\\
\langle J_3^2\rangle&=&\langle J,M \vert J_3^2 \vert J,M\rangle\nonumber\\
&=&M^2\langle J,M \vert  J,M\rangle\nonumber\\
&=&M^2,
\end{eqnarray}
\begin{eqnarray}
\label{offdiag0}
\langle J_1J_2+J_2J_1\rangle&=&\langle J,M \vert J_1J_2+J_2J_1 \vert J,M\rangle\nonumber\\
&=&\frac{1}{4}\langle J,M \vert (J_+ + J_-)(J_+ - J_-)+(J_+ +J_-)(J_+ - J_-) \vert J,M\rangle\nonumber\\
&=&0.
\end{eqnarray}
Similarly, we find that
\begin{equation}
\label{dickeofft} 
\langle J_iJ_j+J_jJ_i\rangle=0 \,\,\,\,\, {\rm for}\,\,\,\,\, i \neq j.
\end{equation}
\subsection{Two qubit state parameters for Dicke state}
We may recall here that the components of the single qubit orientation vector
$s_i$ (Eq.~(\ref{avgspin})) are related to the
first order moments of the collective spin observables
(see Eq.~(\ref{vars1})) through the relation 
%\begin{equation}
%\label{vars11}
$\langle J_i\rangle=\frac{N}{2}s_i.$
%\end{equation}
Thus it is clear from Eq.~(\ref{expecs1}), that
\begin{eqnarray}
\label{avgspinsdicke}
s_1&=&\frac{2}{N}\langle J_1\rangle=0,\nonumber\\
\nonumber\\
s_2&=&\frac{2}{N}\langle J_2\rangle=0,\nonumber\\
\nonumber\\
s_3&=&\frac{2}{N}\langle J_3\rangle
=\frac{2M}{N}.
\end{eqnarray}
In other words, the qubit orientation vector of any random qubit drawn from a  $N$ qubit Dicke state has the form,
\begin{eqnarray}
\label{dicst}
\vec{s}\equiv \left(0,\,0,\,\displaystyle{\frac{2M}{N}}\right).
\end{eqnarray}
As the qubit correlations $t_{ij}$ are related to the collective second moments $\langle J_iJ_j+J_jJ_i\rangle$ through %%@
(see Eq.~(\ref{varcorrt})) 
\begin{eqnarray}
\label{ecpect11}
t_{ij}=\frac{1}{N-1}\left[\frac{2\,\langle J_iJ_j+J_jJ_i\rangle}{N}-
\delta_{i\, j}\right], \nonumber\\
\end{eqnarray}
we can evaluate the matrix elements of $T$ by using Eq.~(\ref{diagonalele})
\begin{eqnarray}
\label{offdiag} 
t_{11}&=&\frac{4[\langle J_1^2\rangle]}{N(N-1)}-\frac{1}{N-1}\nonumber\\
&=&\frac{1}{N(N-1)}\left[\frac{4(N^2+2N-4M^2)}{8}- N\right] \nonumber\\
&=&\frac{N^2-4M^2}{2N(N-1)},\nonumber\\
\nonumber\\
t_{22}&=&\frac{4[\langle J_2^2\rangle]}{N(N-1)}-\frac{1}{N-1}\nonumber\\
&=&\frac{1}{N(N-1)}\left[\frac{4(N^2+2N-4M^2)}{8}-N\right] \nonumber\\
&=&\frac{N^2-4M^2}{2N(N-1)},\nonumber\\
\nonumber\\
t_{33}&=&\frac{4[\langle J_3^2\rangle]}{N(N-1)}-\frac{1}{N-1}\nonumber\\
&=&\frac{1}{N(N-1)}[4M^2-N]\nonumber\\
&=&\frac{4M^2-N}{N(N-1)}.
\end{eqnarray}
Further from Eq.~(\ref{dickeofft}), it can be easily seen that the off diagonal elements
of the correlation matrix corresponding to Dicke state are all zero i.e.,
\begin{equation}
\label{tij0}
t_{ij}=0\,\,\,\, {\rm with}\,\,\, i\neq j.
\end{equation}  
We thus find that the $3\times 3$ real symmetric two qubit correlation matrix $T$ associated for any random pair of %%@
qubits, drawn from a multiqubit Dicke state is explicitly given by,
\begin{eqnarray}
\label{dickeT}
T={\rm diag}\,\,(t_1,\,t_2,\,t_3)
=\left(\begin{array}{ccc}
                     \frac{N^2-4M^2}{2N(N-1)}& 0 & 0 \cr 
                    0 & \frac{N^2-4\,M^2}{2N(N-1)} & 0 \cr
					 0 & 0 & \frac{4M^2-N}{N(N -1)} \cr
                    \end{array}\right).  
\end{eqnarray}
We construct the density matrix characterizing a pair of qubits arbitrarily chosen from a multiqubit Dicke state 
\begin{eqnarray}
 \label{dicke}
 \varrho_{\rm sym}=\left(\begin{array}{llll}
                     a & 0 & 0& 0\cr 
                     0& c & c& 0\cr 
                     0& c & c& 0 \cr 
                     0 & 0 & 0& d\cr 
              \end{array}\right) \, , 
\end{eqnarray}
where,
\begin{eqnarray}
 \label{dickeelmts}
 a&=&\frac{(N+2M)(N-2+2M)}{4N(N-1)}, \nonumber \\
 c&=&\frac{N^2-4M^2}{4N(N-1)},  \nonumber\\
 d&=&\frac{(N-2M)(N-2-2M)}{4N(N-1)}.
\end{eqnarray}
We may note here that the above density matrix belongs to the special class of density matrices 
discussed in Sec~\ref{special class}. Thus the non-local properties of symmetric N-qubit Dicke states 
are characterized either by  subset of three invariants $({\cal I}_3, {\cal I}_4, {\cal I}_5),$ when ${\cal I}_3\neq %%@
0$ or a subset of two invariants $({\cal I}_1, {\cal I}_2),$ when ${\cal I}_3=0.$
We now proceed to evaluate the two qubit local invariants associated with the two qubit density matrix %%@
Eq.~(\ref{dicke}).
\subsection{Local invariants}
\label {Dicke local}
The two-qubit local invariants (Eq.~(\ref{inv})), associated with Dicke state can be readily evaluated and we obtain, 
\begin{eqnarray}
\label{dicinv}
{\cal I}_1&=&{\rm det}\, T=t_1\,t_2\,t_3\nonumber\\
&=&\left(\frac{N^2-4M^2}{2N(N-1)}\right)^2
\,\left(\frac{4M^2-N}{N(N-1)}\right),\nonumber\\
\nonumber\\
\end{eqnarray}
\begin{eqnarray}
{\cal I}_2&=&{\rm Tr}\, (T^2)=t_1^2+t_2^2+t_3^2\nonumber\\
&=&2\left(\frac{N^2-4M^2}{2N(N-1)}\right)^2
+\left(\frac{4M^2-N}{N(N-1)}\right)^2, \nonumber\\
\nonumber\\
{\cal I}_3&=&s^T\, s=s_1^2+s_2^2+s_3^2\nonumber\\
&=&\frac{4M^2}{N^2},\nonumber\\
\nonumber\\
{\cal I}_4&=&s^T\, T\, s=s_1^2\, t_1+s_2^2\, t_2+s_3^2\, t_3\nonumber\\
&=&{\cal I}_3\,\left(\frac{4M^2-N}{N(N-1)}\right),\nonumber\\
\nonumber\\
{\cal I}_5&=&\epsilon_{ijk}\,\epsilon_{lmn}\, s_i\,s_l\, t_{jm}\, t_{kn}\nonumber\\
&=&2\,(s_1^2\, t_2\,t_3+s_2^2\, t_1\, t_3+s_3^2\, t_1\, t_2)\nonumber\\
&=&8\,{\cal I}_3\left(\frac{N^2-4M^2}{4\,N(N-1)}\right)^2, \nonumber\\
\nonumber\\
{\cal I}_{6}&=&\epsilon_{ijk}\, s_i\, (T\, s)_j\,
(T^2\,s)_k\nonumber\\
&=&s_1 s_2 s_3\, 
\left[t_1\, t_2\, (t_2-t_1)+t_2\, t_3\,(t_3-t_2)+t_3\, t_1\,(t_1-t_3)\right] \nonumber\\
&=&0.
\end{eqnarray}
Further, the combination ${\cal I}_4-{\cal I}_3^2$ of invariants, is given by
 \begin{equation}
{\cal I}_4-{\cal I}_3^2= \left(\frac{4M^2-N^2}{N^2(N-1)}\right)
\,{\cal I}_3.
\end{equation}
We now consider three different cases (for different values of 
$M$) and explicitly verify pairwise entanglement of Dicke states through two qubit local invariants.\\
(i)\ When  $M=\pm\frac{N}{2}:$ \\
In this case, the multiqubit dicke state state has the form
\begin{eqnarray}
\label{dickecase1}
\left\vert \frac{N}{2},\, \frac{N}{2}\right\rangle= \left\vert 0_10_2 \cdots 0_N\right\rangle.
\end{eqnarray}
This corresponds to a situation in which all the qubits are 
{\em spin-up} .
The N-qubit Dicke state in which all the qubits are {\em spin-down} is given by. 
\begin{eqnarray}
\label{dickecase1}
\left\vert \frac{N}{2},\, -\frac{N}{2}\right\rangle=\left\vert 1_11_2 \cdots1_N\right\rangle.
\end{eqnarray}
The collective  state, corresponding to this case,
is obviously a uncorrelated product state. The invariants in this case are given by
\begin{eqnarray}
{\cal I}_1&={\cal I}_5=&0,\nonumber\\
 {\cal I}_2&={\cal I}_3=&{\cal I}_4=1,
\end{eqnarray}
which are all non-negative indicating that 
$\left\vert \frac{N}{2},\, \pm\frac{N}{2}\right\rangle$  Dicke states are separable.\\
(ii)\  M=0:\\
$\left\vert \frac{N}{2},\, M=0\right\rangle$ Dicke states are written as 
\begin{eqnarray}
\label{dickecase3}
\left\vert \frac{N}{2},\, 0\right\rangle= \frac{1}{\sqrt{N}}\left\vert 1_11_2\cdots %%@
1_{\frac{N}{2}},0_{\frac{N}{2}+1}0_{\frac{N}{2}+2}\cdots 0_N\right\rangle.
\end{eqnarray}
The invariants in this case are given by
\begin{eqnarray}
\label{M=0}
{\cal I}_3={\cal I}_4={\cal I}_5=0, 
\end{eqnarray}
while the non-zero invariant,
\begin{eqnarray}
{\cal I}_1=-\frac{1}{4}\left(\frac{N}{N-1}\right)^3
\end{eqnarray}
assumes negative value.\\ 
So, the Dicke state $\left|\frac{N}{2},\, 0\right\rangle$, (with even number of
atoms), exhibits pairwise entanglement, which is signalled in terms of collective signature (see Table.~(\ref{tab1})) %%@
$\langle J_i^2\rangle < \frac{N}{4}$\\
\newpage
(iii)\ $M\neq\pm\frac{N}{2},\,0 :$\\
In this case,  the invariant,
${\cal I}_4$ is bound by  $$-\frac{1}{N-1}<{\cal I}_4<1,$$ and  the combination
${\cal I}_4-{\cal I}_3^2$ is always negative, thus revealing pairwise entanglement in  Dicke atoms in this case too. %%@
The corresponding collective signature is given by $ \frac{N}{4}< \langle(\vec{J}\cdot\hat{n}_0)^2\rangle <
\frac{N}{4}+\frac{(N-1)}{N}\, \vert\langle\vec{J}\rangle\vert^2$ 
(see Table.~(\ref{tab1})).
\section{Kitagawa-Ueda state generated by one axis twisting Hamiltonian}

In 1993, Kitagawa and Ueda~\cite{Kit93} had proposed the generation of
correlated $N$-qubit states, which are spin squeezed,  through  the nonlinear Hamiltonian
evolution $H=J_{1}^2\, \chi,$
\begin{equation}
\label{ku}
\left|\Psi_{\rm K-U}\right\rangle= e^{-iHt}\,\left|J, -J\right\rangle;\ \ J=\frac{N}{2},
\end{equation}
referred to  as {\em one-axis twisting mechanism}. 
The $N$ qubit state $\left|J, -J\right\rangle$ is the all {\em spin down} state
$$\vert J, -J\rangle=\vert 1_1,1_2,1_3,...1_N \rangle.$$
 
The one-axis twisting Hamiltonian has been realized in various quantum
systems including quantum optical systems~\cite{opti}, ion traps~\cite{ion},
cavity quantum electro magnetic dynamics~\cite{qemd}.
This effective Hamiltonian
$H=J_{1}^2\, \chi$, has already been employed to produce entangled states of four qubit maximally entangled states in %%@
an ion trap~\cite{Sac00}. Collisional interactions between
atoms in two-component Bose-Einstein condensation are also modeled using this one-axis twisting %%@
Hamiltonian~\cite{Sor101}.

In order to investigate the entanglement properties  for a random pair of qubits drawn from the Kitagawa-Ueda state, %%@
we evaluate the first and second order moments 
of the \,\,\,\,collective spin operators.\\
\noindent{\bf Expectation values of the spin operator $J$: }\\
To evaluate the expectation values $\langle J_i \rangle$ and $\langle J_iJ_j+J_jJ_i \rangle,$ let us first consider %%@
the time dependent operators
$J_3(t)$ under the Hamiltonian evolution:
%, J_+(t), J_-(t)
\begin{eqnarray}
\label{hamiltonian}
J_3(t)=e^{iHt}\,J_3 e^{-iHt}= J_3+[iHt,J_3]+\frac{1}{2!}[iHt,[iHt,J_3]]+...
\end{eqnarray}
(Here we have used the Baker-Campbell-Hausdorf formula
$e^A B e^{-A}=B+[A, B]+\frac{1}{2!}[A,[A,B]]+\frac{1}{3!}[A,[A,[A,B]]]+\cdots$)

The commutators in Eq.~(\ref{hamiltonian}) are given by:
\begin{eqnarray} 
[iHt,J_3]&=&[i\chi tJ_1^2,J_3]=i\chi t\left\{J_1[J_1,J_3]+[J_1,J_3]J_1\right\}\nonumber\\
&=&i \chi t\left\{J_1(-iJ_2)+(-iJ_2)J_1\right\}\nonumber\\
&=&\chi t [J_1,J_2]_+,\nonumber
\end{eqnarray}
where we have denoted $[O_1,O_2]_+=O_1O_2+O_2O_1.$ We further obtain,
\begin{eqnarray} 
[iHt,[iHt,J_3]]&=&[i \chi t J_1^2, \chi t[J_1,J_2]_+]\nonumber\\
&=& i\chi^2 t^2\left\{J_1[J_1,[J_1,J_2]_+] + [J_1,[J_1,J_2]_+]J_1\right\}\nonumber\\
&=&-\chi^2 t^2\left\{J_1^2J_3+2\,J_1J_3J_1+J_3J_1^2\right\}.\nonumber
\end{eqnarray}
Therefore we get,
\begin{eqnarray}
\label{J3}
J_3(t)&=&J_3+\chi t [J_1,J_2]_+ -\frac{1}{2!}\chi^2t^2\left\{J_1^2J_3+2\,J_1J_3J_1+J_3J_1^2\right\}+ \cdots.
\end{eqnarray}
Now we consider the  time dependent operator $J_+(t)$ under the Hamiltonian evolution $H,$
\begin{eqnarray}
\label{hamiltonian1}
J_+(t)=e^{iHt}\,J_+ e^{-iHt}= J_++[iHt,J_+]+\frac{1}{2!}[iHt,[iHt,J_+]]+\cdots.
\end{eqnarray}
The commutators can be computed as follows:
\begin{eqnarray} 
[iHt,J_+]&=&[i\chi t J_1^2,J_+]=i\chi t\left\{J_1[J_1,J_+]+[J_1,J_+]J_1\right\}\nonumber\\
&=&i\chi t\left\{J_1(-J_3)+(-J_3)J_1\right\}\nonumber\\
&=&-i\chi t [J_1,J_3]_+\nonumber
\end{eqnarray}
\begin{eqnarray} 
[iHt,[iHt,J_+]]&=&[i \chi t J_1^2,-i\chi t[J_1,J_3]_+]\nonumber\\
&=& \chi^2 t^2\left\{J_1[J_1,[J_1,J_3]_+] + [J_1,[J_1,J_3]_+]J_1\right\}\nonumber\\
&=&-i\chi^2 t^2\left\{J_1^2J_2+2\,J_1J_2J_1+J_2J_1^2\right\}.\nonumber
\end{eqnarray}
Hence,
\begin{eqnarray}
\label{J+}
J_+(t)&=&J_+-i\chi t [J_1,J_3]_+ -i\chi^2 t^2\left\{J_1^2J_2+2\,J_1J_2J_1+J_2J_1^2\right\}+\cdots.
\end{eqnarray}
Similarly  $J_-(t)$ can be evaluated as
\begin{eqnarray}
\label{hamiltonian2}
J_-(t)=e^{iHt}\,J_- e^{-iHt}= J_-+[iHt,J_-]+\frac{1}{2!}[iHt,[iHt,J_-]]+\cdots
\end{eqnarray}
\begin{eqnarray} 
[iHt,J_-]&=&[i\chi t J_1^2,J_-]=i\chi t\left\{J_1[J_1,J_-]+[J_1,J_-]J_1\right\}\nonumber\\
&=&i \chi t\left\{J_1(J_3)+(J_3)J_1\right\}\nonumber\\
&=&i\chi t [J_1,J_3]_+\nonumber
\end{eqnarray}
\begin{eqnarray} 
[iHt,[iHt,J_-]]&=&[i \chi t J_1^2,i\chi t[J_1,J_3]_-]\nonumber\\
&=&- \chi^2 t^2\left\{J_1[J_1,[J_1,J_3]_+] + [J_1,[J_1,J_3]_+]J_1\right\}\nonumber\\
&=&i\chi^2 t^2\left\{J_1^2J_2+2\,J_1J_2J_1+J_2J_1^2\right\}.\nonumber
\end{eqnarray}
So we obtain,
\begin{eqnarray}
\label{J-}
J_-(t)&=&J_-+i\chi t [J_1,J_3]_+ +i\chi^2 t^2\left\{J_1^2J_2+2\,J_1J_2J_1+J_2J_1^2\right\}+\cdots.
\end{eqnarray}
Using Eqs.~(\ref{J3}),~(\ref{J+}),~(\ref{J-}), the expectation values are evaluated as follows:
\begin{eqnarray}
\label{J3expec}
\langle\Psi_{\rm K-U}|J_3(t)|\Psi_{\rm K-U}\rangle&=&\left\langle J -J\right\vert J_3(t)|J -J\rangle\nonumber\\
&=&\left\langle J -J\right\vert [J_3+\chi t [J_1,J_2]_+ %%@
-\frac{1}{2!}\chi^2t^2\left\{J_1^2J_3+2\,J_1J_3J_1+J_3J_1^2\right\}+\cdots]
\left\vert J -J\right\rangle\nonumber\\  
&=&-\frac{N}{2}\left[1-\frac{1}{2!}\chi^2t^2(N-1)]+\frac{1}{4!}\chi^4t^4(N-1)^2-\cdots
\right]\nonumber\\
&=&-\frac{N}{2}\cos^{N-1} (\chi t), 
\end{eqnarray}
\begin{eqnarray}
\label{J+expec}
\langle\Psi_{\rm K-U}|J_+(t)|\Psi_{\rm K-U}\rangle&=&\langle J -J|J_+(t)|J -J\rangle\nonumber\\
&=&\left\langle J -J\right\vert J_+-i\chi t [J_1,J_3]_+ -i\chi^2 %%@
t^2\left\{J_1^2J_2+2\,J_1J_2J_1+J_2J_1^2\right\}+\cdots
\left\vert J -J\right\rangle\nonumber\\  
&=&0 
\end{eqnarray}
\begin{eqnarray}
\label{J-expec}
\langle\Psi_{\rm K-U}|J_-(t)|\Psi_{\rm K-U}\rangle&=&\langle J -J|J_-(t)|J -J\rangle\nonumber\\
&=&\left\langle J -J\right\vert J_-+i\chi t [J_1,J_3]_+ +i\chi^2 %%@
t^2\left\{J_1^2J_2+2\,J_1J_2J_1+J_2J_1^2\right\}+\cdots
\left\vert J -J\right\rangle\nonumber\\  
&=&0 
\end{eqnarray}
\begin{eqnarray}
\label{J32expec}
\langle\Psi_{\rm K-U}|J_3^2|\Psi_{\rm K-U}\rangle&=&\langle J -J|J_3^2(t)|J -J\rangle\nonumber\\
&=&\left\langle J -J\right\vert \left(J_3+\chi t [J_1,J_2]_+ %%@
-\frac{1}{2!}\chi^2t^2\left\{J_1^2J_3+2\,J_1J_3J_1+J_3J_1^2\right\}+
\cdots\right)^2\left\vert J -J\right\rangle \nonumber\\
&=&\frac{1}{8}[N^2+N+N(N-1)\cos^{N-2} (2\chi t)].
\end{eqnarray}
In a Similar manner, the remaining first and second order expectation values can be obtained. 
The average values of the collective spin observables for Kitagawa-Ueda state are listed below:
\begin{eqnarray}
\label{expecku}
\langle J_1\rangle&=&\langle J_2\rangle=0,\nonumber\\
\langle J_3\rangle&=&-\frac{N}{2}\cos^{N-1}(\chi t)\nonumber\\
\langle J_1^2\rangle&=&\frac{N}{4},\nonumber\\
\langle J_2^2\rangle&=&\frac{1}{8}\left(N^2+N-N(N-1)\cos^{N-2}(2 \chi t)\right)\nonumber\\
\langle J_3^2\rangle&=&\frac{1}{8}\left(N^2+N+N(N-1)\cos^{N-2}(2 \chi t)\right),\nonumber\\
\langle[J_1,J_2]_+\rangle&=&\frac{1}{2}N(N-1)\cos^{N-2}(\chi t)
\sin(\chi t),\nonumber\\
\langle [J_+,J_3]_+\rangle&=&0.
\end{eqnarray}
\subsection{Two qubit state variables}  
The qubit state parameter $s_i$ associated with a random qubit chosen from a multiqubit Kitagawa-Ueda state can  be %%@
written (from Eq.~(\ref{expecku})) as,
\begin{eqnarray}
\label{avgspinsku}
s_1&=&\frac{2}{N}\langle J_1\rangle=0,\nonumber\\
\nonumber\\
s_2&=&\frac{2}{N}\langle J_2\rangle=0,\nonumber\\
\nonumber\\
s_3&=&\frac{2}{N}\langle J_3\rangle=-\cos^{(N-1)}(\chi\, t).
\end{eqnarray}  
Therefore the orientation vector $\vec{s}$ for qubit drawn randomly 
from the Kitagawa-Ueda state is given by
\begin{equation*}
\label{kus}
\vec{s}=\left(0,\, 0,\, -\cos^{(N-1)}(\chi\, t)\right).
\end{equation*}
The two qubit correlation matrix elements are readily obtained from
the second order moments of the collective spin observable Eq.~(\ref{expecku}),
$$t_{ij}=\frac{1}{N-1}\left[\frac{2\,\langle J_iJ_j+J_jJ_i\rangle}{N}-
\delta_{i\, j}\right].$$
Thus we have,
\begin{eqnarray}
t_{11}&=&t_{13}=t_{23}=0,\nonumber\\
t_{12}&=&\cos^{(N-2)}(\chi\, t)\, \sin(\chi\, t),\nonumber \\
t_{22}&=&\frac{1}{2}\, \left( 1-\cos^{(N-2)}(2\chi\, t)\right),\nonumber\\
t_{33}&=&\frac{1}{2}\, \left( 1+\cos^{(N-2)}(2\chi\, t)\right).
\end{eqnarray}
The $3 \times 3$ real two qubit correlation matrix $T$ can be thus written as, 
\begin{eqnarray*}
\label{KUT}
 T=\left(\begin{array}{ccc}
     0& \cos^{(N-2)}(\chi\, t)\, \sin(\chi\, t) & 0 \cr 
    \cos^{(N-2)}(\chi\, t)\, \sin(\chi\, t) & \frac{1}{2}\, 
\left( 1-\cos^{(N-2)}(2\chi\, t)\right)& 0 \cr
	0& 0& \frac{1}{2}\,\left(1+\cos^{(N-2)}(2\chi\,t)\right)\cr
\end{array}\right).  
\end{eqnarray*}
The two qubit density matrix for a pair of qubits arbitrarily chosen from
a multiqubit Kitagawa-Ueda state is given by,
\begin{eqnarray}
 \label{kitagawa}
 \varrho_{\rm sym}=\left(\begin{array}{llll}
                     a & 0 & 0& b\cr 
                     0& c & c& 0\cr 
                     0& c & c& 0 \cr 
                     b & 0 & 0& d\cr 
              \end{array}\right) \, , 
\end{eqnarray}
with the matrix elements given by,
\begin{eqnarray*}
 \label{KUelmts}
 a&=&\frac{3+\cos^{N-2}(2\chi t)-2\cos^{N-1}(\chi t)}{8}, \nonumber \\
 b&=&\frac{1-\cos^{N-2}(2\chi t)}{8},\nonumber\\
  c&=&\frac{N^2-4M^2}{4N(N-1)},  \nonumber\\
 d&=&\frac{3+\cos^{N-2}(2\chi t)+2\cos^{N-1}(\chi t)}{8}.
 \end{eqnarray*}
\subsection{Local invariants}
The two qubit local invariants Eq.~(\ref{inv}) associated with the $N$ qubit Kitagawa-Ueda state are given by
\begin{eqnarray}
 \label{kuinv}
{\cal I}_1&=&{\rm det}\, T\nonumber\\
&=&-\frac{1}{2}\, \cos^{2(N-2)}(\chi\, t)\, \sin^2(\chi\, t)\,\left( 1+\cos^{(N-2)}(2\chi\, t)\right), \nonumber \\
{\cal I}_2&=&{\rm Tr}\, (T^2)\nonumber\\
&=&2\, \cos^{2(N-2)}(\chi\, t)\, \sin^2(\chi\, t)+
\frac{1}{2}\,  \left( 1+\cos^{2(N-2)}(\chi\, t)\right), \nonumber  \\
{\cal I}_3&=&s^T\, s=s_1^2+s_2^2+s_3^2\nonumber\\
&=&\cos^{2(N-1)}(\chi\, t), \nonumber\\
{\cal I}_4&=&s^T\, T\, s\nonumber\\
&=&\frac{1}{2}\, {\cal I}_3\,
\left( 1+\cos^{(N-2)}(2\chi\, t)\right), \nonumber  \\
{\cal I}_5&=&\epsilon_{ijk}\,\epsilon_{lmn}\, s_i\,s_l\, t_{jm}\, t_{kn}\nonumber\\
&=&-2\, {\cal I}_3\, \cos^{2(N-2)}(\chi\, t)\, \sin^2(\chi\, t),\nonumber\\
{\cal I}_{6}&=&\epsilon_{ijk}\, s_i\, (T\, s)_j\,
(T^2\,s)_k\nonumber\\
&=&0.
\end{eqnarray}
We see that pairwise entanglement is manifest through  the negative value of the invariant ${\cal I}_5$. Further, from %%@
Eq.~(\ref{I5spin}) it is evident that ${\cal I}_5<0$ implies that the state is  spin squeezed.
\begin{center}
 \begin{figure}
 \label{kufigure1}
 \centering   
 \includegraphics*[width=5.5in,keepaspectratio]{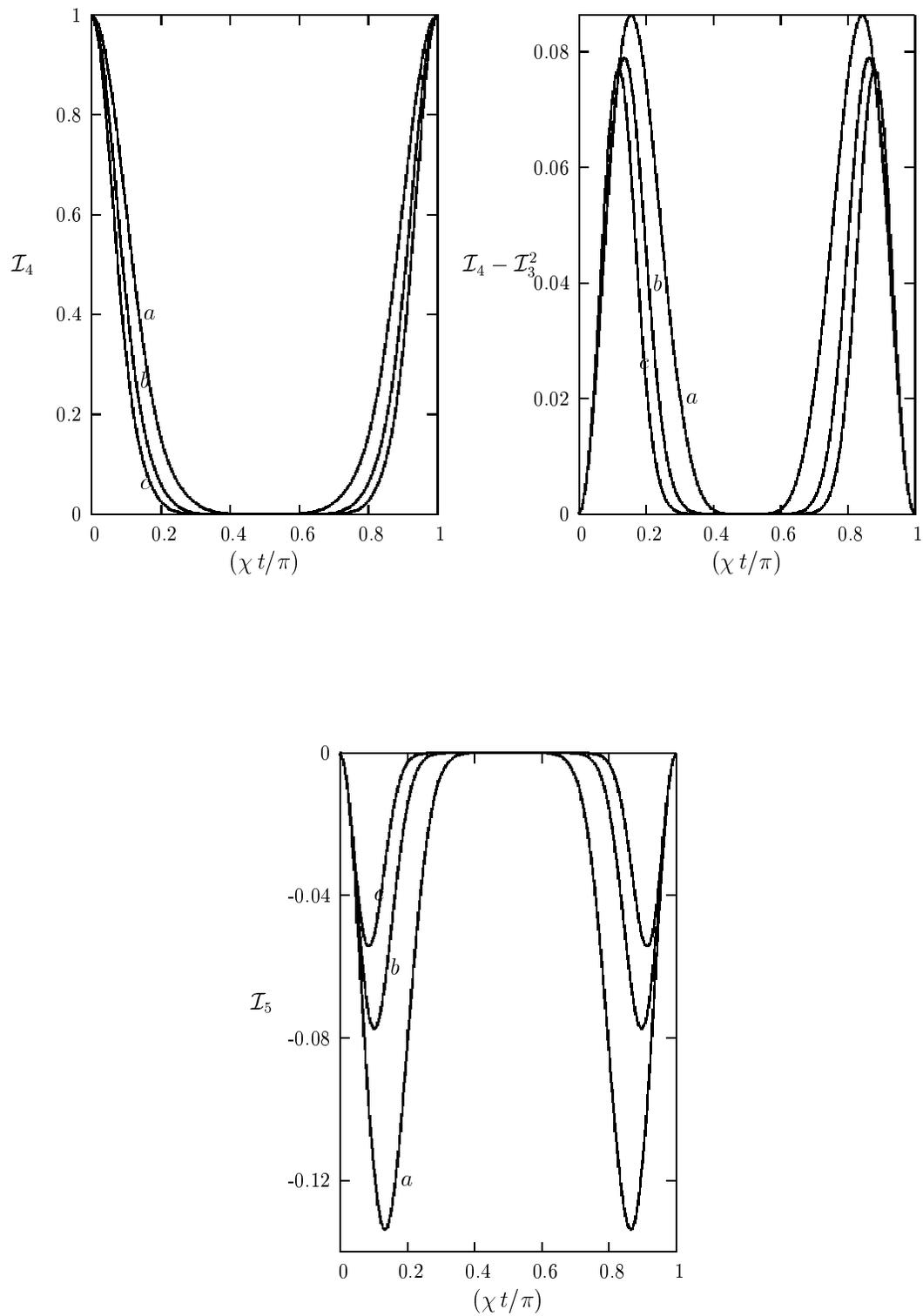}
  \caption{The invariants  ${\cal I}_4,\ {\cal I}_4-{\cal I}_3^2 \  {\rm and}\ {\cal I}_5$, corresponding to a %%@
$N$-qubit Kitagawa-Ueda
state. Curve $a:\, N=4,\, b: \, N=6, \ {\rm and}\  c: \, N=8.$}
\end{figure}
\end{center} 
In Fig. 4.1, we have plotted the two qubit local invariants ${\cal I}_4,\ {\cal I}_4-{\cal I}_3^2 \  {\rm and}\ {\cal %%@
I}_5$, associated with the multiqubit kitagawa-Ueda state for different values of $N$.

\section{Atomic spin squeezed states}
Consider a system of N identical two-level atoms. The atomic squeezed states  are defined by 
\begin{equation}
\label{squeezed states}
\left|\Psi_{M}\right\rangle=A_{0}\, {\rm exp}\left(\theta\, J_{3}\right)\,
{\rm exp}\left(-i\, \frac{\pi}{2}\, J_2\right)\, \left| J=\frac{N}{2},\, M\right\rangle,
\end{equation}
where $\theta$ is conveniently parameterized as $e^{2\theta}=\tanh 2\xi.$ 

The atomic spin squeezed states $|\Psi_{M}\rangle$
are shown~\cite{Agarwal90} to be the eigenvectors of the nonhermitian  operator $R_3,$ given by 
\begin{eqnarray}
\label{R3}
R_3=\frac{J_-\cosh \xi+J_+\sinh \xi}{\sqrt{2\sinh 2 \xi}},
\end{eqnarray} 
where
\begin{equation}
\label{R3psi}
R_3|\Psi_M\rangle=M|\Psi_M\rangle.
\end{equation} 
Here, we concentrate on the state
%In other words, for $M=0,$ we have,
%\begin{equation}
%\label{R3psi}
%R_3|\Psi_0\rangle=0
%\end{equation} 
%here $|\Psi_{0}\rangle$, can be explicitly written as,
\begin{equation}
\label{agarwal}
\left|\Psi_{0}\right\rangle=A_{0}\, {\rm exp}\left(\theta\, J_{3}\right)\,
{\rm exp}\left(-i\, \frac{\pi}{2}\, J_2\right)\, \left| J=\frac{N}{2},\, 0\right\rangle. 
\end{equation}
In the interaction of a collection of even number of $N$  two-level atoms with
squeezed radiation, it has been shown~\cite{Agarwal90} that the steady state
$|\Psi_0\rangle$ of Eq.~(\ref{agarwal}) is the pure atomic squeezed state 
for certain values of external field strength and detuning parameters. \\
It is convenient to express the atomic squeezed state $\left|\Psi_{0}\right\rangle$ as,
\begin{equation}
\label{agarwal1}
\left|\Psi_{0}\right\rangle=A_{0}\,\sum_{M=-J}^J d_{M\,0}^J\left(\frac{\pi}{2}\right)  e^{M\theta}\left| %%@
J=\frac{N}{2},\, M\right\rangle, 
\end{equation}
where the coefficients $d^{J}_{M\, 0}(\frac{\pi}{2})$ are given  by~\cite{Brink}
\begin{equation}
d^{J}_{M\, 0}\left(\frac{\pi}{2}\right)= \frac{J!\, \displaystyle\sqrt{(J+M)!\, (J-M)!}}{2^J}\,
\displaystyle\sum_{p=M}^{J-M}\, \frac{(-1)^p}{ (J-M)!\, p!\, (p-M)!\,  (J+M-p)!}.
\end{equation}

We now proceed to study the pairwise entanglement properties of atomic squeezed state $|\Psi_0\rangle.$\\
\noindent{\bf First and second order moments of the collective spin observable for the Atomic squeezed states:}\\
The first order expectation values of $\langle J_i\rangle$ are evaluated as follows:\\
Consider the expectation value of the nonhermitian operator $R_3$ (see Eq.~(\ref{R3}))
\begin{equation}
\label{expecR3}\langle\Psi_{0}|R_3|\Psi_{0}\rangle=
\frac{1}{\sqrt{2\sinh 2 \xi}}\langle\Psi_{0}|J_-\cosh \xi+J_+\sinh \xi
\left\vert\Psi_{0}\right\rangle.
\end{equation}
However, it is clear from Eq.~(\ref{R3psi}) that
\begin{equation*}
\langle\Psi_0|R_3|\Psi_0\rangle=0.
\end{equation*}
So, we obtain,
\begin{equation}
\label{R3psi1}
\langle\Psi_{0}|J_-\cosh \xi+J_+\sinh \xi
\left\vert\Psi_{0}\right\rangle=0.
\end{equation}
Further, it is evident that,
\begin{equation*}
\langle\Psi_0|R_3^\dag|\Psi_0\rangle=0,
\end{equation*}
which in turn leads to,
\begin{equation}
\label{R3psidag} 
\langle\Psi_{0}|J_+\cosh \xi+J_-\sinh \xi\left\vert\Psi_{0}\right\rangle=0.
\end{equation}
Using Eq.~(\ref{R3psi1}) and Eq.~(\ref{R3psidag}), it is easy to see that~\footnote{since $J_+=\frac{J_1+J_2}{2}$ and %%@
$J_-=\frac{J_1-J_2}{2i},$  
$\langle J_+\rangle=0,\,\langle J_-\rangle=0\Longrightarrow \langle J_1\rangle=\langle J_2\rangle=0.$}
\begin{eqnarray*}
\label{expecj1j2}
\langle J_+\rangle&=&0,\,\,\,\,\,\langle J_-\rangle=0, \nonumber\\
{\rm or}\,\,\,\,\,\,\,
\langle J_1\rangle&=&0,\,\,\,\,\,\,\langle J_2\rangle=0.
\end{eqnarray*}
The first order moment $\langle J_3 \rangle$ can be explicitly evaluated as follows:
We have, 
\begin{eqnarray}
\label{expecj3}
J_3\left|\Psi_{0}\right\rangle&=&A_{0}\,\displaystyle\sum_{M=-J}^J M d_{M\,0}^J\left(\frac{\pi}{2}\right)
e^{(M\theta)}|J\, M\rangle,\nonumber\\
{\rm and\,\,\, therefore} \hskip .5in\nonumber\\
\langle\Psi_{0}|J_3|\Psi_{0}\rangle
&=&A_{0}^2\displaystyle\sum_{M=-J}^J M\,\left[d_{M\,0}^J\left(\frac{\pi}{2}\right)\right]^2\,e^{(2M\theta)}.
\end{eqnarray}

The second order moments $\langle (J_iJ_j+J_jJ_i)\rangle$ of the collective spin operator, are conveniently evaluated %%@
in terms of the expectation values 
 of the non-hermitian operator $R_3^2:$ 
\begin{eqnarray}
\label{R3^2}   
R_3^2&=&\frac{(J_-\cosh \xi+J_+\sinh \xi)\,(J_-\cosh \xi+J_+\sinh \xi)}
{2\sinh 2 \xi}\nonumber\\
&=&\frac{J_-^2\cosh^2 \xi+J_+^2\sinh^2 \xi+(J_-J_++J_+J_-)
\sinh \xi\cos \xi)}
{2\sinh 2 \xi}\nonumber\\
\langle \Psi_{0}|R_3^2|\Psi_{0} \rangle
&=&\langle J_1^2-J_2^2\rangle\coth 2 \xi+
\langle J_1^2+J_2^2\rangle-i\langle [J_1,J_2]_+\rangle\coth 2\xi.
\end{eqnarray}
Since $\langle \Psi_0|R_3^2|\Psi_0 \rangle=0,$ (see Eq.~(\ref{R3psi})) we obtain,
\begin{eqnarray}
\langle J_1^2-J_2^2\rangle\coth 2\xi+
\langle J_1^2+J_2^2\rangle-i\langle [J_1,J_2]_+\rangle\coth 2\xi&=&0.
\end{eqnarray}
In other words, we have,
\begin{equation}
\label{repart=0}
{\rm Re}\,\,(\langle R_3^2 \rangle)=\langle J_1^2-J_2^2\rangle\coth 2\xi+
\langle J_1^2+J_2^2\rangle=0
\end{equation}
and
\begin{equation}
\label{impart=0}
{\rm Im}\,\,(\langle R_3^2 \rangle)=(\langle [J_1,J_2]_+\rangle\coth 2\xi)=0.
\end{equation}
We now evaluate the expectation value of $R_3^\dag R_3$ given explicitly as,
\begin{eqnarray}
\label{r3dagr3}
R_3^\dag R_3&=&\frac{(J_+\cosh \xi+J_-\sinh \xi)(J_-\cosh \xi+J_+\sinh \xi)}{2\sinh 2\xi}\nonumber\\
&=&\frac{J_+J_-\cosh^2\xi+ J_-J_+\sinh^2\xi+(J_+^2 + J_-^2)\sinh \xi\cos \xi}
{2\sinh 2 \xi}\nonumber\\
%{\rm we \,\, have}\nonumber\\
\langle \Psi_0|R_3^\dag R_3|\Psi_0\rangle
&=&\langle J_1^2+J_2^2\rangle\coth 2 \xi+
\langle J_1^2-J_2^2\rangle+\langle J_3\rangle.
\end{eqnarray}
From Eq.~(\ref{R3psi}), it is clear that $\langle R_3^\dag R_3 \rangle=0,$ and therefore we get,
\begin{eqnarray}
\label{r3dagr3=0}
\langle J_1^2+J_2^2\rangle\coth 2\xi+
\langle J_1^2-J_2^2\rangle+\langle J_3\rangle=0\nonumber\\
{\rm i.e.,}\,\,\,\,\,\,\,
-\langle J_1^2-J_2^2\rangle-\langle J_1^2+J_2^2\rangle\coth 2\xi
=\langle J_3\rangle.
\end{eqnarray}
Simplifying the Eqs.~(\ref{repart=0}),~(\ref{r3dagr3=0}), we obtain the expectation values of $J_1^2$ and $J_2^2$ as,
\begin{eqnarray}
\label{j1j2}
\langle J_1^2\rangle&=&-\frac{1}{2}\langle J_3\rangle\,e^{-2\xi}\nonumber\\
\langle J_2^2\rangle&=&-\frac{1}{2}\langle J_3\rangle\,e^{2\xi}.
\end{eqnarray}
Now, to compute the average value of  $J_3^2,$ we use
\begin{equation}
\label{J}
\langle J^2\rangle=\langle J_1^2+ J_2^2 + J_3^2\rangle=J(J+1)
\end{equation}
and obtain,  
\begin{eqnarray}
\label{j3^2}
\langle J_3^2\rangle&=&\langle J^2- J_1^2-J_2^2\rangle\nonumber\\
&=&J(J+1)-\langle J_3\rangle\cosh 2\xi.
\end{eqnarray}
The second order expectation values $\langle\Psi_0|[J_1,J_3]_+|\Psi_0\rangle$ and %%@
$\langle\Psi_0|[J_2,J_3]_+|\Psi_0\rangle$
are determined as follows:

Let us consider $\langle[J_+,J_3]_+\rangle=\langle\psi_0|J_+J_3|\psi_0\rangle+\langle\psi_0|J_3J_+|
\psi_0\rangle.$ 
By computing each term separately, 
\begin{eqnarray}
\label{expecJ+J3}
\langle\Psi_0|J_+J_3|\Psi_0\rangle&=&A_{0}\displaystyle\sum_{M=-J}^J M d_{M\,0}^J\left(\frac{\pi}{2}\right) %%@
d_{M'\,0}^J \left(\frac{\pi}{2}\right)
e^{(M+M')\theta}\sqrt{(J+M')(J-M'+1)}\,\delta_{M', M+1} \nonumber\\
&=&A_{0}\,\displaystyle\sum_{M=-J}^J M d_{M\,0}^J\left(\frac{\pi}{2}\right) d_{M+1\,0}^J \left(\frac{\pi}{2}\right)
e^{(2M+1)\theta}\displaystyle\sqrt{(J+M+1)(J-M)}\nonumber\\
\langle J_+J_3\rangle&=&0,
\end{eqnarray}
since~\footnote{The coefficient $d^{J}_{M\, 0}\left(\frac{\pi}{2}\right)=\frac{\sqrt{(J+M)!(J-M)!}}
{2^J}\,\frac{1}{(\frac{J+M}{2})!(\frac{J-M}{2})!}\,(-1) \frac{J-M}{2}$ for $J+M={\rm even}$
and $d^{J}_{M\, 0}(\frac{\pi}{2})=0$ for $J+M={\rm odd}.$ Therefore, we obviously have $d^{J}_{M 0}d^{J}_{M+1 0}=0.$ } %%@
$d^{J}_{M 0}d^{J}_{M+1 0}=0.$\\
Similarly,  
\begin{eqnarray}
\label{expecJ3J+}
J_3J_+|\Psi_0\rangle&=&A_{0}\displaystyle\sum_{M=-J}^J M d_{M\,0}^J\left(\frac{\pi}{2}\right) d_{M'\,0}^J %%@
\left(\frac{\pi}{2}\right)
e^{(M+M')\theta}\sqrt{(J-M')(J+M'+1)}\,\delta_{M', M-1} \nonumber\\
&=&A_{0}\displaystyle\sum_{M=-J}^J M d_{M\,0}^J\left(\frac{\pi}{2}\right) d_{M-1\,0}^J \left(\frac{\pi}{2}\right)
e^{(2M+1)\theta}\sqrt{(J-M+1)(J+M)}\nonumber\\
\langle J_3J_+\rangle&=&0.
\end{eqnarray}
From Eqs.~(\ref{expecJ+J3},~(\ref{expecJ3J+}), we obtain
\begin{equation}
\langle [J_+,J_3]_+\rangle=0.
\end{equation}
So, we obtain,
\begin{equation}
\label{j1j3j2j3}
\langle [J_1,J_3]_+\rangle=0, \,\,\,\,\,\ \langle [J_2,J_3]_+\rangle=0.
\end{equation}
We next determine the two qubit state parameters
associated with the atomic squeezed systems.
\subsection{Two qubit state parameters} 
The components of the single qubit orientation vector $\vec{s}$ drawn from a  collective atomic system, 
are given by (see Eqs.~(\ref{expecj1j2}),~(\ref{expecj3}))
\begin{eqnarray}
\label{avgspinssquee}
s_1&=&\frac{2}{N}\langle J_1\rangle=0,\nonumber\\
\nonumber\\
s_2&=&\frac{2}{N}\langle J_2\rangle=0,\nonumber\\
\nonumber\\
s_3&=&\frac{2}{N}\langle J_3\rangle=\frac{2}{N}\,A_{0}^2\displaystyle\sum_{M=-J}^J %%@
M\,\left[d_{M\,0}^J\left(\frac{\pi}{2}\right)\right]^2\,e^{(2M\theta)}.
\end{eqnarray}
Thus the average spin vector 
~$\vec{s}$~ for atomic squeezed states of Eq.~(\ref{agarwal})
assumes the form
\begin{eqnarray*}
\label{agarwalst}
\vec{s}&=&\left(0,\, 0, \, \frac{2\,\langle J_3\rangle}{N}\right).
\end{eqnarray*}
The elements of the two qubit correlation matrix which are expressed in terms of the second order moments (see %%@
Eq.~(\ref{varcorrt})) are given by,
\begin{eqnarray*}
%\label{ecpect11}
t_{ij}=\frac{1}{N-1}\left[\frac{2\,\langle J_iJ_j+J_jJ_i\rangle}{N}-
\delta_{i\, j}\right].\nonumber\\
\end{eqnarray*}
The diagonal elements of the correlation matrix $T$ are obtained
using  Eqs.~(\ref{j1j2}),~(\ref{j3^2}) and are given by,
\begin{eqnarray}
\label{diagtii} 
t_{11}&=&\frac{4[\langle J_1^2\rangle]}{N(N-1)}-\frac{1}{N-1}\nonumber\\
&=&\frac{1}{N(N-1)}\left[-\frac{1}{2}\langle J_3\rangle\,e^{-2\xi}- N\right]\nonumber\\
&=&\frac{-2\, \langle J_3\rangle\, e^{-2\xi}-N }{N(N-1)},\nonumber\\
\nonumber\\
t_{22}&=&\frac{4[\langle J_2^2\rangle]}{N(N-1)}-\frac{1}{N-1}\nonumber\\
&=&\frac{1}{N(N-1)}\left[-\frac{1}{2}\langle J_3\rangle\,e^{2\xi}-N\right]\nonumber\\
&=&\frac{-2\, \langle J_3\rangle\, e^{2\xi}-N}{N(N-1)}, \nonumber\\
\nonumber\\
t_{33}&=&\frac{4[\langle J_3^2\rangle]}{N(N-1)}-\frac{1}{N-1}\nonumber\\
&=&\frac{1}{N(N-1)}\left[J(J+1)-\langle J_3\rangle\cosh 2\xi-N\right]\nonumber\\
&=&\frac{4\, \langle J_3\rangle\, \cosh(2\xi)+N^2+N}{N(N-1)}.
\end{eqnarray}
Further, from Eqs.~(\ref{j1j3j2j3}),~(\ref{impart=0}), it is easy to see that  the off-diagonal elements of $T$ are %%@
all zero
\begin{eqnarray}
t_{12}&=&t_{21}=0,\nonumber\\
t_{13}&=&t_{31}=0,\nonumber\\
t_{23}&=&t_{32}=0.   
\end{eqnarray}
Thus, the correlation matrix $T$ has the following structure,
\begin{eqnarray*}
\label{squeezedT}
T={\rm diag\,}(t_i,t_2,t_3)
=\left(\begin{array}{ccc}
                     \frac{-2\, \langle J_3\rangle\, e^{-2\xi}-N }{N(N-1)}& 0 & 0 \cr 
                    0 & \frac{-2\, \langle J_3\rangle\, e^{2\xi}-N}{N(N-1)} & 0 \cr
					 0 & 0 & \frac{4\, \langle J_3\rangle\, \cosh(2\xi)+N^2+N}{N(N-1)} \cr
                    \end{array}\right).  
\end{eqnarray*}
\subsection{Local invariants}
The two qubit local invariants (see Eq.~(\ref{inv})) associated with the 
atomic spin squezed states are listed below:
\begin{eqnarray}
\label{agarwalinv}
{\cal I}_1&=&t_1\,t_2\,t_3,\nonumber\\
{\cal I}_2&=&t_1^2+t_2^2+t_3^2, \nonumber \\
{\cal I}_3&=&s^T\, s=s_1^2+s_2^2+s_3^2\nonumber\\
&=&\frac{4\, \left\langle J_3\right\rangle^2}{N^2},\nonumber\\
{\cal I}_4&=&s^T\, T\, s=s_1^2\, t_1+s_2^2\, t_2+s_3^2\, t_3\nonumber\\
&=&{\cal I}_3\, \left[\frac{4\, \langle J_3\rangle\, \cosh(2|\xi|)+N^2+N}{N(N-1)}\right], \nonumber \\
{\cal I}_5&=&\epsilon_{ijk}\,\epsilon_{lmn}\, s_i\,s_l\, t_{jm}\, t_{kn}\nonumber\\
&=&\frac{{2\,\cal I}_3}{N^2\, (N-1)^2}\, \left(2\, \langle J_3\rangle\, e^{-2|\xi|}+N\right)\,
\left(2\, \langle J_3\rangle\, e^{2|\xi|}+N\right),\nonumber\\
{\cal I}_{6}&=&\epsilon_{ijk}\, s_i\, (T\, s)_j\,
(T^2\,s)_k\nonumber\\
&=&0.
\end{eqnarray}
\begin{center}
 \begin{figure}
 \centering
   \label{fig:squewig}
 \includegraphics*[width=5.5in,keepaspectratio]{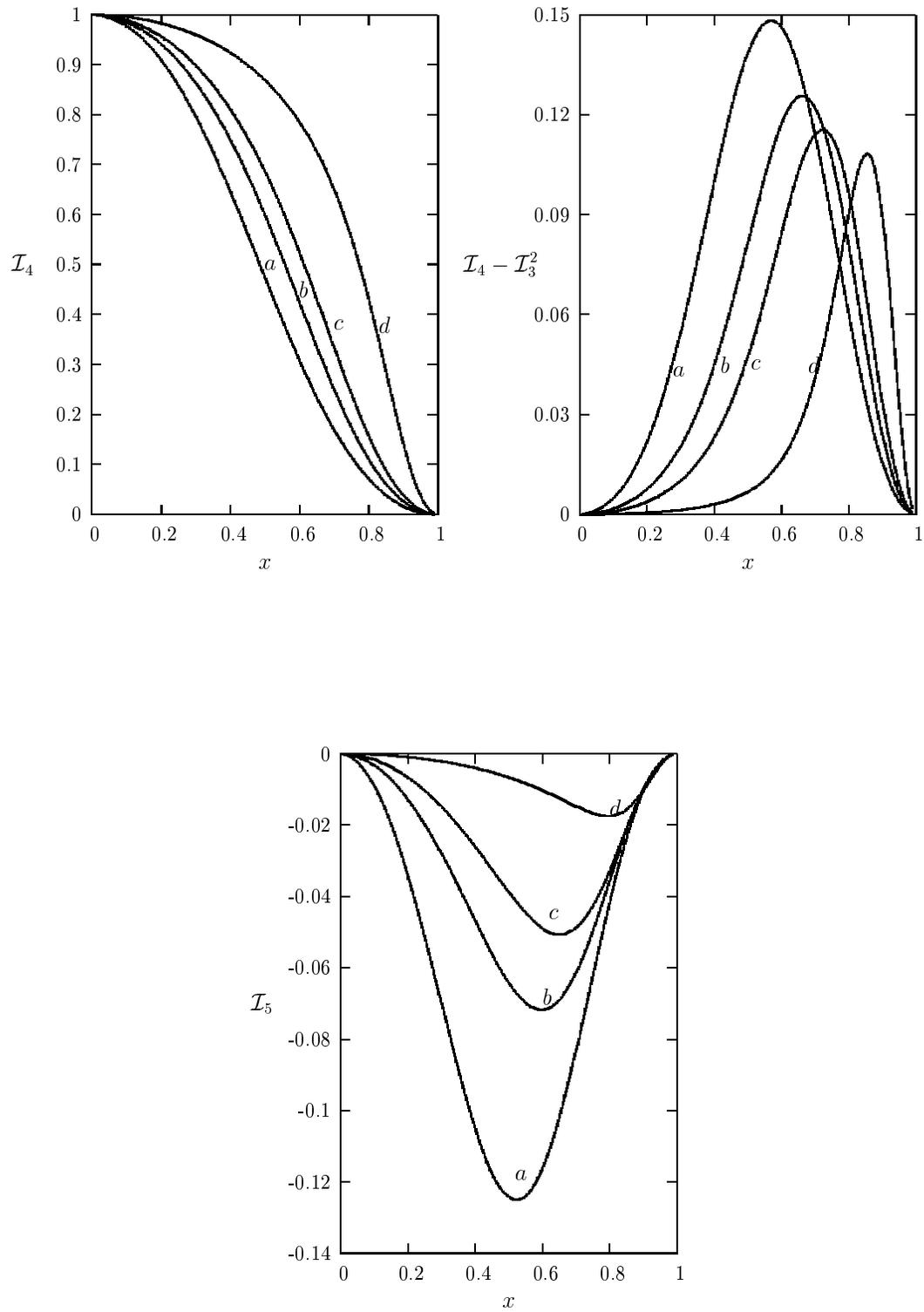}
  \caption{The invariants  ${\cal I}_4,\ {\cal I}_4-{\cal I}_3^2 \  {\rm and}\ {\cal I}_5$,  associated with the %%@
atomic squeezed state of $N$ two level atoms
interacting with  squeezed radiation. Curve $a:\, N=4,\, b: \, N=6, \, c: N=8,\  {\rm and}\  d: \, N=20.$}
\end{figure}
\end{center}
In Fig. 4.2, we have plotted the invariants ${\cal I}_4,\ {\cal I}_4-{\cal I}_3^2 \  {\rm and}\ {\cal I}_5$,  as a %%@
function of
the parameter $x=e^{2\,\theta}$, for different values of $N$. 
These  plots
demonstrate  that the invariant ${\cal I}_5$ is negative, highlighting the pairwise entanglement (spin squeezing) of %%@
the
atomic state.
\newpage
\section{Conclusions}
\label{con}
In this chapter, we have considered few interesting  symmetric
multi-qubit dynamical models like Dicke states, Kitagawa-Ueda state generated by one axis twisting Hamiltonian and %%@
Atomic squeezed state $|\Psi_0\rangle$. The density matrix of  these states have
a specific structure and belong to the special class of symmetric states Eq.~(\ref{cansym})
discussed in 
Chapter.~\ref{c:c2twoqubit}. We have evaluated the two qubit local invariants and investigated the nonlocal properties %%@
associated with the states. 
In each case, the entanglement properties are reflected through the negative values of 
some of the two qubit invariants thus highlighting our separability 
criteria.

%%%%%%%%%%%%%%%%%%%%%%%%%%%%%%%%%%%%%%%%%%%%%%%%%%%%%%%%%%%%%%%%%%%%%%%%%%%%%%
%%%%%%%%%%%%%%%%%%%%%%%%%%%%%%%%%%%%%%%%%%%%%%%%%%%%%%%%%%%%%%%%%%%%%%%%%%%%%%
%%%%%%%%%%%%%%%%%%%%%%%%%%%%%%%%%%%%%%%%%%%%%%%%%%%%%%%%%%%%%%%%%%%%%%%%%%%%%%
\chapter{Constraints on the variance matrix of entangled symmetric qubits}
\label{c:necce and suffi}
\markboth{}{Constraints on the variance matrix of entangled symmetric qubits} 

The nonseparability constraints on the two qubit local invariants
${\cal I}_4<0,\ {\cal I}_5<0$ and ${\cal I}_4-{\cal I}_3^2<0$ derived, in 
Chapters~\ref{c:c2twoqubit} and ~\ref{c:multiqubit states}, serve only 
as sufficient condition for pairwise 
entanglement in symmetric qubits. In this Chapter, we derive necessary and sufficient 
condition for entanglement in symmetric two qubit states by establishing an
equivalence between the Peres-Horodecki criterion~\cite{Peres, horo96} and the negativity of the two qubit covariance %%@
matrix. 
Pairwise entangled symmetric multiqubit states 
{\em necessarily} obey these  constraints. We also  bring out a local invariant 
structure exhibited by these constraints. 

\section{Peres-Horodecki inseparability criterion for CV states}
\label{PHcriterion}
 Peres-Horodecki inseparability  criterion~\cite{Peres, horo96}\, viz., 
{\em positivity under partial transpose} (PPT)  has been extremely 
fruitful in characterizing entanglement for finite dimensional systems. 
It provides necessary and sufficient conditions for $2\times 2$ and $2\times3$ dimensional systems. It is found that %%@
the PPT  criterion is 
significant in the case of infinite dimensional bipartite Continuous Variable (CV)
states too. An important advance came about through an 
identification of  how Peres-Horodecki criterion gets translated elegantly into the properties of 
the second moments (uncertainties) of CV states~\cite{Simon}.  This results in  restrictions~\cite{Simon, duan} on  %%@
the covariance matrix of an entangled bipartite CV state.
In the special case of  two-mode Gaussian states, where the basic entanglement properties are imbibed in the structure %%@
of  its covariance matrix,  the  restrictions on the covariance matrix are found to be necessary and sufficient for %%@
inseparability~\cite{Simon, duan}. 

Here, we construct a two qubit variance matrix (analogous to that of CV states) for
two qubits and derive corresponding inseparability constraints imposed on it.

Let us first recapitulate succinctly  the approach employed by Simon~\cite{Simon}
for  bipartite CV states: 
The basic variables of bi-partite CV states are the conjugate quadratures of two field modes,
\begin{equation}
\label{qudra}
\hat{\xi}=(\hat{q}_1, 
\hat{p}_1, \hat{q}_2, \hat{p}_2),  
\end{equation}  
which satisfy the canonical commutation relations
\begin{equation}
\label{commu}
[\hat{\xi}_\alpha,\hat{\xi}_\beta]= i\,\Omega_{\alpha\beta}, \,\,\,\,\,\,\,
\alpha, \beta=1,2,3,4\,   \\
\end{equation}
where,
\begin{eqnarray}
\Omega&=&\left(
\begin{array}{cc}
J & 0\\
0 & J
\end{array}
\right),\ \ \ \nonumber\\
\nonumber\\
{\rm and}\,\,\,\,\,\,\,
J&=&\left(
\begin{array}{cc}
0 & 1\\
-1 & 0
\end{array}
\right). 
\end{eqnarray}
The matrix $\Omega$ is known as a symplectic matrix. All real linear transformations
applied to the operators ${\xi}_{\alpha\,\beta}$ that obey the commutation realtions Eq.~(\ref{commu}),
form a group known as the symplectic group.

The second moments are embodied in the real
symmetric $4 \times 4$ covariance matrix of a bipartite CV state, which is 
defined through it's elements:
 \begin{equation}
 \label{cvvar}
V_{\alpha\beta}=\frac{1}{2}\langle\{ 
\Delta\hat{\xi}_\alpha, \Delta\hat{\xi}_\beta\}\rangle ,
\end{equation}
 where,
\begin{equation}
\label{deltaxi}
\Delta\hat{\xi}=\hat{\xi}_\alpha-\langle \hat{\xi}_\alpha\rangle,
\end{equation}
and
\begin{equation}
\label{deltaxi1}
\left\{\Delta\hat{\xi}_\alpha, \Delta\hat{\xi}_\beta \right\}= 
 \Delta\hat{\xi}_\alpha\Delta\hat{\xi}_\beta+\Delta\hat{\xi}_\beta
 \Delta\hat{\xi}_\alpha.
\end{equation}

Under canonical transformations, the variables of the two-mode system transform 
as 
$$\hat{\xi}\rightarrow\hat{\xi'}=S\,\hat{\xi},$$
where $S\in Sp\,(4,R)$ corresponds to a real symplectic $4\times 4$ matrix.
Under such transformations, the covariance matrix goes as 
$$V\rightarrow V'= S\,V\,S^T.$$ 
It is convenient to caste the 
covariance matrix in a $2\times 2$ block form:
\begin{eqnarray}
\label{var}
V=\left(
\begin{array}{cc}
A & C\\
C^T & B
\end{array}
\right).
\end{eqnarray}
The entanglement properties hidden in the covariance matrix $V$ remain unaltered under a local $Sp\,(2,R) \otimes %%@
Sp\,(2,R)$ transformation.
Such a local operation transforms  
the blocks $A,\ B,\ C$ of the variance matrix Eq.~(\ref{var}) as
\begin{eqnarray}
\label{trans}
A\rightarrow A'&=&S_1\,A\,S^T_1,\nonumber\\
B\rightarrow B'&=&S_2\,B\,S^T_2,\nonumber \\
C\rightarrow C'&=&S_1\,C\,S^T_2.
\end{eqnarray}
There are four local invariants associated with $V$ given in terms of the blocks
$A,\ B,\ C$:
\begin{eqnarray}
\label{blockinv}
I'_1&=&{\rm det}\,A,\nonumber\\
I'_2&=&{\rm det}\,B, \nonumber\\
I'_3&=&{\rm det}\ C ,\nonumber\\ 
I'_4&=&{\rm Tr}(AJCBC^TJ).
\end{eqnarray}
The  Peres-Horodecki criterion imposes the restriction~\cite{Simon}
\begin{equation}
\label{cvinv} 
I'_1\, I'_2+\left(\frac{1}{4}-\vert I'_3\vert\right)^2-I'_4\geq\,\frac{1}{4}\,(I'_1+I'_2)
\end{equation}
on the second moments of every  separable CV state.\\
The signature of the invariant $I'_3={\rm det}\,C$ has an important consequence: 
{\em Gaussian states with \,$I'_3\geq 0$ 
are necessarily separable, where as those with\, $I'_3< 0$ and violating 
{\rm Eq.~(\ref{cvinv})}
are entangled}. \\ 
In other words, for Gaussian states violation of the 
condition Eq.~(\ref{cvinv}) is both necessary and sufficient for entanglement.

In the next section, we explain a similar formalism for symmetric two qubits by constructing a covariance matrix and %%@
analyzing its inseparability behaviour.

\section{Two qubit covariance matrix}
\label{c block}
The basic variables of a two qubit system are expressed as a operator column(row) as
 $$\hat{\zeta}^T~=~\left(\sigma_{1i},\sigma_{2j}\right),\,\,\,\,i,j=1,2,3.$$
The  $6 \times 6$  real symmetric covariance matrix ${\cal V}$ of a two qubit system  may be  defined through,
\begin{equation} 
\label{variance}
{\cal V}_{\alpha i;\, \beta  j}=\frac{1}{2}\,\langle \{\Delta \hat{\zeta}_{\alpha\, i}, \Delta \hat{\zeta}_{\beta\, j} %%@
\}\rangle,
\end{equation}
with $\alpha,\ \beta=1,2\,; \,\,\,\,\ i,j=1,2,3.$ 
The variance matrix ${\cal V}$ can 
be conveniently written in the $3\times 3$ block form as
\begin{eqnarray}
\label{variance2}
{\cal V}=\left(
\begin{array}{cc}
{\cal A} & {\cal C}\\
{\cal C}^T & {\cal B}
\end{array}
\right),
\end{eqnarray}
where,
\begin{eqnarray}
\label{elem}
{\cal A}_{ij}&=&\frac{1}{2}\left[\langle\{\sigma_{1i},\sigma_{1j}\}\rangle
-\langle\sigma_{1i}\rangle\ \langle\sigma_{1j}\rangle\right] \nonumber \\
 &=&\delta_{ij}- \langle\sigma_{1i}\rangle\ 
\langle\sigma_{1j}\rangle\nonumber\\
&=&\delta_{ij}-s_{i}s_{j}, \nonumber\\
 \nonumber \\
{\cal B}_{ij}&=&\frac{1}{2}\left[\langle\{\sigma_{2i},\sigma_{2j}\}\rangle
-\langle\sigma_{2i}\rangle\ \langle\sigma_{2j}\rangle\right]   \nonumber \\ 
 &=&\delta_{ij}- \langle\sigma_{2i}\rangle\ \langle\sigma_{2j}\rangle\nonumber\\ &=&\delta_{ij}-r_{i}r_{j},\nonumber\\
 \nonumber \\
{\cal C}_{ij}&=&\frac{1}{2}\left[\langle{\sigma_{1i}\sigma_{2j}}\rangle
-\langle\sigma_{1i}\rangle\ \langle\sigma_{2j}\rangle\right]\nonumber\\
&=&t_{ij}-s_{i}r_{j}.
\end{eqnarray}
In other words, we have,
\begin{eqnarray}  
 {\cal A}&=&{\cal I}-s\,s^T,\nonumber\\ 
  {\cal B}&=&{\cal I}-r\,r^T, \nonumber\\
{\cal C}&=&T-s\,r^T.  
\end{eqnarray}
Here  ${\cal I}$ denotes a $3\times 3$ identity matrix and $s_i,\,r_i$ and $t_{ij}$ are the state parameters of an %%@
arbitrary two qubit state 
(see Eqs.~(\ref{twoqubit})~-~(\ref{corrt})). 

In the case of symmetric states, considerable simplicity ensues as a result of   
Eqs.~(\ref{symm}),~(\ref{Tsymm})  and the covariance matrix of Eq.~(\ref{variance2})
assumes the form:
\begin{eqnarray}
\label{symmvariance}
{\cal V}=\left(
\begin{array}{cc}
{\cal A} & {\cal C}\\
{\cal C}^T & {\cal A}
\end{array}
\right).
\end{eqnarray} 
where,
$${\cal A}~=~{\cal B}~=~{\cal I}~-~ss^T,$$ 
$$ {\rm and}\,\,\,\,\,\,\,\,\,\,{\cal C}~=~T~-~ss^T.$$ 
Explicitly, 
\begin{eqnarray}
\label{blockA}
{\cal A}=\left(
\begin{array}{ccc}
1-s_1^2 & -s_1s_2 & -s_1s_3\\
-s_1s_2 & 1-s_2^2 & -s_2s_3\\
-s_1s_3 & s_2s_3  & 1-s_3^2
\end{array}
\right),
\end{eqnarray}
and
\begin{eqnarray}
\label{blockc}
{\cal C}=\left(
\begin{array}{ccc}
t_{11}-s_1^2 & t_{12}-s_1s_2 & t_{13}-s_1s_3\\
t_{12}-s_1s_2 & t_{22}-s_2^2  & t_{23}-s_2s_3\\
t_{13}-s_1s_3 & t_{23}-s_2s_3  & t_{33}-s_3^2
\end{array}
\right).
\end{eqnarray}

We now establish an important property exhibited by  the 
off-diagonal block ${\cal C}$ of the covariance matrix of a symmetric two qubit state.
\section{Inseparability constraint on the covariance matrix}
{\bf Lemma}: {\it For every  separable symmetric state,  ${\cal C}~=~T-ss^T~$ is a positive semidefinite matrix.}\\ 
\\
{\bf Proof}: 
Consider a separable symmetric state of two qubits  
\begin{equation}
\label{sep}
\rho_{\rm (sym-sep)}=\sum_w\, p_{w}\, \rho_w\otimes\rho_w,\,\,\,
\displaystyle\sum_w p_w =1;\ \ 0\leq p_w\leq 1.
\end{equation}
The state variables $s_i$ and $t_{ij}$ associated with a separable symmetric state have the following structure:
\begin{eqnarray}
\label{sepst}
s_i&=&{\rm Tr}\,\left(\rho_{\rm (sym-sep)}\, \sigma_{\alpha\, i}\right)
=\displaystyle\sum_w p_w\, s_{wi},\nonumber \\ 
t_{ij}&=&{\rm Tr}\,\left(\rho_{\rm (sym-sep)}\,
 \sigma_{1i} \sigma_{2j}\right)
=\displaystyle\sum_w p_w\, s_{wi}\, s_{wj}.\,\,\,\,\,\,
\end{eqnarray}
Let us now evaluate  the quadratic form $n^T(T-ss^T)\,n$ where $n\, (n^T)$ denotes any  arbitrary real three %%@
componental column (row),  
in a separable symmetric state:
\begin{eqnarray}
\label{proof}
n^T(T-ss^T)\,n&=&\sum_{i,j}\, (t_{ij}-s_i\,s_j)\, n_i\, n_j\hskip 1in  \nonumber\\
 &=&\sum_{i,j}\left[\displaystyle\sum_{w}  p_w\,s_{wi}\,s_{wj}-\sum_{w} p_w \,
s_{wi}\, \sum_{w^\prime} p_{w^\prime}\, 
s_{wj^\prime}\right]\, n_i\, n_j\nonumber\\
\hskip 0.2in &=&\sum_w p_w\, (\vec{s}\cdot\hat{n})^2-\left(\sum_w p_w\, 
(\vec{s}\cdot\hat{n})\right)^2 ,\hskip 0.8in  
\end{eqnarray}
which has the structure $\langle A^2\rangle - \langle A\rangle^2$ and is 
therefore a positive semi-definite quantity.$\Box$
 
This lemma establishes the fact that the off diagonal block ${\cal C}$ of 
the covariance matrix is  {\em necessarily}   positive semidefinite for 
separable symmetric states. And therefore, ${\cal C}<0$  serves as a sufficient 
condition for inseparability in  two-qubit symmetric states.

We now investigate the inseparability constriant $T-ss^{\rm T}<0$ in the case of a pure entangled two qubit state. \\
An arbitrary pure two qubit state can be written in a Schmidt decomposed form  
\begin{eqnarray}
\label{schm}
\vert\Phi\rangle&=&\kappa_1\,\vert 0_1\, 0_2\rangle+\kappa_2\,\vert 
1_1\,1_2\rangle,\hskip 0.2in \kappa_1^2+\kappa_2^2=1\\
{\rm where,}\hskip .5in\nonumber\\
0&<&\kappa_2\, \leq \kappa_1<1,\nonumber
 \end{eqnarray}
are the Schmidt coefficients. 
The two qubit state can be written in the $4 \times 4$ matrix form using the basis $\{\vert 0_1\,0_2\rangle, \vert %%@
0_1\,1_2\rangle, \vert 1_1\,0_2\rangle,\vert 1_1\,1_2\rangle\}:$
\begin{eqnarray}
\label{blockc}
\rho=\vert\Phi\rangle\langle\Phi\vert=\left(
\begin{array}{cccc}
\kappa_1^2 & 0 & 0 & 2\,\kappa_1\kappa_2\\
0 & 0 & 0 & 0\\
0 & 0 & 0 & 0\\
2\,\kappa_1\kappa_2& 0 & 0 & \kappa_2^2
\end{array}
\right).
\end{eqnarray}

The $3\times 3$ real symmetric correlation matrix $T$ may be readily obtained using Eq.~(\ref{corrt}) as, 
\begin{eqnarray}
\label{Tblock}
T=\left(
\begin{array}{ccc}
2\kappa_1 \kappa_2,\ & 0 & 0\\
0 & -2\kappa_1 \kappa_2,  & 0\\
0 & 0  & 1
\end{array}
\right), \,\,\,\,\,\, {\rm Tr}\,T=1.
\end{eqnarray}
The average qubit orientation (Eq.~(\ref{avgspin}))  has the form
\begin{equation}
\label{spinsch}
s=\left(0,\, 0,\,\kappa_1^2-\kappa_2^2\right)=r.
\end{equation}
From Eq.~(\ref{Tblock}) and Eq.~(\ref{spinsch}), it is clear that an arbitrary 
two qubit pure state is symmetric in the Schmidt basis. 

The $3 \times 3$ matrix ${\cal C}=T-ss^{\rm T}$ takes the form
\begin{eqnarray}
\label{Cblock}
{\cal C}=T-ss^{\rm T}=\left(
\begin{array}{ccc}
2\kappa_1 \kappa_2,& 0 & 0\\
0 & -2\kappa_1 \kappa_2,  & 0\\
0 & 0  & 4\, \kappa_1^2\,\kappa_2^2
\end{array}
\right).
\end{eqnarray}
It can be clearly seen that ${\cal C}<0,$ for all entangled pure 
two-qubit states. \\
In other words, the condition ${\cal C}<0$ is both {\em necessary and sufficient} for pure entangled two-qubit states.

Interestingly, non-positivity of ${\cal C}$ completely characterizes inseparability in an arbitrary symmetric two %%@
qubit state, which will be proved in the Sec.~(\ref{insepcriteria}).  

\section{Complete characterization of inseparability in mixed two qubit symmetric states}
\label{insepcriteria}
We prove the following theorem:\\
{\bf Theorem:} The off-diagonal block ${\cal C}$ of covariance matrix of an entangled two qubit mixed state is %%@
\underbar {necessarily} non-positive.
 
{\bf Proof}: An arbitrary two qubit symmetric state, characterized by the density matrix Eq.~(\ref{symrho}), with the %%@
state parameters obeying the permutation symmetry requirements Eqs.~(\ref{symm}),~(\ref{Tsymm}) has the following %%@
matrix form:
\begin{equation}
\label{symmetricrho}
\rho_{\rm sym}=\frac{1}{4}
\left(\begin{array}{cccc} 1+2\, s_3 + t_{33} & A^*  & A^*  & (t_{11}-t_{22})-2i\, t_{12} \cr 
A  & (t_{11}+t_{22}) & (t_{11}+t_{22}) & B^*   \cr 
A  & (t_{11}+t_{22}) & (t_{11}+t_{22}) & B^*   \cr 
(t_{11}-t_{22})+2\, i t_{12}& B & B & 1 - 2\, s_3 + t_{33}   
\end{array}  \right)
\end{equation}
in the standard  two-qubit basis $\left\{|0_1\, 0_2\rangle\, , |0_1\, 1_2\rangle\, , |1_1\, 0_2\rangle\, , |1_1\, %%@
1_2\rangle\right\}.$\\
Here, we have denoted,
\begin{eqnarray*}
A&=&(s_{1}+i\, s_2)+ (t_{13}+i\, t_{23})\\
B&=&(s_{1}+i\, s_2)- (t_{13}+i\, t_{23}).
\end{eqnarray*}
Note that the two qubit basis is related to the total angular momentum basis $\vert J, M\rangle$ with $J=1, 0\, ; \, %%@
-J\leq M\leq J$ as follows: 
\begin{eqnarray}
\label{symbasis}
\vert 0_1\, 0_2\rangle &=& \vert 1, 1\rangle, \nonumber \\ 
\vert 1_1\, 1_2\rangle &=& \vert 1, -1\rangle, \hskip 1in   \nonumber \\ 
 \vert 0_1\, 1_2\rangle &=&\frac{1}{\sqrt 2}( \vert 1, 0\rangle + \vert 0, 0\rangle ),\nonumber \\ 
\vert 1_1\, 0_2\rangle &=& \frac{1}{\sqrt 2}( \vert 1, 0\rangle - \vert 0, 0\rangle ) ,  
\end{eqnarray}
and the following unitary matrix, 
\begin{equation}
\label{CG}
U=\left(\begin{array}{cccc} 1 & 0 & 0 & 0 \cr 
0 & \frac{1}{\sqrt{2}} & \frac{1}{\sqrt{2}} & 0 \cr 
0 & 0 & 0 & 1 \cr 
0 & \frac{1}{\sqrt{2}} & -\frac{1}{\sqrt{2}} & 0 
\end{array}\right),
\end{equation}
transforms the two qubit density matrix $\rho_{\rm sym}$ of Eq.~(\ref{symmetricrho}) to the angular momentum basis %%@
Eq.~(\ref{symbasis}):
\begin{eqnarray}
 U\, \rho_{\rm sym}\, U^\dag =\left(\begin{array}{cc}\rho_S & 0 \cr 0 & 0 \end{array} \right),\hskip 2in\nonumber\\
{\rm where},\hskip 5in\nonumber\\
\rho_S=\frac{1}{4}\left(\begin{array}{ccc} 1+2\, s_3+t_{33} & \sqrt{2}\, 
A^*  & (t_{11}-t_{22})-2i\, t_{12} \cr 
\sqrt{2}\, A  & 2(t_{11}+t_{22}) & \sqrt{2}\, B^*  \cr 
(t_{11}-t_{22})+2i\, t_{12} & \sqrt{2}\, B & 1 - 2\, s_3+t_{33} 
\end{array}  \right).\hskip .7in  
\end{eqnarray}

So, an arbitrary two qubit symmetric state always gets restricted to the 3 dimensional maximal angular momentum %%@
subspace spanned by 
$\left\{\vert J_{\rm max}=1, M\rangle  -1\leq M\leq 1\right\}.$ 
However, the partial transpose of $\rho_{\rm sym}$, does not get restricted to the symmetric subspace with $J_{\rm %%@
max}=1$, when transformed to the total angular momentum basis Eq.~(\ref{symbasis}).

Under the Partial transpose (PT) operation (say, on the second qubit),  
the Pauli spin matrices of second qubit change as 
$$\sigma_{21}\rightarrow\sigma_{21},\  \sigma_{22}\rightarrow -\sigma_{22}, 
\ \sigma_{23}\rightarrow\sigma_{23}.$$
When this PT operation, is followed by a local rotation about the 2-axis by an angle $\pi$, 
the spin operators of the second qubit completely reverse their signs: 
$$\sigma_{2i}\rightarrow -\sigma_{2i}.$$ 
Thus, PT map on the symmetric density operator Eq.~(\ref{symrho})
\begin{equation}
\label{symrho5}
\rho_{\rm sym}=\frac{1}{4}\left(I\otimes I+ \sum_{i=1}^3\,s_i\,(\sigma_{1i}+\sigma_{2i})\,+\sum_{i,j=1}^3 %%@
\sigma_{1i}\sigma_{2j}t_{ij}\right), 
 \end{equation}
leads to
 \begin{equation}
\rho_{\rm sym}^{{\rm T}_2}=\frac{1}{4}\, \left(I\otimes I+\sum_{i=1}^3(\sigma_{1i}\, s_{i}  
          -\sigma_{2i}\, s_{i})
          -\sum_{i=1}^3\sigma_{1i}\, \sigma_{2j}\,t_{ij}\,\right).  
\label{rhopt} 
\end{equation}
(Here ${\rm T}_2$ corresponds to partial transpose map on the second qubit).

We thus obtain, 
\begin{equation}
\label{rhopt}
\rho_{\rm sym}^{{\rm T}_2}=\frac{1}{4}\left( \begin{array}{cccc} (t_{11}+t_{22}) & a^*  & b^*   
  & -(t_{11}-t_{22})+2i\, t_{12} \cr 
a  & 1 + 2\, s_3 + t_{33}  & t_{33}-1 & -a^*   \cr 
b  & t_{33}-1  & 1-2s_3+t_{33} & -b   \cr 
-(t_{11}-t_{22})-2\, i t_{12}& -a & -b^* &  (t_{11}+t_{22})   
\end{array}\right),  
\end{equation} 
in the basis $\left\{|0_1\, 0_2\rangle\, , |0_1\, 1_2\rangle\, , |1_1\, 0_2\rangle\, , |1_1\, 1_2\rangle\right\}.$
Here we have denoted,
\begin{eqnarray*}
a&=&-(s_1+i\, s_2)-(t_{13}+i\, t_{23}),\\
b&=&(s_1+i\, s_2)-(t_{13}-i\, t_{23}).
\end{eqnarray*}
Now a unitary transformation Eq.~(\ref{symbasis}) which corresponds to a basis change Eq.~(\ref{CG}) gives;
\begin{eqnarray}
\label{rhoptJ}
\bar\rho_{\rm sym}^{{\rm T}_2}=U\, \rho^{{\rm T}_2}\, U^\dag \hskip 3in\nonumber\\
\nonumber\\
=\frac{1}{4}\left( \begin{array}{cccc} (t_{11}+t_{22}) & -\sqrt{2}\,t_{13}  & -(t_{11}-t_{22})+2i\, t_{12}  & \bar a %%@
\cr 
-\sqrt{2}t_{13} & 2\,t_{33}  & \bar b^* & 2\,s_3   \cr 
-(t_{11}-t_{22})-2i\, t_{12}   & \bar b  & (t_{11}+t_{22}) & \sqrt{2}s_{1}  \cr 
\bar a^* & 2\,s_3 & \sqrt{2}s_{1} &  4 
\end{array}  \right),  
\end{eqnarray} 
where, 
\begin{eqnarray*}
\bar a&=&\sqrt{2}(-s_1+is_2+it_{23})\\
\bar b&=&\sqrt{2}(t_{13}+is_2+it_{23}).
\end{eqnarray*}
It may therefore be seen that 3 dimensional subspace spanned by $J_{\rm max}=1$
of total angular momentum of a symmetric two qubit state does not restrict itself to a $3 \times 3$ block form.

Interestingly, a further change of basis defined by, 
\begin{eqnarray}
\label{subspace}
\vert X\rangle&=&\frac{-1}{\sqrt 2}(\vert 1,1\rangle-\vert 1, -1\rangle),\nonumber\\ 
\vert Y\rangle&=&\frac{-i}{\sqrt 2}(\vert 1,1\rangle+ \vert 1, -1\rangle),\nonumber\\ 
\vert Z\rangle&=&\vert 1, 0\rangle, 
\end{eqnarray} 
which corresponds to a unitary transformation 
\begin{equation}
\label{U2}
U'=\frac{1}{\sqrt 2}\left(\begin{array}{cccc} -1 & 0 & 1 & 0 \cr 
-i & 0 & -i & 0 \cr 
0 & \sqrt{2} & 0 & 0 \cr 
0 & 0 & 0 & \sqrt{2} 
\end{array}\right)
\end{equation}
on $\bar\rho_{\rm sym}^{{\rm T}_2},$ leads to the following elegant structure 
\begin{eqnarray}
\label{ptsym}
\bar\rho_{\rm sym}^{{\rm T'}_2}&=&U'\, \bar\rho_{\rm sym}^{{\rm T}_2}\, U'^\dag\nonumber\\
&=&\frac{1}{2}\left( \begin{array}{cccc} 
t_{11} & t_{12} & t_{13} &s
_1 \cr 
t_{12} & t_{22} & t_{23} &s_2\cr
t_{13} & t_{23} &t_{33} & s_3 \cr 
s_1 & s_2 & s_3 & 1
\end{array}  \right)\nonumber\\  
%\end{eqnarray}
%which has an elegant block structure for the
%PT symmetric density matrix:  
%\begin{equation}
%\label{eleblock}
&=&\frac{1}{2}\left(\begin{array}{cc} T & s \cr s^T & 1\end{array}   \right)
\end{eqnarray}  

Now a congruence~\footnote{Note that the congruence operation does not alter the positivity (negativity) of the %%@
eigenvalue structure of the matrix.}
operation 
$L\,\bar\rho_{\rm sym}^{{\rm T'}_2}\, L^\dag$ with
$$L=\left(\begin{array}{cc} {\cal I} & -s\cr 0 & 1\end{array} \right),$$ 
gives,
\begin{eqnarray}
L\,\bar\rho_{\rm sym}^{{\rm T'}_2}\, L^\dag
&=&\frac{1}{2}\left( \begin{array}{cccc} 
t_{11}-s_1^2 & t_{12} & t_{13} &0 \cr 
t_{12} & t_{22}-s_2^2 & t_{23} &0\cr
t_{13} & t_{23} &t_{33}-s_3^2 & 0 \cr 
0 & 0 & 0 & 1
\end{array}  \right)\nonumber
\end{eqnarray}
or
\begin{eqnarray}
\label{t-ss}  
L\,\bar\rho_{\rm sym}^{{\rm T'}_2}\, L^\dag&=&\frac{1}{2}\left(\begin{array}{cc} T-ss^T & 0 \cr 0 & 1\end{array} %%@
\right).
\end{eqnarray}
It is therefore evident that {\em \underbar {negativity of $T-ss^T$ necessarily implies} \underbar{negativity of the %%@
partially} \underbar{ transposed arbitrary} \underbar{two qubit symmetric density matrix}} as,
$$\rho_{\rm sym}^{{\rm T}_2}<0\Leftrightarrow\bar\rho_{\rm sym}^{{\rm T}_2}<0\Leftrightarrow L\,\bar\rho_{\rm %%@
sym}^{{\rm T'}_2}\, L^\dag<0\Leftrightarrow {\cal C}=T-ss^{\rm T}<0;$$
i.e., non-positivity of ${\cal C}=T-ss^{\rm T}$ necessarily implies that
partially transposed two qubit symmetric density matrix $\rho_{\rm sym}^{{\rm T}_2}$ is negative.
In other words, ${\cal C}<0$ captures Peres's inseparability  criterion on symmetric two qubit state completely. Hence %%@
the theorem. $\Box$

In the next section, we  explore  how negativity of the matrix 
${\cal C}$ reflects itself on the structure of the local invariants associated with the two qubit state.
\section{Local invariant structure}
\label{local}
The off-diagonal block ${\cal C}$ of the covariance matrix is a real 
$3 \times 3$ symmetric matrix and so, can be diagonalized by  an orthogonal matrix $O$ i.e.,
$$O{\cal C}O^\dag={\cal C}^d=(c_1,c_2,c_3).$$
The orthogonal transformation corresponds to identical unitary transformation $U \otimes U$ on the qubits.

We denote the eigenvalues of the off-diagonal block  ${\cal C}$ of the covariance matrix Eq.~(\ref{variance2}) 
by $c_1,\ c_2,\, {\rm and} \  c_3$.  Restricting   ourselves  to identical 
local unitary transformations, we define  three local invariants, which 
completely determine the eigenvalues $c_1,c_2,c_3$ of 
${\cal C}=T-ss^{\rm T}$: 
\begin{eqnarray}
\label{newinv}
\bar{\cal I}_1&=&\det\, ({\cal C})=c_1\,c_2\,c_3, \nonumber \\ 
 \bar{\cal I}_2&=&{\rm Tr}\, ({\cal C})=c_1+c_2+c_3,\nonumber \\
  \bar{\cal I}_3&=&{\rm Tr}\, ({\cal C}^2)=c_1^2+c_2^2+c_3^2. 
\end{eqnarray}  
The invariant  $\bar{\cal I}_2$ may be rewritten as
\begin{equation}
\label{ci2} 
\bar{\cal I}_2={\rm Tr}\,(T~-~s\,s^T)~=~1-s_0^2,
\end{equation} 
since ${\rm Tr}\,(T)~=~1$ for a symmetric state. 
Here, we have denoted $${\rm Tr}\, (s\,s^T)=s_1^2+s_2^2+s_3^2=s_0^2.$$

Another useful invariant, 
which is a combination of the invariants defined through Eq.~(\ref{newinv}),  may be 
constructed as 
\begin{equation}   
\label{aninv}
 \bar{\cal I}_4=\frac{\bar{\cal I}_2\,^2-\bar{\cal I}_3}{2}=c_1\,c_2+c_2\,c_3+c_1\,c_3. 
 \end{equation}
 Positivity of the single qubit reduced density operator demands   
 $s_0^2\leq 1$  and leads in turn to the observation that the invariant 
 $\bar{\cal I}_2$ is  positive for all symmetric states. Thus, all the three eigen values 
 $c_1,\, c_2,\, c_3$  of ${\cal C}$ can never  assume negative values for 
 symmetric qubits  and at most two of them can be negative.

We consider three distinct cases encompassing all pairwise  entangled symmetric 
states. 

Case (i): Let one of the eigenvalues $c_1=0$ and of the remaining two, let $c_2<0$ and $c_3>0$.

Clearly, the invariant $\bar{\cal I}_1=0$ in this case. 
But we have 
\begin{eqnarray}
\label{cond2}
\bar{\cal I}_4=c_2\, c_3 < 0,
\end{eqnarray}
which leads to a local invariant condition for two-qubit entanglement.

\noindent Case (ii): Suppose any two eigenvalues say, $c_1, c_2$, are  negative  and the 
third one $c_3$ is positive. 

Obviously, $\bar{\cal I}_1>0$ in this case. But the invariant $\bar{\cal I}_4$ 
assumes negative value:
\begin{eqnarray}
\label{cond3}
\bar{\cal I}_4=c_1\, \bar{\cal I}_2-c_1^2+c_2\, c_3<0 \\ \nonumber
\end{eqnarray}
as each term in the right hand side  is negative. In other words, 
 $\bar{\cal I}_4<0$ 
gives the criterion for bipartite entanglement in this case too.

\noindent Case (iii): Let  $c_1<0$;   $c_2\  {\rm and}\  c_3$ be positive.

In this case we have 
\begin{eqnarray}
\label{cond1}
\bar{\cal I}_1<0,  
\end{eqnarray}
giving the inseparability  criterion in terms of a local invariant. 

The new set of local invariants (see Eqs.~(\ref{newinv}),~(\ref{aninv})) associated
with the off-diagonal block ${\cal C}$ of the covariance matrix
can be related to the symmetric two qubit local invariants given by Eq.~(\ref{inv}).
In the following discussion, we restrict ourselves to the identical 
local unitary transformations $U\otimes U$ (Eq.~(\ref{tperp}))
which transform the state vectors to the following form:
\begin{eqnarray}
\label{tperp1}
\vec{s}&=&(0, 0, s_0),\nonumber\\
\nonumber\\
{\rm and}\ \ \  T&=&\left(\begin{array}{lll}t_\perp^{(+)} & 0 & t''_{13}\\
 0 & t_{\perp}^{(-)}& t''_{23}\\
 t''_{13} & t''_{23} & t'_{33}
\end{array}\right).
\end{eqnarray} 
We note that
\begin{eqnarray}
\label{CI1} 
\det\, ({\cal C})&=&\det\, (T-ss^T)\nonumber\\
&=&t_\perp^+t_\perp^-(t'_{33}-s_0^2)-(t''_{23})^2t_\perp^+ -(t''_{13})^2t_\perp^-.\nonumber\\
&=&{\rm det}\, T-s_0^2 t_\perp^+t_\perp^-.\nonumber\\
&=&{\cal I}_1-\frac{{\cal I}_5}{2},
\end{eqnarray}
where we have used Eqs.~(\ref{twoinv}),~(\ref{ss2I5}). We thus have,
\begin{eqnarray}
\label{CI11} 
\bar{\cal I}_1={\cal I}_1-\frac{{\cal I}_5}{2}.
\end{eqnarray}
We further find that,
\begin{eqnarray}
\label{CI2}
\bar{\cal I}_2={\rm Tr}\, ({\cal C})&=&{\rm Tr}\, (T-ss^T)\nonumber \\
&=& {\rm Tr}\, (T)-{\rm Tr}\,(ss^T)\nonumber \\
&=& 1-s^{\rm T} s\nonumber\\
&=& 1-{\cal I}_3.
\end{eqnarray}
The invariant $\bar{\cal I}_3,$ can be written as
\begin{eqnarray}
\label{CI3}
\bar{\cal I}_3={\rm Tr}\, ({\cal C}^2)&=&{\rm Tr}\,[(T-ss^T)^2]\nonumber\\
&=& {\rm Tr}\,[(T^2)+(ss^{\rm T}ss^{\rm T})-Tss^{\rm T}
-ss^{\rm T}T]\nonumber\\
&=& {\rm Tr}\, (T^2)+(s^T\,s)^2-2\,s^{\rm T}Ts.\nonumber
\end{eqnarray}
From Eq.~(\ref{inv}), we have $\bar{\cal I}_3$ given by,
\begin{eqnarray}
\label{CI31}
\bar{\cal I}_3&=&{\cal I}_2+{\cal I}_3^2-2{\cal I}_4.
\end{eqnarray}

Now, we  proceed to explore how this basic structure ${\cal C}<0$ 
reflects itself  via  collective second moments of a symmetric $N$ qubit system.
\section{Implications of ${\cal C}<0$ in symmetric $N$ qubit systems} 
Collective observables are expressible in terms of total  angular momentum operator as
\begin{equation}
\label{ang} 
\vec{J}=\sum_{\alpha=1}^N\frac{1}{2}\,\vec{\sigma}_\alpha
\end{equation}
where $\vec{\sigma}_\alpha$ denote  the Pauli spin operator of the 
$\alpha^{\rm th}$ qubit.

The collective  correlation matrix involving first and second moments of 
$\vec{J}$ may be defined as,
\begin{equation} 
\label{vn} 
V^{(N)}_{ij}=\frac{1}{2}\langle J_i J_j+J_j J_i\rangle - \langle J_i\rangle \langle J_j \rangle. 
\end{equation}
Using Eq.~(\ref{vars1}) and Eq.~(\ref{vart}), we can express the first and second order moments  $\langle %%@
J_i\rangle,\break \langle J_i J_j+J_j J_i\rangle$
in terms of the two qubits state parameters. After simplification, we obtain, 
\begin{equation}
V^{(N)}=\frac{N}{4}\left(\begin{array}{lll}1-t_{11}+N\,(t_{11}-s_1^2) &
 (N-1)\,t_{12} & (N-1)\,t_{13}\\
 (N-1)\,t_{12} & 1-t_{22}+N\,(t_{22}-s_2^2)& (N-1)\,t_{23}\\
 (N-1)\,t_{13}& (N-1)\,t_{23}& 1-t_{33}+N\,(t_{33}-s_0^2)
\end{array}\right). 
\end{equation}
We can now express $V^{(N)}$ in the following compact form,
\begin{equation}
\label{varmatrix}
V^{(N)}=\frac{N}{4}\left({\cal I}-ss^T+ (N-1)\, (T-ss^T)\right) 
\end{equation} 
where ${\cal I}$ is a $3\times 3$ identity matrix; $s^{\rm T}=(s_1,s_2,s_3)$
and $T$ denotes the two qubit correlation matrix (see Eq.~(\ref{corrmat})).
We simplify Eq.~(\ref{varmatrix}) further.\\
By shifting the second term i.e., $\frac{N}{4}\, ss^{\rm T}$ to the left hand side,
we obtain, 
\begin{eqnarray}
\label{varmatrix2}
V^{(N)}+\frac{N}{4}\, ss^{\rm T}=\frac{N}{4}\left({\cal I}+ (N-1)\,
 {\cal C}\right),\,\,\,\,\,\,{\cal C}=T-ss^{\rm T}  
\end{eqnarray}
Expressing $s_i$ in terms of the collective observables, we have,
\begin{equation}
\label{subs} 
s_i=\frac{2}{N}\,\langle J_i\rangle=\frac{2}{N}\,S_i.
\end{equation}
Now, substituting Eq.~(\ref{subs}) in Eq.~(\ref{varmatrix2}), we obtain,
\begin{equation}
\label{varmatrix3}
V^{(N)}+\frac{1}{N}\, SS^{\rm T}=\frac{N}{4}\left({\cal I}+ (N-1)\, {\cal C}\right). 
\end{equation}
For all symmetric separable states we have established that
${\cal C} \geq 0$ (see our theorem in Sec.~\ref{insepcriteria}).
We thus obtain the following constraint on the collective correlation matrix $V^{(N)}:$
\begin{equation} 
\label{con}
V^{(N)}+\frac{1}{N}\, SS^{\rm T} < \frac{N}{4}\, {\cal I}.
\end{equation}
Pairwise entangled symmetric multiqubit states {\em necessarily} satisfy the above condition.

Note that under identical local unitary transformations 
~$U\otimes~ U\otimes ~\ldots ~\otimes~ U$~ on the qubits, the variance matrix $V^{(N)}$ and the average spin $S$ %%@
transform as 
\begin{eqnarray} 
\label{tran2}
V^{(N)'}&=&O\, V^{(N)}\, O^T,\nonumber\\
{\rm and}\,\,\,\,\,\,\,\,\, S'&=&O\, S, 
\end{eqnarray}
where $O$ is a $3 \times 3$ real orthogonal rotation matrix corresponding  to 
the  local unitary transformation $U$ on all the qubits. Thus, the $3\times 3$ real symmetric matrix %%@
$V^{(N)}+\frac{1}{N}\, SS^T$ can 
always be diagonalized by a suitable  identical local unitary transformation on all the qubits. In other words, %%@
(\ref{con}) is a local invariant condition and it essentially implies:
\begin{center}
  {\em The symmetric $N$ qubit system is pairwise entangled iff the least eigen value of the real symmetric matrix %%@
~$~V^{(N)}~+~\frac{1}{N}~\,~ SS^T$~
 is less than $N/4$.} 
\end{center}
\section{Equivalence between the generalized spin squeezing inequalities and negativity of
 ${\cal C}$}
Let us consider the generalized spin squeezing inequalities of Ref.~\cite{Kor} 
\begin{eqnarray}
\label{spin}
\frac{4\langle\Delta J_k^2\rangle}{N}<1-\frac{4\langle J_k\rangle^2}{N^2},
\end{eqnarray}
where $J_k=\vec{J}\cdot\hat{k},$ with $\hat{k}$ denoting an arbitrary unit 
vector, and $$\langle\Delta J_k^2\rangle=\langle J_k^2\rangle-\langle 
J_k\rangle^2.$$ 
Expressing $\langle J_k\rangle$ and $\langle J_k^2\rangle$ in terms of the two qubit state variables $s_i$ and %%@
$t_{ij}$  we have,
\begin{equation} 
\label{spin2} 
\langle J_k \rangle=\frac{1}{2}\ \sum_{\alpha=1}^N\sum_{i=1}^3 
\left\langle\sigma_{\alpha i}\right\rangle
k_i=\frac{N}{2}\ (\vec{s}\cdot\hat{k})
\end{equation}
\begin{eqnarray}
\label{spin3}
  \langle J_k^2 \rangle&=& \frac{N}{4}+\frac{1}{4}\,\sum_{i,j}
 \sum_{\alpha,\beta\neq\alpha}\langle\sigma_{\alpha i} \sigma_{\beta 
j}\rangle\ k_i k_j\nonumber\\
&=&\frac{N}{4}+\frac{1}{2}\,\sum_{i,j}\sum_{\alpha=1}^N 
\sum_{\beta>\alpha=1}^N\langle\sigma_{\alpha i} \sigma_{\beta j}\rangle\ 
k_ik_j\nonumber\\
&=&\frac{N}{4}+\frac{1}{2}\sum_{i,j}\sum_{\alpha,\beta>\alpha}t_{ij}\ 
k_ik_j\nonumber\\
&=&\frac{N}{4}+\frac{N(N-1)}{4}\,\sum_{i,j}t_{ij}\ k_ik_j\nonumber\\
{\rm i.e.,}\ \langle J_k^2 \rangle&=&\frac{N}{4}\ \left(1+(N-1)\,k^T T 
k\right).
\end{eqnarray}
Using Eqs.~(\ref{spin2}),~(\ref{spin3}) in the generalized spin squeezing inequalities
given in Eq.~(\ref{spin}), we obtain,
\begin{eqnarray}
\label{equa}
\frac{4}{N}[\langle J_k^2\rangle-\langle 
J_k\rangle^2]&<&1-\frac{4 \langle J_k^2\rangle}{N^2}\nonumber\\
\frac{4}{N}\left[\frac{N}{4}\ \left(1+(N-1)\,k^T T 
k\right)-\frac{N^2}{4}\ (\vec{s}\cdot\hat{k})^2\right]
&<&1-\frac{4}{N^2}\,\left(\frac{N^2 (\vec{s}\cdot\hat{k})^2}{4}\right)\nonumber\\
\left[(1+(N-1)\,k^T T k)- N(\vec{s}\cdot\hat{k})^2\right]
&<&1-(\vec{s}\cdot\hat{k})^2\nonumber\\
(N-1)\,k^T T k&<&(N-1)\,(\vec{s}\cdot\hat{k})^2\nonumber\\
k^T(T-ss^T)k&<&0,\nonumber\\
{\rm or}\ \ \ \
{\cal C}<0.
\end{eqnarray}
Thus we find that the generalized spin squeezing inequality 
is equivalent to the condition ${\cal C}=T-ss^{\rm T}<0.$
\newpage
\section{Conclusions}
We have constructed a two qubit variance matrix and have shown here that the off-diagonal block of the variance matrix  %%@
${\cal C}$ 
of a separable symmetric two qubit state is a positive semidefinite quantity.  
An equivalence between the Peres-Horodecki 
criterion and the negativity of the covariance matrix ${\cal C}$ is  
established, showing that the covariance matrix criterion is both necessary  and sufficient for entanglement in %%@
symmetric two qubit states.
Thus symmetric two-qubit states satisfying the condition ${\cal C}<0$ are identified as inseparable. 
Further, the inseparability constraint ${\cal C}<0$ is shown to be equivalent to 
the recently proposed~\cite{Kor} generalized spin squeezing inequalities for pairwise
entanglement in symmetric $N$-qubit states. 
An elegant local invariant structure exhibited by these constraints on the two qubit covariance matrix has also been %%@
discussed.	%%%%%%%%%%%%%%%%%%%%%%%%%%%%%%%%%%%%%%%%%%%%%%%%%%%%%%%%%%%%%%%%%%%%%%%%%%%%%%
%%%%%%%%%%%%%%%%%%%%%%%%%%%%%%%%%%%%%%%%%%%%%%%%%%%%%%%%%%%%%%%%%%%%%%%%%%%%%%
%%%%%%%%%%%%%%%%%%%%%%%%%%%%%%%%%%%%%%%%%%%%%%%%%%%%%%%%%%%%%%%%%%%%%%%%%%%%%%
\chapter{Summary}
\label{c:conclusion}

\markboth{}{Summary}

Quantum correlated multiqubit states offer promising
possibilities in low-noise \break spectroscopy~\cite{Wineland94}, high precision
interferometry~\cite{Yur86,Kit91,Gio04} and in the implementation of quantum
information protocols~\cite{Brie}. Multiqubit states which are symmetric under interchange of particles (qubits) form %%@
an important class due to their experimental significance~\cite{Sor101,Sac00, Exptl1} as well as the mathematical %%@
simplicity
and elegance associated with them.

Individual qubits (two-level atoms) in a multiqubit system  
are not accessible in the macroscopic ensemble and therefore only collective
measurements are feasible. Any characterization of entanglement
requiring individual control of qubits cannot be 
experimentally implemented. For example, spin squeezing~\cite{Kit93}, i.e., reduction
of quantum fluctuations in one of the spin components
orthogonal to the mean spin direction below the fundamental
noise limit N/4 is an important collective signature of entanglement
in symmetric N qubit systems and is a consequence of
two-qubit pairwise entanglement~\cite{Sor101,Xwang03,ARU1}. 

Spin squeezing is one of the important quantifying signatures of quantum correlations in multiqubit systems. Spin %%@
squeezed atomic states are produced routinely in several laboratories~\cite{Kuz97, Kuz98} today. However, it is %%@
important to realize that spin squeezing
does not capture  quantum correlations {\em completely} and it serves only as a sufficient condition for pairwise %%@
entanglement in symmetric N qubit states. Investigations on other collective signatures~\cite{Kor, ARU1, ARU2} of %%@
entanglement gain their significance in this context. 

In this thesis, we have investigated the pairwise entanglement properties of symmetric multiqubits obeying permutation %%@
symmetry by employing two qubit local invariants. We have shown that a subset of 6 invaraints  $\{ {\cal I}_1\ -\  %%@
{\cal I}_6\}$, 
of a more general set of 18 invariants proposed by Makhlin~\cite{Mak02}, completely characterizes pairwise %%@
entanglement of the collective state. For a specific case of 
symmetric two-qubit system, which is realized in  several physically interesting examples like, even and odd spin %%@
states~\cite{Xi03}, 
Kitagawa - Ueda state generated by one-axis twisting Hamiltonian~\cite{Kit93},  
Atomic spin squeezed states~\cite{Agarwal90} etc, a subset of three independent invariants is sufficient to %%@
characterize the non-local properties completely. 
{\em For symmetric separable states, we have  proved that the entanglement invariants
${\cal I}_1,\, {\cal I}_4,\, {\cal I}_5$ and ${\cal I}_4-{\cal I}_3^2$ assume non-negative values.}

Based on  {\em negative} values of the  invariants
${\cal I}_1,\, {\cal I}_4,\, {\cal I}_5$ and ${\cal I}_4-{\cal I}_3^2,$
we have proposed a detailed classification scheme,  for pairwise entanglement in symmetric multiqubit system,   Our %%@
scheme also relates appropriate
collective {\em non-classical} features, which  can be
identified in each case of pairwise entanglement. Further, we have expressed collective features of entanglement, such %%@
as {\em spin squeezing}, in terms of these invariants. More specifically, we have shown that a symmetric multi-qubit %%@
system is spin squeezed {\em iff}\,  one of the entanglement
invariant is {\em negative}. Moreover, our invariant criteria are shown to be related to the family of generalized %%@
spin squeezing inequalities~\cite{Kor} (two qubit entanglement) involving
collective first and second order moments of total
angular momentum operator. 

Further, we have established an equivalence between the Peres-Horodecki 
criterion and the negativity of the off diagonal block ${\cal C}$ of the two qubit covariance matrix thereby showing %%@
that our condition is both {\em necessary  and sufficient} for  entanglement in symmetric two qubit %%@
states~\cite{ARU3}. Pairwise entangled symmetric multiqubit states necessarily obey these  constraints. We have also %%@
brought out an elegant local invariant structure exhibited by our constraints.

	%%%%%%%%%%%%%%%%%%%%%%%%%%%%%%%%%%%%%%%%%%%%%%%%%%%%%%%%%%%%%%%%%%%%%%%%%%%%%%
%%%%%%%%%%%%%%%%%%%%%%%%%%%%%%%%%%%%%%%%%%%%%%%%%%%%%%%%%%%%%%%%%%%%%%%%%%%%%%
%%%%%%%%%%%%%%%%%%%%%%%%%%%%%%%%%%%%%%%%%%%%%%%%%%%%%%%%%%%%%%%%%%%%%%%%%%%%%%

   \appendix

\addappheadtotoc
%%%%%%%%%%%%%%%%%%%%%%%%%%%%%%%%%%%%%%%%%%%%%%%%%%%%%%%%%%%%%%%%%%%%%%%%%%%%%%
%%%%%%%%%%%%%%%%%%%%%%%%%%%%%%%%%%%%%%%%%%%%%%%%%%%%%%%%%%%%%%%%%%%%%%%%%%%%%%
%%%%%%%%%%%%%%%%%%%%%%%%%%%%%%%%%%%%%%%%%%%%%%%%%%%%%%%%%%%%%%%%%%%%%%%%%%%%%%
\chapter{Pure and mixed density operators}
\label{apdx01}
\markboth{}{Pure and mixed density operators}

Consider a quantum system characterized by  a state $|\psi\rangle$ in the  Hilbert space ${\cal H}.$ Using a complete %%@
orthonormal basis $\left\{|u_n\rangle\right\}$ satisfying
\begin{itemize}
\item{orthonormality: $\langle u_m|u_n\rangle=\delta_{mn}$}
\item{completeness: $\sum_n |u_n\rangle\langle u_n|=I,$}
\end{itemize}
where $I$ is the unit operator.

We can expand  $|\psi\rangle$ as follows: 
\begin{equation}
\label{psi}
|\psi\rangle=\sum_n c_n|u_n\rangle,\,\,\,\,\,\,\,\, c_n=\langle u_n|\psi\rangle.
\end{equation}  
The expansion coefficients $c_n$ satisfy the normalization condition 
$$\sum_n \mid c_n \mid^2=\sum_n \mid\langle u_n|\psi\rangle\mid^2=1.$$
Let us evaluate the expectation value of an observable $A$ in the state  $|\psi\rangle$
\begin{eqnarray}
\label{expectation}
\langle A \rangle&=&\langle\psi| A |\psi \rangle\nonumber\\
&=& \sum_m\sum_n\langle\psi|u_m\rangle \langle u_m|A|u_n\rangle\langle u_n|\psi\rangle\nonumber\\
&=&\sum_m\sum_n\langle u_n|\rho|u_m\rangle\langle u_m|A|u_n\rangle\nonumber\\
&=&\sum_m\sum_n\rho_{nm} A_{mn}\nonumber\\
{\rm or}\,\,\,\,\,\,\,\,\,\,\,
\langle A \rangle&=&{\rm Tr}\,(\rho A),
\end{eqnarray}
where, 
\begin{equation}
\label{density matrix}
\rho=|\psi\rangle\langle\psi|
\end{equation}
is the {\em density operator} associated with the quantum system
(elements of which are denoted by $\langle u_n|\rho|u_m\rangle=\rho_{nm}$
in the given basis $\left\{|u_n\rangle\right\}$). 

%$$\rho_{mn}=\sum_m\sum_n\langle u_m|\psi\rangle\langle\psi|u_n\rangle.$$\\
The density operator satisfies the following properties:
\begin{enumerate}
\item{$\rho$ is a hermitian matrix
\begin{equation} 
\label{1rho}
\rho^*_{mn}=\rho_{nm}.
\end{equation}}
\item{$\rho$ is positive semi definite,
\begin{equation}
\label{2rho}
\rho\geq 0.
\end{equation}}
\item{$\rho$ has unit trace,
\begin{equation}
\label{3rho}
{\rm Tr}\,\rho=1.
\end{equation}}
\end{enumerate}
From the structure of $\rho=|\psi\rangle\langle\psi|,$
 (see Eq.~(\ref{density matrix})) it is clear that, 
\begin{equation}
\label{rho2=rho}
\rho^2=\rho,
\end{equation}
i.e., $\rho$ is idempotent operator.

Density operators satisfying Eq.~(\ref{density matrix}) form a subclass of the more general class of density operators %%@
and are termed as {\em pure} density operators in contrast to mixed states.

Mixed density operators are, in general, a convex mixture of pure states $|\psi_i\rangle:$
\begin{eqnarray}
\label{mixedrho}
\rho=\sum_i p_i |\psi_i\rangle\langle\psi_i|,\,\,\,\,\,\,0\leq p_i\leq 1,\,\sum_ip_i=1.
\end{eqnarray}
\noindent{\bf Number of real independent parameters characterizing the density operators}:\\
Writing a pure density operator $\rho$~(\ref{density matrix}) in the basis 
$\left\{|u_n\rangle\right\}$ explicitly, we obtain,
\begin{equation}
\label{purerho1}
\rho=\left(\begin{array}{cccccc}
                     |c_1|^2 & c_1c^*_2 & . &. &. & c_1c^*_n \cr 
					 c_2c^*_1 & |c_2|^2 & . &. &. & c_2c^*_n  \cr 
					 . & . & . &. &. &. \cr 
					 . & . & . &. &. &. \cr 
                      . & . & . &. &. &.\cr 
					 c_nc^*_1 & c_nc^*_2 & . &. &. & |c_n|^2 \cr 
                    \end{array}\right),   
\end{equation}   
i,.e, the density matrix given above is specified completely by the complex coefficients $c_n,$ which are constrained %%@
by the normalization condition  $\sum_n \mid c_n \mid^2=1.$ In other words, the density matrix 
of Eq.~(\ref{density matrix}) is completely characterized by $(2n-1)$ \underbar{real} parameters.

However, the more general class of mixed density operators satisfying the properties %%@
(Eqs.~(\ref{1rho})~-~~(\ref{3rho})) are characterized by
$n^2-1$ real parameters as shown below:

The density operator $\rho$ of a mixed quantum state may be explicitly written in the matrix form (in a suitable %%@
complete orthonormal basis $\left\{|u_n\rangle\right\})$ as, 
\begin{eqnarray}
\label{mixedrho1}
\rho=\left(\begin{array}{cccccc}
                     \rho_{11} & \rho_{12} & . &. &. &\rho_{1n} \cr 
					 \rho_{12}^* & \rho_{22} & . &. &. &\rho_{2n} \cr 
					 . & . & . &. &. &. \cr 
					 . & . & . &. &. &. \cr 
                      . & . & . &. &. &.\cr 
					 \rho_{n1}^* & \rho_{n2}^* & . &. &.&\rho_{nn} \cr 
                    \end{array}\right),\,\,\,\,\,\,\, 
					{\rm Tr}\,(\rho)=\sum_{n}\rho_{nn}=1.  
\end{eqnarray}
We count the number of parameters as follows:
\begin{itemize}  
\item {diagonal elements of $\rho$ are real and are constrained by the unit trace condition. So, we have $(n-1)$ real %%@
parameters specifying the diagonal elements.}
\item{There are $\frac{n(n-1)}{2}$ independent complex parameters $\rho_{nm}=\rho^*_{mn}, n\neq m,$ which fix the %%@
off-diagonal elements of $\rho.$ This leads to $n(n-1)$ real parameters.

 So, the total number of independent real parameters characterizing a mixed quantum state $\rho$ are given by %%@
$(n-1)+n(n-1)=n^2-1$}.
\end{itemize}
\noindent{\bf Single qubit density matrices:}\\
There are three real parameters specifying a mixed density operator of  
single qubit system which is represented by a $2\times 2$ hermitian    
matrix of unit trace as,
\begin{equation}
\label{singlequbit}
\rho=\left(\begin{array}{cc}
                     \rho_{11} & \rho_{12} \cr 
					 \rho_{12}^* & 1-\rho_{11} \cr 
               \end{array}\right).   
\end{equation}   
The Pauli spin matrices,   
\begin{equation}
\label{sigma1}
 \sigma_1=\left(\begin{array}{ll}
                    0 & 1 \cr 
                    1 & 0  \cr 
                    \end{array}\right), \,  
	\sigma_2=\left(\begin{array}{ll}
                    0 & -i \cr 
                    i & 0 \cr 
                    \end{array}\right), \, 
	\sigma_3=\left(\begin{array}{ll}
                    1 & 0 \cr 
                    0 & -1  \cr 
                    \end{array}\right) \,   				
\end{equation}			  
together with the identity matrix
\begin{equation}
\label{identity}
I=\left(\begin{array}{ll}
                    1 & 0 \cr 
                    0 & 1  \cr 
                    \end{array}\right)
\end{equation} 
provide a matrix basis for any $2 \times 2$ matrices.

In the case of spin-$\frac{1}{2}$ quantum states, we can always express 
\begin{eqnarray}
\label{singlequ1}
\rho=\frac{1}{2}[I+\sigma_1\, s_{1}+\sigma_2\, s_{2}+\sigma_3\, s_{3}],
\end{eqnarray} 
where, it is readily seen that
\begin{eqnarray}
\label{avgspin2}
s_1=\langle \sigma_{1}\rangle={\rm Tr}\,(\rho\,\sigma_{1}), \nonumber\\
s_2=\langle \sigma_{2}\rangle={\rm Tr}\,(\rho\, \sigma_{2}), \nonumber\\
s_3=\langle \sigma_{3}\rangle={\rm Tr}\,(\rho\, \sigma_{3}). 
\end{eqnarray}
\noindent{\bf Two qubit quantum system:}\\ 
There are $n^2-1=15$ parameters characterizing a two qubit system.
A natural operator basis $\left\{I \otimes I,\,\sigma_{1i},\,\sigma_{2i},\,\sigma_{1i}\sigma_{2j}\right\}$
is employed generally to expand the two qubit density matrix:
\begin{equation}
\label{twoqubitope}
\rho=\frac{1}{4}\left(I\otimes I+ \sum_{i=1}^3 s_i\, \sigma_{1i}+\sum_{i=1}^3\sigma_{2i}\, r_i
     +\sum_{i,j=1}^3 t_{ij}\,\sigma_{1i}\sigma_{2j}\right) \, , 
 \end{equation}
 with
\begin{eqnarray}
\label{sigma1i}
\sigma_{1i}&=&\sigma_i\otimes I\nonumber\\
\sigma_{2i}&=&I \otimes  \sigma_i.
\end{eqnarray} 
The single qubit state parameters $s_i$ and $r_i$ are given by
\begin{eqnarray}
\label{avgspin3}
s_i&=&{\rm Tr}\,(\rho\, \sigma_{1i}) \nonumber\\
r_i&=&{\rm Tr}\, (\rho\, \sigma_{2i}), \,\,\,\, i=1,2,3, 
\end{eqnarray} 
and the remaining two qubit correlation parameters are evaluated as follows: 
\begin{eqnarray} 
\label{corrt12}
t_{ij}={\rm Tr}\,(\rho\, \sigma_{1i}\sigma_{2j}).
\end{eqnarray}

   %%%%%%%%%%%%%%%%%%%%%%%%%%%%%%%%%%%%%%%%%%%%%%%%%%%%%%%%%%%%%%%%%%%%%%%%%%%%%%
%%%%%%%%%%%%%%%%%%%%%%%%%%%%%%%%%%%%%%%%%%%%%%%%%%%%%%%%%%%%%%%%%%%%%%%%%%%%%%
%%%%%%%%%%%%%%%%%%%%%%%%%%%%%%%%%%%%%%%%%%%%%%%%%%%%%%%%%%%%%%%%%%%%%%%%%%%%%%
\chapter{Peres PPT criterion}
\label{apdx02}
\markboth{}{Peres PPT criterion}

Any bipartite state $\rho$ defined on the Hilbert space $H$ is separable 
if it can be expressed in the convex product form
\begin{equation}
\label{product}
\rho_{\rm sep}=\sum_w p_w \rho_w^{(1)}  \otimes \rho_w^{(2)} \,\,\,\,{\rm where} \,\,0\leq p_w \leq 1\,\, {\rm and} %%@
\,\,\sum_w p_w=1. 
\end{equation}
Here, $\rho_w^{(1)}$ and $\rho_w^{(2)}$ are the density operators defining the 
systems 1 and 2 respectively. Quantum systems which cannot be written in the form
Eq.~(\ref{product}) are called {\em entangled states}. The fundamental question
in entanglement theory is the following: Given a composite quantum state
how do we test if the state is entangled? 

In this context, Peres~\cite{Peres} has
shown that the partially transposed density matrix of a separable state 
is a positive definite matrix. 
In other words, negative eigenvalues of a partially transposed density matrix 
necessarily imply entanglement in a quantum state.

To demonstrate this explicitly, let us write the matrix elements of the separable state given in Eq.~(\ref{product}) %%@
in the standard basis.
\begin{eqnarray}
\label{matrixelem}
\langle m\mu\vert\rho_{\rm sep}\vert n\nu\rangle&=&\sum_w p_w \langle m\vert \rho_w^{(1)}\vert n\rangle\, \langle %%@
\mu\vert \rho_w^{(2)}\vert \nu\rangle\nonumber\\
{\rm or}\,\,\,\,\,\,\,\,\,\,\, (\rho_{\rm sep})_{m\mu;n\nu}&=&\sum_w %%@
p_w\,(\rho^{(1)}_w)_{mn}\,(\rho^{(2)}_w)_{\mu\nu}.
\end{eqnarray}   
Here, Latin indices refer to the system 1 and Greek indices correspond to the system 2. 

The partial transposition (PT) of any density matrix $\rho$ with respect to one of the subsystems,
(say system 2) is defined as,
$$\rho^{T_2}_{m\mu;n\nu}=\rho_{m\nu;n\mu}.$$
Thus, the partial transpose of $\rho_{\rm sep}$ (Eq.~(\ref{matrixelem})) is given by 
\begin{eqnarray}
\label{ptmatrixelem}
(\rho_{\rm sep}^{{\rm T}_2})_{m\mu;n\nu}&=&(\rho_{\rm sep})_{m\nu;n\mu}\nonumber\\
&=&\sum_w p_w\,(\rho^{(1)}_w)_{mn}\,(\rho^{\left(2\right)}_w)_{\nu\mu}.
\end{eqnarray}
Using the hermiticity property, $\rho^\dag_i=\rho_i$ (see Eq.~(\ref{1rho})) of the density matrices, we obtain,
\begin{eqnarray}
\label{tmatrixelem}
(\rho_w^{(2)})^T&=&(\rho_w^{(2)})^*,
\end{eqnarray}
and thus,
\begin{eqnarray}
\label{ptrhosep}
(\rho_{\rm sep}^{{\rm T}_2})_{m\mu;n\nu}&=&\sum_w p_w\,(\rho^{(1)}_w)_{mn}\,(\rho^{(2)^*}_w)_{\nu\mu}\nonumber\\
{\rm or}\,\,\,\,\,\,\,\,\,\,\rho_{\rm sep}^{{\rm T}_2}&=&\sum_w p_w\,(\rho^{(1)}_w)\otimes(\rho^{(2)}_w)^*.
\end{eqnarray} 
Since $\left\{(\rho^{(2)}_w)^*\right\}$ correspond to {\em physical} density matrices, $\rho^{{\rm T}_2}_{\rm sep}$ is %%@
again a physically valid separable
density matrix.

 Therefore, the PT operation preserves the trace,
hermiticity and also the positive semi definiteness of a separable state whereas the {\em last property need not be %%@
respected by an entangled state}. 

Horodecki et. al~\cite{horo96} showed that Peres' positivity under partial transpose (PPT) is both necessary and %%@
sufficient for entanglement in 
$2 \times 2$ and $2 \times 3$ systems. However, for higher dimensions, there exist {\em bound entangled states} which %%@
are PPT states, though they are inseparable. Negative under partial transpose (NPT) is a sufficient condition for %%@
entanglement in higher dimensional composite quantum systems.

	 %%%%%%%%%%%%%%%%%%%%%%%%%%%%%%%%%%%%%%%%%%%%%%%%%%%%%%%%%%%%%%%%%%%%%%%%%%%%%%
%%%%%%%%%%%%%%%%%%%%%%%%%%%%%%%%%%%%%%%%%%%%%%%%%%%%%%%%%%%%%%%%%%%%%%%%%%%%%%
%%%%%%%%%%%%%%%%%%%%%%%%%%%%%%%%%%%%%%%%%%%%%%%%%%%%%%%%%%%%%%%%%%%%%%%%%%%%%%

\chapter{A Complete set of 18 invariants for an arbitrary two qubit state}
\label{apdx03}
\markboth{}{Arbitrary two qubit invariants}

%\linespread{2.2}

An arbitrary two qubit density matrix (see Eq.~(\ref{twoqubit})), specified by 15 real parameters \break $\{ s_i,\ %%@
r_i, \  t_{ij}\},$ is  characterized by a {\em complete} set of 18 polynomial invariants~\cite{Mak02}. These local %%@
invariants are given in terms of the state 
parameters associated with  an arbitrary two qubit system. Any two density matrices $\rho_1$ and $\rho_2$ are said to %%@
be locally equivalent
if and only if all the 18 invariants (see Table.1) have identical values for these states. 
To illustrate that these 18 invariants form a complete set, we may chose to work in a basis in which the $3\times 3$ %%@
real correlation matrix $T$ (which is nonsymmetric, in general) is diagonal.
Such a singular value decomposition of $T$ can be achieved by proper rotations $O^{(1)},\,O^{(2)} \in SO(3,R):$ 
\begin{eqnarray}
\label{tdiag}
T^d=O^{(1)}\, T\,  O^{(2)\, T}={\rm diag}(t_1,\,t_2,\,t_3).
\end{eqnarray}
We may note here that the diagonal elements of the correlation matrix $T^d$ are not the  eigenvalues of $T$ as %%@
$O^{(1)}\,T\,O^{(2)^{\rm T}}$ is not similarity transformation. However,
\begin{eqnarray}
\label{Ttranst}
T^d\,T^{d^{\rm T}}&=&O^{(1)}\,T\,O^{{(2)}T}\,O^{(2)}\,T^{\rm T}\,O^{{(1)}^{\rm T}}, \nonumber \\
&=&O^{(1)}\,T\,T^{\rm T}\,O^{{(1)}^{\rm T}},\nonumber \\
&=&\left(\begin{array}{ccc}
                     t_1^2 & 0 & 0 \cr 
                     0 & t_2^2 & 0\cr
					 0 & 0 & t_3^2 \cr
                    \end{array}\right),   
\end{eqnarray}
and 
$$T^{d^{\rm T}}\,T^d=O^{(2)}\,T^{\rm T}\,T\,O^{{(2)}^{\rm T}}.$$ 
In other words, $(t_1^2,\,t_2^2,\,t_3^2)$ are the eigenvalues of real symmetric matrix $T\,T^{\rm T}$ as well as %%@
$T^{\rm T}\,T.$
In order to determine the eigenvalues $(t_1^2,\,t_2^2,\,t_3^2)$ the following
polynomial quantities may be employed:
\begin{eqnarray}
\label{diagT}
{\rm det}\,(T\,T^{\rm T})&=&t_1^2\,t_2^2\,t_3^2, \nonumber\\
{\rm Tr}\,(T^{\rm T}T)&=&t_1^2+t_2^2+t_3^2, \nonumber\\
{\rm Tr}\,[(T^{\rm T}T)^2]&=&t_1^4+t_2^4+t_3^4.
\end{eqnarray}
Note that ${\rm det}\,(T^{\rm T}\,T ),\,{\rm Tr}\,(T^{\rm T}T),\,{\rm Tr}\,(T^{\rm T}T)^2$
are invariant under local unitary transformations on the two qubit state.
It is easy to see that
\begin{eqnarray}
\label{3invar}
{\rm det}\,(T\,T^{\rm T})&=&{\rm det}\,(T){\rm det}\,(T^{\rm T})=({\rm det}\,(T))^2, \nonumber\\
&=&{\rm det}\,(O^{(1)}\,T^d\,O^{(2)^{\rm T}}){\rm det}\,(O^{(2)}\,T^{d^{\rm T}},\,
O^{(1)^{\rm T}}),\nonumber\\
&=&{\rm det}(T^d)\,{\rm det}(T^d),\nonumber\\
&=&({\rm det}(T^d))^2.
\end{eqnarray}
Thus $\vert\,{\rm det}(T)\,\vert$ itself may be used in the first line of Eq.~(\ref{diagT})
instead of ${\rm det}\,(T\,T^{\rm T}).$ So, Makhlin chooses the first three
elements of his set of local invariants as,
\begin{eqnarray}
\label{invT}
I_1&=&{\rm det}\,(T)=t_1\,t_2\,t_3,\\
I_2&=&{\rm Tr}\,(T^{\rm T}T)=t_1^2+t_2^2+t_3^2,\\
I_3&=&{\rm Tr}\,[(T^{\rm T}T)^2]=t_1^4+t_2^4+t_3^4
\end{eqnarray}
The diagonal form of $T$ viz., $(t_1,\,t_2,\,t_3)$ can be determined using the invariants 
$I_{1-3}$  up to a simultaneous sign change for any two of them. Now, a local rotation on the first qubit  %%@
$R(\pi)^i\otimes I$ (where $R(\pi)^i$ is the $\pi$ rotation about the axis $i=1,2,3$), may be used to fix the signs of %%@
$t_1,\,t_2,\,t_3.$ It is then convenient to adopt the convention, (i) if $I_1\geq 0$, elements of $T^d$ are all %%@
positive 
and (ii) $t_1,\,t_2,\,t_3,$ are all negative, when $I_1<0.$ 

Let us restrict to two qubit
states with a fixed diagonal correlation matrix $T$ ( which is achieved through appropriate local unitary operations %%@
on the qubits).

To determine the absolute values of the state parameters, $s_1,\,s_2,\,s_3,$ the invariants $I_{4-6}$ of Table %%@
\ref{tab:1} are used: 
\begin{eqnarray}
\label{s1s2s3}
I_4&=&s^{\rm T}\, s=s_1^2+s_2^2+s_3^2, \nonumber \\
I_5&=&s^{\rm T}\, T\, T^{\rm T}\, s=s_1^2\, t_1^2+s_2^2\, t_2^2+s_3^2\, t_3^2, \nonumber \\
I_6&=&s^{\rm T}\, (T\, T^{\rm T})^2\, s= s_1^2\, t_1^4+s_2^2\, t_2^4+s_3^2\, t_3^4
\end{eqnarray}
(Here, $T$ is considered to be nondegenerate, i.e., $t_1\neq t_2\neq t_3$). 
The absolute values of state parameters $r_1,\,r_2,\,r_3,$ can be determined via the 
invariants $I_{7-9}$ of Table \ref{tab:1}: 
\begin{eqnarray}
\label{r1r2r3}
I_7&=&r^{\rm T}\, r
=r_1^2+r_2^2+r_3^2, \nonumber \\
I_8&=&r^{\rm T}\, T\, T^{\rm T}\, r=r_1^2\, t_1^2+r_2^2\, t_2^2+r_3^2\, t_3^2,  \nonumber \\
I_9&=&r^{\rm T}\, (T\, T^{\rm T})^2\, r= r_1^2\, t_1^4+r_2^2\, t_2^4+r_3^2\, t_3^4.
\end{eqnarray}
Now, the invariant, $I_{10}\,\,(I_{11})$ is useful to fix the overall sign of $s_1,\, s_2,$ and $s_3$\,\,($r_1,\, %%@
r_2,$ and $r_3$):  
\begin{eqnarray}
\label{signs}
I_{10}&=&\epsilon_{ijk}\, s_i\, (T\, T^{\rm T}\, s)_j\,
([T\,T^{\rm T}]^2\,s)_k \nonumber\\
&=&\left(t_1^4\,(t_3^2-t_2^2)+t_2^4\,(t_1^2-t_3^2)+t_3^4\,(t_2^2-t_1^2)\right)s_1\,s_2\,s_3, \nonumber\\
%\end{eqnarray}
%\begin{eqnarray}
%\label{signr}
I_{11}&=&\epsilon_{ijk}\, r_i\,(T^{\rm T}\, T\, r)_j\, ([T^{\rm T}\,T]^2\,r)_k \nonumber\\
&=&\left(t_1^4\,(t_3^2-t_2^2)+t_2^4\,(t_1^2-t_3^2)+t_3^4\,(t_2^2-t_1^2)\right)r_1\,r_2\,r_3. 
\end{eqnarray}
Furthermore, the relative signs between $s_ir_i$ are determined using the invariants
$I_{{12}-{14}},$ 
\begin{eqnarray}
I_{12}&=&s^{\rm T}\,T\, r=s_1\,r_1t_1+s_2\,r_2 t_2+s_3\,r_3 t_3,  \nonumber\\
I_{13}&=&s^{\rm T}\,T\,T^{\rm T}\,T\, r=s_1r_1t_1^3+s_2r_2t_2^3+s_3r_3t_3^3, \nonumber\\
I_{14}&=&\epsilon_{ijk}\,\epsilon_{lmn}\, s_i\,r_l\, t_{jm}\, t_{kn}=s_1r_1\,t_2\,t_3+s_2r_2\, t_1\, t_3+s_3r_3\, %%@
t_1\, t_2,
\end{eqnarray}
which provide three linear constraints on $s_1r_1,\,s_2r_2$ and $s_3r_3.$

Next, the individual signs of $(s_1,\,s_2,\,s_3)$ and $(r_1,\,r_2,\,r_3)$
can also be determined when, atleast two components, say, $s_1,\,s
_2$ are nonzero, and $s_3=0.$ In this case, the signs of $s_1$ and $s_2$ 
can be made positive with the help of local rotations~\footnote{Note that the diagonal form of $T$ remains unchanged
under $R(\pi)^i\otimes R(\pi)^i$ with $i=1,2,3.$} $(R(\pi)^1\otimes R(\pi)^1),
\,(R(\pi)^2\otimes R(\pi)^2)$. 
So, with $s_1,\,s_2>0$ and $s_3=0$ the invariant $I_{15}$ of Table \ref{tab:1} has the form
\begin{eqnarray}
\label{I15}
I_{15}&=&s_1s_2t_3r_3[t_2^2-t_1^2],
\end{eqnarray}
thus fixing the sign of $r_3,$ provided $t_3\neq 0.$
If $t_3=0,$ the invariant $I_{17}$ ( which is evaluated for $s_1,\,s_2>0$ and $s_3=t_3=0$) 
\begin{eqnarray}
I_{17}&=&s_1s_2t_1t_2r_3[t_2^2-t_1^2] 
\end{eqnarray}
is utilized for determining the sign of $r_3.$

A similar argument is applicable using - the invariants $I_{{11},{16},{18}}$  for fixing the sign of $s_3\,(s_3\neq %%@
0)$, when $r_1$ and $r_2$ are nonzero and $r_3=0.$

  %%%%%%%%%%%%%%%%%%%%%%%%%%%%%%%%%%%%%%%%%%%%%%%%%%%%%%%%%%%%%%%%%%%%%%%%%%%%%%
%%%%%%%%%%%%%%%%%%%%%%%%%%%%%%%%%%%%%%%%%%%%%%%%%%%%%%%%%%%%%%%%%%%%%%%%%%%%%%
%%%%%%%%%%%%%%%%%%%%%%%%%%%%%%%%%%%%%%%%%%%%%%%%%%%%%%%%%%%%%%%%%%%%%%%%%%%%%%
   \backmatter

\chapter{List of Publications}
\label{publications}
\markboth{}{List of Publications}

\noindent \underbar{\bf Journals:}
\bigskip

\begin{enumerate}

\item{{\bf Non-local properties of a symmetric two-qubit system} 

\medskip 
 A. R. Usha Devi, {\bf M. S. Uma},  R. Prabhu and Sudha
\smallskip 

{\em J. Opt. B: Quantum Semiclass. Opt.}  {\bf 7} (2005) S740-S744. } 

\item{{\bf Local invariants and pairwise entanglement in symmetric multi-qubit system}

\medskip 
 A. R. Usha Devi, {\bf M. S. Uma}, R. Prabhu and Sudha
\smallskip 

{\em Int. J. Mod. Phys. B.} {\bf 20} (2006) 1917-1933.}

\item{{\bf Non-classicality of photon added coherent and thermal radiations}

\medskip 

A. R. Usha Devi, R. Prabhu and {\bf M. S. Uma}
\smallskip 

{\em Eur. Phys. J. D} {\bf 40} (2006) 133-138.}

\item{{\bf Constraints on the uncertainties of entangled symmetric qubits} 

\medskip 
A. R. Usha Devi, {\bf M. S. Uma}, R. Prabhu and A. K. Rajagopal
\smallskip 

{\em Phys. Lett. A} {\bf 364} (2007) 203-207.}

\end{enumerate}
\newpage

\noindent \underbar{\bf Conferences/Symposia/Workshop}

\bigskip
\begin{enumerate}

\item{{\bf Pairwise entanglement properties of a symmetric multi-qubit system}

\medskip

A. R. Usha Devi, {\bf M. S. Uma}, R. Prabhu and Sudha
\smallskip

XVI-DAE-BRNS High Energy Physics Symposium held at Saha Institute of Nuclear Physics, Kolkata, India, during 
29th November to 3rd December 2004. }

\item{{\bf Non-local properties of a symmetric two-qubit system}

\medskip 

A. R. Usha Devi, {\bf M. S. Uma}, R. Prabhu and Sudha
\smallskip

Seventh International Conference on Photoelectronics, Fiber Optics and Photonics held at International 
School of Photonics, Cochin University of Science and Technology, Kochi, India, during 9-11 December 2004. }

\item{{\bf Separability, negativity, concurrence and local invariants of symmetric two qubit states}

\medskip

A. R. Usha Devi, {\bf M. S. Uma}, R. Prabhu and Sudha
\smallskip

International Conference on Squeezed States and Uncertainty Relations (ICSSUR'05) held at Besancon, France,  during 
2-6 May 2005. }

\item{{\bf Nonclassicality of photon added Gaussian light fields}

\medskip

A. R. Usha Devi, R. Prabhu and {\bf M. S. Uma}
\smallskip

Second International Conference on Current Developments in Atomic, Molecular
and Optical Physics with Applications (CDAMOP'06) held at Delhi University, New Delhi, during 
21-23 March 2006.}

\end{enumerate}

\end{document}